\documentclass[12pt,a4paper]{article}

\usepackage[final]{graphicx}
\usepackage{amsmath}

\renewcommand{\theequation}{\arabic{section}.\arabic{equation}}%
\newcommand{\req}[1]{(\ref{#1})}

\newcommand{\eps}{\epsilon}

\usepackage{amssymb}
\newcounter{comment}

{\refstepcounter{comment}%
\begin{quote}
\ttfamily\small$\blacksquare$ \textbf{\underline{Comment} $\sharp$\thecomment:}}%
{\end{quote}}

{%\refstepcounter{comment}
\begin{quote}
\ttfamily\small$\blacktriangleright$ \textbf{\underline{Reply} $\sharp$\thecomment:}}%
{\end{quote}}
\begin{document}

\begin{titlepage}

\centerline{\large \bf Nucleon Form Factors to Next-to-Leading Order
} \vspace{2mm}
\centerline{\large \bf with Light-Cone Sum Rules   }

\vspace{7mm}

\centerline{\bf K.~Passek-Kumeri{\v c}ki$^{a,b}$,
G. Peters$^a$}

\vspace{4mm} \centerline{\it $^a$Institut f\"ur Theoretische
Physik, Universit\"at Regensburg} \centerline{\it D-93040
Regensburg, Germany}

\vspace{5mm} \centerline{\it $^b$Theoretical Physics Division,
Rudjer Bo{\v s}kovi{\'c} Institute} \centerline{\it P.O.Box 180,
HR-10002 Zagreb, Croatia}

\vspace{5mm}

\centerline{\bf Abstract}

\vspace{0.5cm}

\noindent
We have calculated 
the leading-twist next-to-leading order (NLO),
i.e., ${\cal O}(\alpha_s)$, 
correction
to the light-cone sum rules prediction for the
electromagnetic form factors of the nucleon.
We have used
the Ioffe nucleon interpolation current
and worked in $M_N=0$ approximation, with $M_N$ being the mass of the 
nucleon. 
In this approximation, only the Pauli form factor $F_2$ 
receives a correction
and the calculated correction is quite sizable (cca 60\%).
The numerical results for the proton form factors
show the improved agreement with the experimental data.
We also discuss the problems encountered when going away from $M_N=0$
approximation at NLO, as well as, gauge invariance of the perturbative 
results. This work presents the first step towards the NLO accuracy 
in the light-cone sum rules for baryon form factors.

\vspace*{12mm}

\noindent  {\bf PACS numbers:} 
%12.38.-t,
13.40.Gp,
14.20.Dh,
12.38.Lg,
12.38.Bx

\vspace{3mm}


\begin{flushleft}
{\small
%Last updated May, 9th, 2008 
%by K.P-K.
%by G.P. 
}
\end{flushleft}

\end{titlepage}

\tableofcontents

%%%%%%%%%%%%%%%%%%%%%%%%%%%%%%%%%%%%%%%%%%%%%%%%%%%%%%%%%%%%%%%%%%%%%%%%%%%%%%%
%%%%%%%%%%%%%%%%%%%%%%%%%%%%%%%%%%%%%%%%%%%%%%%%%%%%%%%%%%%%%%%%%%%%%%%%%%%%%%%
\section{Introduction}
%%%%%%%%%%%%%%%%%%%%%%%%%%%%%%%%%%%%%%%%%%%%%%%%%%%%%%%%%%%%%%%%%%%%%%%%%%%%%%%
%%%%%%%%%%%%%%%%%%%%%%%%%%%%%%%%%%%%%%%%%%%%%%%%%%%%%%%%%%%%%%%%%%%%%%%%%%%%%%%

% Uvod
Exclusive processes offer challenging ground
for quantum chromodynamics (QCD) and,
especially the ones involving hadron form factors,
provide us with valuable
insight in the internal structure of composite particles. 
The simplest probe is the photon and thus obtained
electromagnetic form factors characterise 
hadron's spatial charge and current distributions.

% BL picture, GPDs 
The framework for analyzing
exclusive processes at large-momentum transfer 
%(so--called, hard processes)
within the context
of perturbative QCD (pQCD)
has been developed in the late seventies
\cite{ChernyakZ77,ChernyakZ80,Radyushkin77,EfremovR80,EfremovR80a,LepageB79,LepageB80,DuncanM80,DuncanM80a}.
It was demonstrated, to all orders in perturbation theory,
that exclusive amplitudes involving
large-momentum transfer, i.e., so-called, hard-scattering
amplitudes, factorize into a convolution
of a process-independent and perturbatively incalculable
soft part, i.e., distribution amplitude 
(one for each hadron involved in the amplitude), with
a process-dependent and perturbatively calculable
elementary hard-scattering amplitude.
In the leading-twist approximation of the standard hard-scattering approach, 
hadron is regarded
as consisting only of valence Fock states, and transverse quark momenta
are neglected (collinear approximation) as well as quark masses.
In this picture each hard gluon exchange brings factor 
$\alpha_s/\pi$, while higher-twist effects are suppressed
by $1/Q^2$, with $Q^2$ being the characteristic large scale of the process
(i.e., in the case of electromagnetic form factors, that is the virtuality 
of the photon).
Although this pQCD approach 
undoubtedly represents an adequate and efficient tool for analyzing
exclusive processes at very large momentum transfer,
its applicability at experimentally accessible
momentum transfers has been long debated and attracted much attention.
Even in a moderate energy region (a few GeV)
soft contributions (resulting from the competing, so-called, 
Feynman mechanism) can still be substantial, 
although the estimation of their size is model dependent.
Recently, the concept of generalized parton distributions (GPDs)
\cite{MuellerRGDH94,Radyushkin96,Ji96} has been introduced
to describe the soft part in various exclusive processes
(like deeply-virtual Compton scattering, 
deeply-virtual electroproduction of mesons \ldots) and make a connection
between inclusive and exclusive processes and corresponding
characteristic quantities like parton distribution functions 
(PDFs) and form factors.
Although more general, that approach 
(for details, see reviews \cite{Diehl03,BelitskyR05})
is basically similar to previously described pure pQCD approach, 
only at, for example, leading-twist the hard-scattering part does not 
involve all Fock state partons
but instead one uses the so-called ''hand-bag'' picture.

% sum rules
The QCD sum rule approach
\cite{ShifmanVZ79,ShifmanVZ79a}
applied to the pion form factor supports the conclusion
that the soft contributions are dominant
at moderate momentum transfers up to
$Q^2 \approx 2-3$ GeV$^2$ \cite{NesterenkoR82,IoffeS83}.
The application of the method at higher $Q^2$ faces 
the problem of an ill-behaved series in $Q^2/M_B^2$,
where $M_B^2$ is Borel parameter.
Moreover, for nucleon form factors the QCD sum rule 
approach only works in the region of small momentum
transfers $Q^2 < 1$ GeV$^2$ \cite{BelyaevK93,CastilloDL03}.
One can find in the literature 
many approaches and attempts to circumvent these problems.

% LCSR
The light-cone sum rule (LCSR) approach
\cite{BalitskyBK89,ChernyakZ90},
adopted also in this work, can be regarded as
a successful technique which combines
sum rule principles with pure perturbative QCD approach
advocated earlier. 
The domain of validity extends above few GeV$^2$.
In LCSR approach the ''soft'' contributions to the form factors
are calculated in terms of the same DAs that enter the pQCD
calculation and there is no double counting.
Hence, although LCSRs do involve a certain model dependence,
the important advantage of this approach lies in the fact
that it is fully consistent with pQCD.
% and apart from DAs no other nonperturbative parameters enter.
In the last years it has been widely applied
to mesons, see \cite{Braun98,ColangeloKh00} for reviews.
Moreover in Refs. \cite{BraunFMS00,BraunLMS01}
LCSR approach was introduced for the description of
nucleon DAs and nucleon form factors,
and further analysis followed in Refs. 
\cite{LenzWS03,BraunLW06} and \cite{WangWY06,WangWY06a}. 
The weak decay $\Lambda_b \to p l \nu_l$ was considered
in \cite{HuangW04} and the $N \gamma \Delta$ transition form
factor was worked out in \cite{BraunLPR05}.

As explained in detail in, for example, Ref. \cite{BraunLW06},
the basic object of LCSR approach to, say, nucleon form factors,
is a correlation function
expressed in terms of the matrix element of the time ordered
product of the current of interest (in our case, electromagnetic
current) and a suitable nucleon interpolation current.
The matrix element is taken not between the vacuum states
but between vacuum and nucleon state $\left|N(P) \right >$, which represents
the second nucleon in the process. 
When both the virtuality of the photon $q^2=-Q^2$ and the 
momentum flowing through the nucleon interpolation current
vertex $P'^2=(P-q)^2$ are large and negative,
one can use the operator product expansion (OPE)
on the light cone, i.e., one employs
pQCD to evaluate the Wilson coefficients, while the matrix elements
of the relevant composite operators correspond to the appropriate
moments of the nucleon DAs. This procedure is quite analogous to
the determination of the hard-scattering amplitudes in the pure pQCD
approach discussed above.
Furthermore, in order to access the nucleon form factors, 
one then makes use of the dispersion relation in $P'^2$ and define 
the nucleon form factor contribution through the, so-called, interval of duality 
or continuum threshold $s_0$. The usual application of Borel
transformation facilitates further the calculation.

There are some important features of this approach to be stressed.
The leading-order (LO) contribution to the form factor is a purely soft 
contribution which is represented as a sum of terms ordered by twist%
\footnote{In this work under twist we assume light-cone 
and not geometric twist.} 
of the operators, i.e., nucleon DAs. 
Moreover, in contrast to the pQCD hard-scattering approach,
the contributions of higher-twist DAs are not suppressed by $Q^2$ but by
powers of $P'^2$, i.e., by powers of Borel parameter 
$M_B^2 \approx 1-2$ GeV$^2$. 
Thus their role is more pronounced. 
%even at higher $Q^2$.
Furthermore, the LCSR expansion  contains terms generating 
also the asymptotic pQCD contributions.
For the pion form factor the hard-scattering contributions appear 
at order $\alpha_s$, 
and in Refs. \cite{BraunKM99,BijnensK02} it was explicitly
demonstrated that  they are correctly reproduced.
For the nucleon form factors they appear at order $\alpha_s^2$.

% NLO 
Let us now turn to the next-to-leading (NLO) contributions.
It is well known that, unlike in QED, the
leading-order predictions in pQCD
do not have such predictive power, and that higher-order corrections
are important.
Still, although the LO predictions within the hard-scattering approach
(as well as, GPD based approach)
have been obtained for many exclusive processes,
only a few processes have been analyzed at the NLO -- 
see the detailed account 
in, for example, 
Ref.  \cite{Passek04}, and additionally
Refs. \cite{IvanovSK04,Mueller05,KumerickiMPS06}.
Similarly, as it was stressed in, for example,
\cite{BraunLMS01} the LO LCSRs may not be sufficiently accurate.
The radiative gluon corrections to LCSRs
were calculated for number of processes involving
mesons, i.e., pion form factor \cite{BraunKM99},
pion transition form factor \cite{SchmeddingY99},
the decay $B \to \pi e \nu$ \cite{BaganBB98,BallZ01},
$B$ to $\pi$ ($K$, $\eta$) form factor
\cite{BelyaevKR93,KhodjamirianRWY97,BallZ04,DuplancicKMMO08},
and $B$ to light-vector meson ($\rho$, $\omega$, $K^*$, $\phi$)
form factors
\cite{BallB98,BallZ04a}.
The radiative corrections to nucleon form factors 
have not been evaluated either in hard-scattering picture
nor in the LCSR approach.

% what we do here
In this work we took a task of calculating the NLO
corrections to LCSRs for nucleon form factors.
We follow closely Ref. \cite{BraunLW06} and extend 
the formalism to NLO calculation. 
Even at LO the LCSR formalism for baryons is considerably more 
cumbersome than for mesons.
As we shall show, at NLO this is even more pronounced.
For example,
while for mesons 
at next-to-leading twist, i.e., twist-3, 
the use of asymptotic DAs 
along with Windzura-Wilczek approximation
ensured the cancellation of the collinear singularities
without explicitly knowing evolution kernels 
(see, for example, \cite{DuplancicKMMO08}),
the nature of nucleon form factor calculation,
nucleon DAs and corresponding  asymptotic forms is 
quite different and does not enable similar simple cancellations. 
Actually, as we shall show, 
at the moment,
only the approximation $M_N=0$, with $M_N$ being the nucleon mass,
allows fully consistent NLO calculation.
%In such an approximation only the leading twist, 
%that is twist-3 in nucleon case, contributes
%and the nucleon DAs kernels needed for the check of the
%cancellation of the collinear singularities are known.
%By switching-on the nucleon mass, which is, of course necessary
%in order to determine higher-twists, we are stuck with
%mixing of the contributions corresponding to different twists.
%This mixing can be seen even at LO through the check of gauge
%invariance with respect to photon.  
Hence, in this work we calculate NLO corrections to leading twist,
that is twist-3 in nucleon case, in $M_N=0$ approximation,
and outline the problems encountered when going away from this
approximation. We furthermore analyze the numerical
importance of such NLO corrections applying them to 
the assessment of proton form factors.

% plan of the paper
The paper is organized as follows.
In Sec. \ref{sec:def} we
give necessary definitions, 
introduce LCSR formalism
and explain the preliminaries.
The general LO formulas and detailed calculation
of twist-3 and twist-4 contribution to the correlation
function are presented in Sec. \ref{sec:LO}.
The gauge invariance of these LO results is also discussed.
The calculation of the NLO corrections to the correlation function
is explained in detail in Sec. \ref{sec:NLO}.
Section \ref{sec:LCSR} is devoted to developing needed LCSRs,
and numerical results are presented and discussed in Sec. \ref{sec:num}.
We summarize and conclude in Sec. \ref{sec:concl}.
There are five appendices devoted to some more technical details
or summary of analytical results.
In App. \ref{sec:appFEYN} the Feynman rules derived for the leading-twist
calculation and employed in LO and NLO calculation are listed.
The discussion of $\gamma_5$ ambiguity relevant for our NLO calculation
is given in App.  \ref{sec:gamma5-app}.
Appendix \ref{sec:appBfunM0} is devoted to the summary 
of the analytical NLO results used in numerical calculations.
The imaginary parts of selected functions needed in evaluating LCSRs
are derived in App. \ref{sec:appIm},
while the selected LO twist-3 and twist-4 contributions
and corresponding LCSRs are reanalyzed and listed in App. \ref{sec:LOMN}.
%%%%%%%%%%%%%%%%%%%%%%%%%%%%%%%%%%%%%%%%%%%%%%%%%%%%%%%%%%%%%%%%%%%%%%%%%%%%%%%
%%%%%%%%%%%%%%%%%%%%%%%%%%%%%%%%%%%%%%%%%%%%%%%%%%%%%%%%%%%%%%%%%%%%%%%%%%%%%%%
\section{Definitions and preliminaries}
\label{sec:def}
%%%%%%%%%%%%%%%%%%%%%%%%%%%%%%%%%%%%%%%%%%%%%%%%%%%%%%%%%%%%%%%%%%%%%%%%%%%%%%%
%%%%%%%%%%%%%%%%%%%%%%%%%%%%%%%%%%%%%%%%%%%%%%%%%%%%%%%%%%%%%%%%%%%%%%%%%%%%%%%

%%%%%%%%%%%%%%%%%%%%%%%%%%%%%%%%%%%%%%%%%%%%%%%%%%%%%%%%%%%%%%%%%%%%%%%%%%%%%%%
\subsection{Nucleon electromagnetic form factors}
%%%%%%%%%%%%%%%%%%%%%%%%%%%%%%%%%%%%%%%%%%%%%%%%%%%%%%%%%%%%%%%%%%%%%%%%%%%%%%%

The nucleon electromagnetic form factors are defined through
the matrix element of the electromagnetic current by
\begin{eqnarray}
\left< N(P')| j_{\mu}^{\rm em}(0) | N(P) \right>
&=& \bar{N}(P') \,
\left[
\gamma_{\mu} \, F_1(Q^2) - i
\frac{\sigma_{\mu \nu} q^{\nu}}{2 M_N} \,
F_2(Q^2) 
\right] \,
N(P) \, ,
\label{eq:emff}
\end{eqnarray}
where
$F_1$ and $F_2$ are Dirac and Pauli electromagnetic form factors, respectively.
Here, 
%\begin{equation}
$
P'=P-q
$
%\end{equation}
is the outgoing nucleon momentum, while $P$ and $q$ are
incoming nucleon and photon momenta, respectively.
Furthermore, 
%\begin{equation}
$
q^2=-Q^2
$
%\end{equation}
with $Q^2 \ge 0$ 
(for spacelike regime we are interested in, while the sign changes 
in the timelike region), and for on-shell nucleons $P^2=P'^2=M_N^2$.

The Sachs form factors, i.e., 
electric and magnetic form factors $G_{\rm E}$ and $G_{\rm M}$,
are related to $F_1$ and $F_2$ by
\begin{eqnarray}
G_{\rm E}(Q^2) &=&  F_1(Q^2) - \frac{Q^2}{4 M_N^2} F_2(Q^2)
\, ,
  \nonumber  \\
G_{\rm M}(Q^2) &=&  F_1(Q^2) + F_2(Q^2)
\, .
\end{eqnarray}
At $Q^2=0$ they are normalized to proton and neutron
electric charges and anomalous magnetic moments: 
\begin{eqnarray}
&& G_{\rm E}^p(0)=F_1^p(0)=1 \, , \nonumber \\
&& G_{\rm E}^n(0)=F_1^n(0)=0 
\, ,
\end{eqnarray}
and
%\begin{eqnarray}
%G_{\rm M}^p(0)= \mu_p=2.79 \, , & & F_2^p(0)=\kappa_p=1.79 \, ,\nonumber \\
%G_{\rm M}^n(0)= \mu_n=-1.91 \, , & & F_2^n(0)=\kappa_n=-1.91 
%\, ,
%\end{eqnarray}
\begin{equation}
\begin{array}{ll}
G_{\rm M}^p(0)= \mu_p=2.79 \, , &  
F_2^p(0)=\kappa_p=1.79 \, , \\
G_{\rm M}^n(0)= \mu_n=-1.91 \, , & 
F_2^n(0)=\kappa_n=-1.91 
\, ,
\end{array}
\end{equation}
respectively.
%\begin{equation}
%\begin{array}{lll}
%G_{\rm E}^p(0)=F_1^p(0)=1 \, , & 
%G_{\rm M}^p(0)= \mu_p=2.79 \, ,  &
%F_2^p(0)=\kappa_p=1.79 \, , \\
%G_{\rm E}^n(0)=F_1^n(0)=0 \, , &
%G_{\rm M}^n(0)= \mu_n=-1.91 \, , &  
%F_2^n(0)=\kappa_n=-1.91 
%\end{array}
%\end{equation}

%%%%%%%%%%%%%%%%%%%%%%%%%%%%%%%%%%%%%%%%%%%%%%%%%%%%%%%%%%%%%%%%%%%%%%%%%%%%%%%
\subsection{Correlation function}
%%%%%%%%%%%%%%%%%%%%%%%%%%%%%%%%%%%%%%%%%%%%%%%%%%%%%%%%%%%%%%%%%%%%%%%%%%%%%%%

\begin{figure}
%\centerline{\epsfxsize5cm\epsffile{figLCSR.eps}}
\centerline{\includegraphics{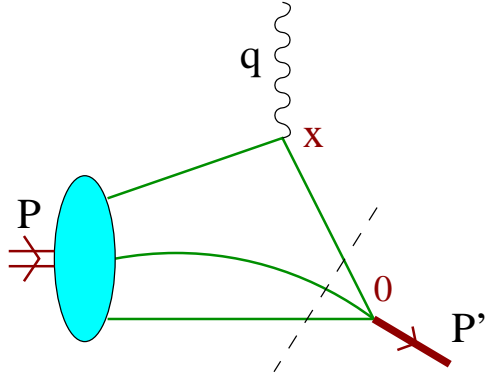}}
\caption{
Schematic structure of the light-cone sum rule for nucleon form factors
(Ref. \protect\cite{BraunLW06}).
The corresponding correlator function is given in 
\protect\req{eq:correlator}, and $P'=P-q$.
}
\label{f:correlator}
\end{figure}

Correlation function, 
the basic object used in LCSR approach 
%in order to access nucleon form factors
(see Fig. \ref{f:correlator} for a schematic representation),
is defined by
%%%%%%%%%%%%%%%%%%%%%%%%%%%%
%%%%%%%%%%%%%%%%%%%%%%%%%%%%%
\begin{equation}
       T_{\mu} (P,q) 
= i\! \int\! d^4 x \, e^{i q x}
\left\langle 0 \right| T\left[\eta(0) 
       j_{\mu}(x) 
\right] 
\left| N(P) \right \rangle
\, ,
\label{eq:correlator}
\end{equation}
with $P$ and $q$ being the nucleon and 
photon momentum, 
respectively.
Here $j_\mu$ is the electromagnetic current 
\begin{equation}
j_{\mu}(x) =
      e_u 
\bar{u}(x) \gamma_{\mu} 
u(x) 
+
      e_d 
\bar{d}(x) \gamma_{\mu} 
d(x) 
\, ,
\label{eq:current}
\end{equation}
and in this work we choose the interpolating nucleon current of the form
\begin{equation}
\eta(x) = \varepsilon^{abc} 
\left[u^a(x)\; \Gamma_1  u^b(x)\right]\; \Gamma_2 \; d^c(x)\,,
\label{eq:nucleon-curr}
\end{equation}
where $a,b,c$ are colour indices.
Here the generic expression
\req{eq:nucleon-curr} actually corresponds to the proton current,
while the neutron current is obtained by replacing 
$u \leftrightarrow -d$.
Specially, for Ioffe current \cite{Ioffe81}
\begin{equation}
\Gamma_1=C\gamma_\lambda \, ,
\qquad
\Gamma_2=\gamma_5 \gamma^\lambda
\, ,
\label{eq:Ioffecurr}
\end{equation}
with $C$ being the charge conjugation matrix. 
Other choices for $\eta(x)$ were discussed in, 
for example, Ref. \cite{BraunLW06}, and Ioffe current was singled out
as a most promising candidate for reliable sum rules.

%%%%%%%%%%%%%%%%%%%%%%%%%%%%%%%%%%%%%%%%%%%%%%%%%%%%%%%%%%%%%%%%%%%%%%%%%%%%%%%
\subsection{Nucleon matrix element of the three-quark operator}
%%%%%%%%%%%%%%%%%%%%%%%%%%%%%%%%%%%%%%%%%%%%%%%%%%%%%%%%%%%%%%%%%%%%%%%%%%%%%%%

Nucleon distribution amplitudes refer to nucleon-to-vacuum matrix elements of
nonlocal operators built of quark and gluon fields at light-like separations,
$x^2=0$. 

We are interested in the three-quark matrix element
\begin{equation}
4 \left \langle 0 \right |
\varepsilon^{abc} u^{a'}_{\alpha}(a_1 x) [a_1 x, a_0 x]_{a'a}
               u^{b'}_{\beta}(a_2 x) [a_2 x, a_0 x]_{b'b}
               d^{c'}_{\gamma}(a_3 x) [a_3 x, a_0 x]_{c'c}
\left | N(P,\lambda ) \right \rangle
\end{equation}
where
\begin{equation}
[x,y]={\cal P} \mbox{exp}[ i g \int_0^1 dt (x-y)_{\mu} A^{\mu}(t x - (1-t) y) ]
\end{equation}
is a gauge factor which we will suppress in the following,
while $\alpha$,$\beta$,$\gamma$ are Dirac indices and $P^2=M_N^2$ 
with $M_N$ being the mass of the nucleon, i.e., proton mass for the
three-quark matrix element considered above. Neutron case proceeds
equivalently 
($u \leftrightarrow d$).
Actually, in this work all results are expressed in terms of proton
quantities but the neutron case is easy to derive from these
(in the end results, it comes basically 
to the replacement $e_u \leftrightarrow e_d$).

For the general Lorentz decomposition 
we introduce a convenient shorthand notation \cite{Wittmann}
\begin{equation}
4 \left \langle 0 \right |
\varepsilon^{abc} u_\alpha^a(a_1 x) 
               u_\beta^b(a_2 x) 
               d_\gamma^c(a_3 x) 
\left | N(P,\lambda) \right \rangle
= \sum \limits_i {\cal F}^{(i)}(\{a_k\},P\cdot x) \; X^{(i)}_{\alpha \beta} Y^{(i)}_{\gamma}
\, ,
\label{eq:gLdecomp}
\end{equation}
where 
${\cal F}^{(i)}\in \{{\cal S}_1,{\cal S}_2,{\cal P}_1,{\cal P}_2,
{\cal V}_1,\ldots,{\cal V}_6,
{\cal A}_1,\ldots,{\cal A}_6,
{\cal T}_1,\ldots,{\cal T}_8\}$ are invariant functions of $P\cdot x$,
while $X^{(i)}_{\alpha \beta}$ and $Y^{(i)}_\gamma$
are Dirac structures which can be read from Ref. \cite{BraunLW06}.

The structures $Y^{(i)}_\gamma$ contain nucleon spinor 
$N_{\gamma}$ and we note that
\begin{equation}
(X^{(i)})^T=\left\{
\begin{array}{l}
X^{(i)} \mbox{ for } {\cal F}^{(i)} \in \{{\cal V}_j, {\cal T}_j \} \\
- X^{(i)} \mbox{ for } {\cal F}^{(i)} \in \{{\cal S}_j, {\cal P}_j, {\cal A}_j \}  
\end{array}
\right.
\, .
\label{eq:XT}
\end{equation}
The functions ${\cal F}^{(i)}$ are not of a  definite twist
and they satisfy
\begin{equation}
{\cal F}^{(i)}(a_1 P \cdot x, a_2 P \cdot x, a_3 P \cdot x)
=\left\{
\begin{array}{l}
{\cal F}^{(i)}(a_2 P\cdot x, a_1 P \cdot x, a_3 P \cdot x)
\mbox{ for } {\cal F}^{(i)} \in \{{\cal V}_j, {\cal T}_j \} \\
-{\cal F}^{(i)}(a_2 P\cdot x, a_1 P \cdot x, a_3 P \cdot x)
\mbox{ for } {\cal F}^{(i)} \in \{{\cal S}_j, {\cal P}_j, {\cal A}_j \}  
\end{array}
\right.
\, .
\label{eq:calFsym}
\end{equation}
Additional, ${\cal O}(x^2)$ terms can be added to
\req{eq:gLdecomp}
but will not be explicitly considered here 
(see Ref. \cite{BraunLW06} instead).

\subsubsection{Light-cone kinematics}
%%%%%%%%%%%%%%%%%%%%%%%%%%%%%%%%%%%%%%%%%%%%%%%%%%%%%%%%%%%%%%%%%%%%%%%%%%%%%%%

For the twist classification it is convenient to go to infinite momentum 
frame and introduce
a light-like vector $z_\mu$ by the condition
\begin{equation}
       q\cdot z =0\,,\qquad z^2 =0
\label{eq:z}
\end{equation}
as well as the second light-like vector 
\begin{equation}
p_\mu = P_\mu  - \frac{1}{2} \, z_\mu \frac{M_N^2}{P\cdot z}\,,~~~~~ p^2=0\,, 
\label{eq:vectors}
\end{equation}
so that $P \to p$ if the nucleon mass can be neglected, $M_N \to 0$.
The projector onto the directions orthogonal to $p$ and $z$,
is given by
\begin{equation}
       g^\perp_{\mu\nu} = g_{\mu\nu} -\frac{1}{pz}(p_\mu z_\nu+ p_\nu z_\mu).
\end{equation}
In turn, $a_\perp$ denotes the generic component of $a_\mu$ orthogonal to
$z$ and $p$,
and thus the photon momentum can be written as 
\begin{eqnarray}
q_{\mu}=q_{\bot \mu}+ z_{\mu}\frac{P\cdot q}{P\cdot z}\, ,
\end{eqnarray}
where the use has been made of $p\cdot q=P\cdot q$ and
$p\cdot z=P\cdot z$.

Assume for a moment that the nucleon moves in the positive 
${\bf e_z}$ direction, then $p^+$ and $z^-$ are the only nonvanishing 
components of $p$ and $z$, respectively. 
The infinite momentum frame can be visualised
as the limit $p^+ \sim Q \to \infty$ with fixed $P\cdot z = p \cdot z \sim 1$ 
where $Q$ is the large scale in the process.
Expanding the matrix element in powers of $1/p^+$ introduces
the power counting in $Q$. In this language,  twist counts the
suppression in powers of $p^+$. 

Similarly, 
the nucleon spinor $N_\gamma(P,\lambda)$ has to be decomposed 
in ``large'' and ``small'' components as
%\begin{eqnarray}
%N_\gamma(P,\lambda) &=& 
%\frac{1}{2 p\cdot z} \left(\!\not\!{p}\! \!\not\!{z} +
%\!\not\!{z}\!\!\not\!{p} \right) N_\gamma(P,\lambda)
%\nonumber\\&=& N^+_\gamma(P,\lambda) + N^-_\gamma(P,\lambda)\,,
%\label{eq:spinor}
%\end{eqnarray}
\begin{equation}
N_\gamma(P,\lambda) = 
(\Lambda^+ + \Lambda^-) N_\gamma(P,\lambda)
= N^+_\gamma(P,\lambda) + N^-_\gamma(P,\lambda)\,,
\label{eq:spinor}
\end{equation}
where we introduce two projection operators
\begin{equation}
\Lambda^+ = \frac{\!\not\!{p}\! \!\not\!{z}}{2 p\cdot z} \quad ,\quad
\Lambda^- = \frac{\!\not\!{z}\! \!\not\!{p}}{2 p\cdot z}
\label{eq:project}
\end{equation}
that project onto the ``plus'' and ``minus'' components of the spinor.
Using the explicit expressions for $N(P)$ it is easy to see 
that $\Lambda^+N = N^+ \sim \sqrt{p^+}$ while 
$\Lambda^-N = N^- \sim 1/\sqrt{p^+}$.

\subsubsection{Twist decomposition}
\label{sec:twistdecomp}
%%%%%%%%%%%%%%%%%%%%%%%%%%%%%%%%%%%%%%%%%%%%%%%%%%%%%%%%%%%%%%%%%%%%%%%%%%%%%%%

The twist decomposition of the nucleon-to-vacuum matrix element 
can be written in a form
\begin{equation}
4 \left \langle 0 \right |
\varepsilon^{abc} u_\alpha^a(a_1 x) 
               u_\beta^b(a_2 x) 
               d_\gamma^c(a_3 x) 
\left | N(P,\lambda) \right \rangle
= \sum \limits_i F^{(i)}(\{a_k\},P \cdot x) \; W^{(i)}_{\alpha \beta} V^{(i)}_{\gamma}
\, ,
\label{eq:twdecomp}
\end{equation}
where 
$F^{(i)}\in \{S_1,S_2,P_1,P_2,
V_1,\ldots,V_6,
A_1,\ldots,A_6,
T_1,\ldots,T_8\}$ represent now nucleon distribution amplitudes (DAs) 
and functions of definite twist:\\[0.2cm]
\begin{equation}
\begin{tabular}{l|l}
twist-3& $V_1$, $A_1$, $T_1$ \\
twist-4& $S_1$, $P_1$, $V_2$, $V_3$, $A_2$, $A_3$, $T_2$, $T_3$, $T_7$ \\
twist-5& $S_2$, $P_2$, $V_4$, $V_5$, $A_4$, $A_5$, $T_4$, $T_5$, $T_8$ \\
twist-6& $V_6$, $A_6$, $T_6$ 
\end{tabular}\\[0.2cm]
\label{eq:twist}
\end{equation}
The Dirac structures $W^{(i)}_{\alpha \beta}$ and $V^{(i)}_\gamma$
can be read from Ref. \cite{BraunLW06}, 
and $V^{(i)}_\gamma$ contain the $N_\gamma^+$ or $N_\gamma^-$ projections 
of the nucleon spinor.

The functions ${\cal F}^{(i)}$ and $F^{(i)}$ are related by
\cite{BraunLW06}
\begin{equation}
\begin{array}{ll}
{\cal S}_1=S_1 & 
2 P\cdot x {\cal S}_2=S_1 - S_2 \\
{\cal P}_1=P_1 & 
2 P\cdot x {\cal P}_2=P_2 - P_1 \\
{\cal V}_1=V_1 & 
2 P\cdot x {\cal V}_2=V_1 - V_2 - V_3 \\
2 {\cal V}_3=V_3 & \ldots \\ 
\ldots & \\
{\cal A}_1=A_1 & 
2 P\cdot x {\cal A}_2=-A_1 + A_2 - A_3 \\
2 {\cal A}_3=A_3 & \ldots \\ 
\ldots & \\
{\cal T}_1=T_1 & 
2 P\cdot x {\cal T}_2=T_1 + T_2 - 2 T_3 \\
\ldots & 
\end{array}
\label{eq:calFtoF1}
\end{equation}
and each DA $F^{(i)} \in V_j,A_j,T_j,S_j,P_j$ can be 
represented as 
\begin{equation}
F^{(i)}(\{a_k\}, P \cdot x) 
= \int \! {\cal D} x\, e^{-i P \cdot x \sum_j x_j a_j} 
              F(\{x_i\})\,,
\label{eq:fourier}
\end{equation}
where the functions $F^{(i)}(x_i)$ depend on the dimensionless
variables $x_i,\, 0 < x_i < 1, \sum_i x_i = 1$ which 
correspond to the longitudinal momentum fractions 
carried by the quarks inside the nucleon.  
The integration measure is defined as 
\begin{equation}
{}\hspace*{-3mm}\int\! {\cal D} x  = \!\int_0^1\!\!\! d x_1 d x_2 d x_3\, 
\delta (x_1 + x_2 + x_3 - 1)\,.\hspace*{5mm}\phantom{-}
\label{eq:integration}
\end{equation}
Analogously to \req{eq:calFsym}, the functions $F^{(i)}(\{x_k\})$
poses the following symmetry properties 
\begin{equation}
F^{(i)}(x_1, x_2, x_3)
=\left\{
\begin{array}{l}
F^{(i)}(x_2, x_1, x_3)
\mbox{ for } F^{(i)} \in \{V_j, T_j \} \\
-F^{(i)}(x_2, x_1, x_3)
\mbox{ for } F^{(i)} \in \{S_j, P_j, A_j \}  
\end{array}
\right.
\, .
\label{eq:Fsym}
\end{equation}

According to \req{eq:calFtoF1}, the functions ${\cal F}^{(i)}$ can be written 
in terms of nucleon DAs $F^{(i)}$ as
\begin{equation}
{\cal F}^{(i)}(\{a_k\}, P\cdot x) =
\frac{1}{(2 P\cdot x)^n} f(F^{(i)}(\{a_k\}, P\cdot x))
\, , \quad n\in\{0,1,2\}
\,,
\label{eq:calFtoF2}
\end{equation}
where $f$ is a linear combination of $F^{(i)}$.
In addition, from 
\req{eq:gLdecomp} and Ref. \cite{BraunLW06}
one can read off the dependence of Dirac structures
$X^{(i)}$ and $Y^{(i)}$ on $x$-coordinates:
\begin{equation}
X^{(i)} Y^{(i)} \sim
 1 \mbox{ or } x_{\kappa} \mbox{ or } x_{\kappa} x_{\rho}
\, .
\label{eq:casesXiYi}
\end{equation}
Taking into account \req{eq:calFtoF2} and \req{eq:casesXiYi} the terms 
from \req{eq:gLdecomp} can be classified according to\\ 
\begin{equation}
\begin{tabular}{l|l}
$n=0$ and $X^{(i)} Y^{(i)} \sim 1$ &
${\cal S}_1$, ${\cal P}_1$,
${\cal V}_1$, ${\cal V}_3$,
${\cal A}_1$, ${\cal A}_3$,
${\cal T}_1$, ${\cal T}_3$ \\
$n=1$ and $X^{(i)} Y^{(i)} \sim x_{\kappa}$  &
${\cal S}_2$, ${\cal P}_2$,
${\cal V}_2$, ${\cal V}_4$, ${\cal V}_5$,
${\cal A}_2$, ${\cal A}_4$, ${\cal A}_5$,
${\cal T}_2$, ${\cal T}_4$, ${\cal T}_5$, ${\cal T}_7$ \\
$n=2$ and $X^{(i)} Y^{(i)} \sim x_{\kappa} x_{\rho}$  &
${\cal V}_6$,
${\cal A}_6$,
${\cal T}_6$, ${\cal T}_8$ 
\end{tabular}\\
\label{eq:cases}
\end{equation}
After ${\cal F}^{(i)}$ functions are replaced by $F^{(i)}$
and the Fourier transform \req{eq:fourier} is employed,
one ends up with corresponding three types of integrals which in LO take the form:
\begin{eqnarray}
1. &&
\int d^4 x \; \int \frac{d^4 k_{\rm in}}{(2 \pi)^4}\; e^{i(q+k_{\rm in})\cdot x}
\int {\cal D}u \; e^{-i P \cdot x \sum_j u_j a_j} 
\;
[ F^{(i)}(\{u_k\}) \ldots ] \nonumber\\ 
2. &&
\int d^4 x \;  \int \frac{d^4 k_{\rm in}}{(2 \pi)^4}\;e^{i(q+k_{\rm in})\cdot x}
\int {\cal D}u \; e^{-i P \cdot x \sum_j u_j a_j} 
\;
\frac{x_{\kappa}}{(2 P\cdot x)}
\;
\label{eq:type2}
[ F^{(i)}(\{u_k\}) \ldots ] \nonumber\\ 
3. &&
\int d^4 x \;  \int \frac{d^4 k_{\rm in}}{(2 \pi)^4} \;e^{i(q+k_{\rm in})x}
\int {\cal D}u \; e^{-i P \cdot x \sum_j u_j a_j} 
\;
\frac{x_{\kappa} x_{\rho}}{(2 P\cdot x)^2}
\;
[ F^{(i)}(\{u_k\}) \ldots ] 
\, , \quad
\label{eq:types}
\end{eqnarray}
with $k_{in}$ being the momentum of the quark propagator.
Similar but slightly more complicated integrals appear at NLO.

The first integral in \req{eq:types}, 
corresponding to the first case in \req{eq:cases},
simplifies trivially to
\begin{eqnarray}
 &&
 \int \frac{d^4 k_{\rm in}}{(2 \pi)^4}\
\int {\cal D}u \; 
(2 \pi)^4 \delta^{(4)}(q+k_{\rm in}-P \sum_j u_j a_j)
[ F^{(i)}(\{u_k\}) \ldots ]  
\nonumber \\ & & \quad
= \int {\cal D}u \; 
[ F^{(i)}(\{u_k\}) \ldots ]_{k_{\rm in}=P \sum_j u_j a_j-q}  
\; \, ,
\label{eq:Int1}
\end{eqnarray}
It is convenient to introduce the notation
\begin{equation}
A(u_1,u_2,u_3) \otimes B(u_1,u_2,u_3)=
\int {\cal D}u \, A(u_1,u_2,u_3) \, B(u_1,u_2,u_3)
\, .
\label{eq:convol}
\end{equation}
and thus such contribution to the correlation function
\req{eq:correlator} 
can be expressed, 
both at LO and higher-orders, by a convolution
\begin{equation}
T_{\mu}= \sum \limits_i {\cal M}_{\mu}^{(i)}(\{u_k\}) \otimes F^{(i)}(\{u_k\})
\, .
%\, ,
\label{eq:convolI}
\end{equation}

In order to evaluate the other two integrals 
from \req{eq:types}
one employs 
{\allowdisplaybreaks \begin{eqnarray}
\int d^4 x \;  
x_{\kappa} \;
e^{i(q+k_{\rm in}-P\sum_j u_j a_j)\cdot x } 
 & = &
-i \frac{\partial}{\partial k_{\rm in}^{\kappa}}
\left( 
(2 \pi)^4 \delta^{(4)}(q+k_{\rm in}-P \sum_j u_j a_j)
\right)
\, ,
\nonumber \\
\int d^4 x \;  
x_{\kappa} x_{\rho}\;
e^{i(q+k_{\rm in}-P\sum_j u_j a_j)\cdot x } 
 & = &
\frac{\partial}{\partial k_{\rm in}^{\kappa}}
\frac{\partial}{\partial k_{\rm in}^{\rho}}
\left(
(2 \pi)^4 \delta^{(4)}(q+k_{\rm in}-P \sum_j u_j a_j)
\right)
\, ,  
\qquad
\label{eq:int23}
\end{eqnarray}
}
as well as
partial integration%
\footnote{ 
Employing partial integration
in the second and third term of \req{eq:types} 
one gets
\begin{eqnarray*}
\lefteqn{\int \! {\cal D} u
\, e^{-i Px \sum_j u_j a_j} F^{(i)}(\{u_k\})
\frac{1}{(2 P\cdot x)}}
\nonumber \\  & \rightarrow &
\frac{i}{2} (a_l - a_n) \; 
\int_0^1 d u_l \; 
\int_1^{u_l} d {v}_l \;
\int_0^{1-{v}_l} d {u}_m \;
\nonumber \\ & & \times
e^{-i P\cdot x \left( u_l (a_l-a_n)+u_m(a_m-a_n)+a_n \right)} 
  F^{(i)}(\{{v}_l,u_m,1-{v}_l-u_m\}) \, ,
\label{eq:int2}
 \\[0.2cm] & &
\mbox{all possible $l,m,n \in \{1,2,3\}$ permutations for which  $a_l-a_m\ne0$}
\nonumber
\end{eqnarray*}
and
\begin{eqnarray*}
\lefteqn{\int \! {\cal D} u
\, e^{-i Px \sum_j u_j a_j} F^{(i)}(\{u_k\})
\frac{1}{(2 P\cdot x)^2}}
\nonumber \\  & \rightarrow &
-\frac{1}{4} (a_l - a_n) \; 
\int_0^1 d u_l \; 
\int_1^{u_l} d {v}_l \;
\int_0^{v_l} d {w}_l \;
\int_0^{1-{w}_l} d {u}_m \;
\nonumber \\ & & \times
e^{-i P\cdot x \left( u_l (a_l-a_n)+u_m(a_m-a_n)+a_n \right)} 
  F^{(i)}(\{{w}_l,u_m,1-{w}_l-u_m\}) \, ,
\label{eq:int3}
\\[0.2cm] & &
\mbox{all possible $l,m,n \in \{1,2,3\}$ permutations for which  $a_l-a_m\ne0$}
\nonumber
\end{eqnarray*}
Note that 
the surface terms that should vanish in Borel transformation 
have been already neglected.},
but we will leave out here further details.

%%%%%%%%%%%%%%%%%%%%%%%%%%%%%%%%%%%%%%%%%%%%%%%%%%%%%%%%%%%%%%%%%%%
\subsection{Derivation of LCSR}
%%%%%%%%%%%%%%%%%%%%%%%%%%%%%%%%%%%%%%%%%%%%%%%%%%%%%%%%%%%%%%%%%%%

Equating the correlation function \req{eq:correlator} results in
several invariant functions that can be separated 
by the appropriate projections. 
Lorentz structures that are most useful for writing the LCSRs are usually 
those containing the maximum power of the large momentum $p^+ \sim p \cdot z$. 
Hence, for the Ioffe current, we define the invariant functions
${\cal A}$ and ${\cal B}$ by
\begin{eqnarray}
  z^{\mu} \Lambda_+ T_{\mu} &=& (p \cdot z)
  \left\{ M_N \mathcal{A} + 
    \!\not\!q_\perp \mathcal{B} \right\}N^+(P)\,,
\label{eq:projections}
\end{eqnarray}
where $\mathcal{A}$ and  $\mathcal{B}$ 
depend on the Lorentz-invariants $Q^2=-q^2$ and $P'^2 = (P-q)^2$.

%\subsubsection{Correlation function in terms of form factors}
\subsubsection{Correlation function versus form factors}

Next we relate the correlation function \req{eq:correlator}
to nucleon form factors.
The correlation function can be written as 
\begin{equation}
T_{\mu}(P,q)= \frac{1}{M_N^2 - P'^2} \sum_s
\left\langle 0 \right| \eta(0) \left| N(P',s) \right \rangle
\left< N(P', s)| j_{\mu}^{\rm em}(0) | N(P) \right>
+ \ldots
\end{equation}
where the leading term is the nucleon contribution
and the dots stand for higher resonances.

Inserting \req{eq:emff} and the matrix element of 
the Ioffe current 
\begin{equation}
\left\langle 0 \right| \eta(0) \left| N(P) \right \rangle
= \lambda_1 \, M_N \, N(P)
\, ,
\label{eq:Ioffematr1}
\end{equation}
using
$\sum_s N(P) \bar{N}(P)=\not \! P + M_N$,
and taking the projection suitable for our calculation,
one obtains
\begin{equation}
\Lambda_+ \, z_{\mu} \, T^{\mu}
=
\frac{\lambda_1}{M_N^2-P'^2}
\left\{
2 (P\cdot z) \,
M_N \, F_1(Q^2) \, N^+(P)
+
 (P\cdot z) \,
 F_2(Q^2) \,
\not \! q_{\perp} \, N^+(P)
\right\}
+
\, \cdots
\label{eq:lcsr1}
\end{equation}
Now, by comparing \req{eq:lcsr1} and \req{eq:projections}
one gets
\begin{eqnarray}
\frac{2 \lambda_1}{M_N^2-P'^2} F_1(Q^2) 
+ \cdots &=& {\cal A}(Q^2,P'^2)
\, , \nonumber \\
\frac{\lambda_1}{M_N^2-P'^2} F_2(Q^2)
+ \cdots &=& {\cal B}(Q^2,P'^2)
\, . 
\label{eq:lcsr2}
\end{eqnarray}

\subsubsection{Light-cone sum rules}
%\subsubsection{Dispersion relation}

We can calculate ${\cal A}$ and ${\cal B}$ perturbatively 
in terms of quarks and gluons.

Furthermore, (see, for example, Ref. \cite{ColangeloKh00})
we can formally write a dispersion relation%
\footnote{Here we use a non-subtracted dispersion relation
for which the condition
\begin{displaymath}
{\rm lim}_{s \to \infty} 
\begin{array}{c}{\cal A}\\{\cal B}\end{array}
(Q^2,s)=0
\end{displaymath}
has to be satisfied.
}
\begin{eqnarray}
{\cal A}(Q^2,P'^2)&=&\frac{1}{\pi}
\int_0^\infty ds \, \frac{{\rm Im} {\cal A}(Q^2,s)}{s-P'^2}
\, , \nonumber \\[0.2cm]
{\cal B}(Q^2,P'^2)&=&\frac{1}{\pi}
\int_0^\infty ds \, \frac{{\rm Im} {\cal B}(Q^2,s)}{s-P'^2}
\, .
\label{eq:disp}
\end{eqnarray}

Now, 
by making use of the quark-hadron duality
(analogously to, for example, Ref. \cite{Melic02})
the effects of higher-resonances
cancel between the left and right hand side of
\req{eq:lcsr2} 
and 
one ends up with 
the sum rules
\begin{eqnarray}
\frac{2 \lambda_1}{M_N^2-P'^2} F_1(Q^2) &=& 
\frac{1}{\pi}
\int_0^{s_0} ds \, \frac{{\rm Im} {\cal A}(Q^2,s)}{s-P'^2}
\, , \nonumber \\[0.2cm]
\frac{\lambda_1}{M_N^2-P'^2} F_2(Q^2) &=& 
\frac{1}{\pi}
\int_0^{s_0} ds \, \frac{{\rm Im} {\cal B}(Q^2,s)}{s-P'^2}
\, , 
\label{eq:LCSRa}
\end{eqnarray}
where $s_0$ is a convenient mass cut-off, 
which in our case corresponds to the continuum threshold
taken at Roper resonance,
$s_0 \approx (1.5 {\rm GeV})^2$ \cite{BraunLW06},
and which eliminates contributions other than nucleon 
(continuum subtraction).

%\subsubsection{Borel transformation}

In practice, one can imagine to perform a power expansion of expression
\req{eq:LCSRa} in the variable $P'^2$.
To improve the convergence of this expansion
one then employs Borel transformation 
\begin{equation}
\mbox{B}_{X \to M_B^2}[F(X)]=
\lim_{n \to \infty} \frac{(-X)^n}{\Gamma(n)}
\left[\frac{d^n}{dX^n} F(X) \right]_{|X|=n M_B^2,\, X \to \infty} 
%\, .
\, ,
\end{equation}
%In the following we give a table containing a few example functions
%(\cite{ColangeloKh00},\cite{Bakulev06}):\\\\
%\begin{equation}
%\begin{tabular}{l|l|l|l|l|l}
%$F(X)$ & $C$ & $C \ln(X/\mu^2)$& $X^n$ & $1/X^n$ & $1/(s-X)$ \\
%\hline
%$B[F(X)]$    & $0$ & $-C$ & 0 & $1/(\Gamma(n) M_B^{2n})$ & $e^{-s/M_B^2}/M_B^2$
%\end{tabular}
%\, .
%\label{eq:tabBorel}
%\end{equation}
%The last entry will be of particular use in further calculation.
and of particular use in further calculation is a Borel transformation
(see, for example, \cite{ColangeloKh00,Bakulev06})
\begin{equation}
\mbox{B}_{X \to M_B^2}[1/(x-X)]=e^{-x/M_B^2}/M_B^2
\, .
\label{eq:Borel-examp}
\end{equation}

%\subsubsection{Light-cone sum rules}

Hence, taking a Borel transformation of the left and right hand-side 
of Eqs. \req{eq:LCSRa}
in variables $X=P'^2$, one ends up 
with the sum rules
\begin{eqnarray}
 F_1(Q^2) &=& 
\frac{1}{2 \lambda_1 \pi }
\int_0^{s_0} ds \; e^{(-s+M_N^2)/M_B^2} \; {\rm Im} {\cal A}(Q^2,s)
\, , \nonumber \\[0.2cm]
 F_2(Q^2) &=& 
\frac{1}{\lambda_1 \pi }
\int_0^{s_0} ds \; e^{(-s+M_N^2)/M_B^2} \; {\rm Im} {\cal B}(Q^2,s)
\, . 
\label{eq:LCSRb}
\end{eqnarray}
For the Borel mass $M_B$ which can be viewed as a matching scale
of hadronic and partonic part of the calculation, 
we, as in Ref. \cite{BraunLW06}, take $M_B^2=2$ GeV$^2$.

We note here that in this draft we adopt the derivation of the sum rules
based on the dispersion relations \req{eq:disp} 
and Borel transformation %given by the last entry in \req{eq:tabBorel}, 
\req{eq:Borel-examp}
which then lead to \req{eq:LCSRb}.
After calculating the needed imaginary parts and performing the necessary
integrals it can be shown that this approach is equivalent to the one 
employed in Ref. \cite{BraunLW06} but perhaps more suitable for NLO
calculations.

The imaginary parts of the functions of interest are listed in 
App. \ref{sec:appIm}.

%%%%%%%%%%%%%%%%%%%%%%%%%%%%%%%%%%%%%%%%%%%%%%%%%%%%%%%%%%%%%%%%%%%%%%%%%%%%%%%
%%%%%%%%%%%%%%%%%%%%%%%%%%%%%%%%%%%%%%%%%%%%%%%%%%%%%%%%%%%%%%%%%%%%%%%%%%%%%%%
\subsection{$M_N$ dependence}
\label{sec:Mdependence}
%%%%%%%%%%%%%%%%%%%%%%%%%%%%%%%%%%%%%%%%%%%%%%%%%%%%%%%%%%%%%%%%%%%%%%%%%%%%%%%
%%%%%%%%%%%%%%%%%%%%%%%%%%%%%%%%%%%%%%%%%%%%%%%%%%%%%%%%%%%%%%%%%%%%%%%%%%%%%%%

%%%%%%%%%%%%%%%%%%%%%%%%%%%%%%%%%%%%%%%%%%%%%%%%%%%%%%%%%%%%%%%%%%%%%%%%%%%%%%%

Due to calculational difficulties at NLO order, it is convenient
to perform an expansion in mass $M_N$, i.e., schematically, 
the contributions to the correlation function
can be presented by 
\begin{equation}
\frac{A + M_N B + C M_N^2 +...}{a + b M_N^2} =
\frac{A}{a} + M_N \frac{B}{a} + {\cal O}(M_N^2) 
\, .
\label{eq:Mexpansion}
\end{equation}

The first term corresponds to taking $M_N=0$, while the first two
terms correspond to taking $M_N^2=0$ while retaining $M_N$ proportional
terms in the calculation.
In calculating NLO contribution we will try to investigate these two cases.
In loop calculations additional $\ln(M_N^2)$ terms appear,
and thus in the $M_N^2=0$ approximation collinear singularities.

The $M_N$-dependence of the general Lorentz decomposition of the nucleon matrix element
\req{eq:gLdecomp} 
can be summarized as
\begin{equation}
\begin{tabular}{l|l}
$M_N^0$-proportional $X^{(i)} Y^{(i)}$ &
${\cal V}_1$, ${\cal A}_1$, ${\cal T}_1$ \\
$M_N^1$-proportional  $X^{(i)} Y^{(i)}$  &
${\cal S}_1$, ${\cal P}_1$,
${\cal V}_2$, ${\cal V}_3$, 
${\cal A}_2$, ${\cal A}_3$, 
${\cal T}_2$, ${\cal T}_3$, ${\cal T}_4$ \\
$M_N^2$-proportional $X^{(i)} Y^{(i)}$  &
${\cal S}_2$, ${\cal P}_2$,
${\cal V}_4$, ${\cal V}_5$, 
${\cal A}_4$, ${\cal A}_5$,
${\cal T}_5$, ${\cal T}_6$, ${\cal T}_7$ \\
$M_N^3$-proportional  $X^{(i)} Y^{(i)}$  &
${\cal V}_6$,
${\cal A}_6$,
${\cal T}_8$ 
\end{tabular}
\label{eq:Mdep}
\end{equation}
Obviously for $M_N=0$ only the terms proportional to
${\cal F}^{(i)} \in \{{\cal V}_1,{\cal A}_1,{\cal T}_1 \}$
will contribute to $T_\mu$. 
But for $M_N \ne 0$ the contributions proportional to
${\cal F}^{(i)} \in \{{\cal V}_1,{\cal A}_1,{\cal T}_1 \}$
will contain also $M_N^n$-proportional ($n \ne 0$) terms.

%%%%%%%%%%%%%%%%%%%%%%%%%%%%%%%%%%%%%%%%%%%%%%%%%%%%%%%%%%%%%%%%%%%%%%%%%%%%%%%
%%%%%%%%%%%%%%%%%%%%%%%%%%%%%%%%%%%%%%%%%%%%%%%%%%%%%%%%%%%%%%%%%%%%%%%%%%%%%%%
\section{LO contributions}
\setcounter{equation}{0}
\label{sec:LO}
%%%%%%%%%%%%%%%%%%%%%%%%%%%%%%%%%%%%%%%%%%%%%%%%%%%%%%%%%%%%%%%%%%%%%%%%%%%%%%%

As a preparation and necessary ingredient of the NLO calculation,
in this section we discuss in detail LO contributions
to correlator function $T_\mu$ \req{eq:correlator}.
We present the complete twist-3 and twist-4 results 
and devote the last part of the section
to the analysis and discussion of gauge invariance.

\subsection{General structure and properties of LO contributions}
%%%%%%%%%%%%%%%%%%%%%%%%%%%%%%%%%%%%%%%%%%%%%%%%%%%%%%%%%%%%%%%%%

Using general Lorentz decomposition of the nucleon
matrix element of the three-quark operator given in Eq. \req{eq:gLdecomp}
and interpolating nucleon current of the form \req{eq:nucleon-curr}, 
one obtains
the general form of the LO contribution 
to the correlation function \req{eq:correlator}
\begin{eqnarray}
T_{\mu}^{\rm LO} 
&=&
-\frac{1}{4} 
\int d^4 x e^{i q\cdot x}
\int \frac{d^4 k_{\rm in}}{(2 \pi)^4} e^{i k_{\rm in}\cdot x}
\sum \limits_{i} 
\nonumber \\ & & \times \left\{
e_u \;
{\cal F}^{(i)}(P\cdot x,0,0) \; 
{\rm Tr} \left[ X^{(i)} \Gamma_1 
\frac{\not \! k_{\rm in} + m_u}{k_{\rm in}^2 - m_u^2}
\gamma_{\mu} 
\right] \; 
\Gamma_2 Y^{(i)}_{\gamma}
\right. \nonumber \\ & & \left. \quad +
e_u  \;
{\cal F}^{(i)}(0,P\cdot x,0) \; 
{\rm Tr} \left[\left( X^{(i)}\right)^T \Gamma_1 
\frac{\not \! k_{\rm in} + m_u}{k_{\rm in}^2 - m_u^2}
\gamma_{\mu} 
\right] \; 
\Gamma_2 Y^{(i)}_{\gamma}
\right. \nonumber \\ & & \left.  \quad +
e_d  \;
{\cal F}^{(i)}(0,0,P\cdot x) \; 
{\rm Tr} \left[\left( X^{(i)}\right)^T \Gamma_1 \right] \; 
\Gamma_2 
\frac{\not \! k_{\rm in} + m_d}{k_{\rm in}^2 - m_d^2}
\gamma_{\mu} 
Y^{(i)}_{\gamma}
\right \}
\, ,
\nonumber \\
\label{eq:LOgen}
\end{eqnarray}
and in following we neglect quark masses $m_u=m_d=0$.
Taking into account the form of $X^{(i)}$ 
it is easy to see that only the terms proportional to 
${\cal F}^{(i)} \in \{ {\cal V}_j, {\cal A}_j \}$
contribute if Ioffe current,
i.e., $\Gamma_1=C\gamma_\lambda$ and $\Gamma_2=\gamma_5 \gamma^\lambda$,
is taken for interpolating nucleon current.
Namely, the matrices $X^{(i)}$ 
(see Ref. \cite{BraunLW06})
proportional to
${\cal F}^{(i)} \in \{ {\cal V}_j, {\cal A}_j \}$
and
${\cal F}^{(i)} \in \{ {\cal S}_j, {\cal P}_j, {\cal T}_j\}$
consist of odd and even number of $\gamma$ matrices, respectively,
and hence the traces from \req{eq:LOgen}
vanish in latter case.

The contribution corresponding to the first case
of \req{eq:cases} 
takes the form
\begin{eqnarray}
T_{\mu}^{{\rm LO}} 
&=&
-\frac{1}{4} 
\sum \limits_{i} \int {\cal D}u \, N^{(i)} \, F^{(i)}(\{u_k\})
\nonumber \\ & & \times \left\{
e_u \; 
{\rm Tr} \left[ X^{(i)} \Gamma_1 
\frac{u_1 \not \! P - \not \! q}{(u_1 P - q)^2}
\gamma_{\mu} 
\right] \; 
\Gamma_2 Y^{(i)}_{\gamma}
\right. \nonumber \\ & & \left. \quad +
e_u \; 
{\rm Tr} \left[\left( X^{(i)}\right)^T \Gamma_1 
\frac{u_2 \not \! P - \not \! q}{(u_2 P - q)^2}
\gamma_{\mu} 
\right] \; 
\Gamma_2 Y^{(i)}_{\gamma}
\right. \nonumber \\ & & \left.  \quad +
e_d \; 
{\rm Tr} \left[\left( X^{(i)}\right)^T \Gamma_1 \right] \; 
\Gamma_2 
\frac{u_3 \not \! P - \not \! q}{(u_3 P - q)^2}
\gamma_{\mu} 
Y^{(i)}_{\gamma}
\right \}
\, ,
\nonumber \\
\label{eq:LOcase1}
\end{eqnarray}
and using \req{eq:convolI} we can write
\begin{eqnarray}
T_{\mu}^{{\rm LO}} &=&
\sum \limits_{i} {\cal M}^{{\rm LO},(i)}(\{u_k\}) \,\otimes \, F^{(i)}(\{u_k\})
\, ,
\label{eq:LOcase1convol}
\end{eqnarray}
where from comparison  with \req{eq:LOcase1} the definition
of ${\cal M}^{{\rm LO},(i)}(\{u_k\})$ 
is obvious.
As explained in Sec. \ref{sec:twistdecomp}, for the other two cases
listed in \req{eq:cases} slightly more involved formulas appear.

For Ioffe current \req{eq:Ioffecurr},  
there are two twist-3 contributions
($F^{(i)} \in \{ V_1, A_1 \}$)
calculable by
\req{eq:LOcase1} 
with $N^{(i)}=1$,
and steaming from
\begin{equation}
\begin{tabular}{l|l|l}
${\cal F}^{(i)}$ &
$X^{(i)}_{\alpha \beta}$ &
$Y^{(i)}_{\gamma}$ \\ \hline
%%%%%%%%%%%%%%%%%%%%%%%%%%
${\cal V}_{1}=V_1$ &
$(\not \! P C)_{\alpha \beta}$ &
$(\gamma_5 N)_{\gamma}$ \\
%%%%%%%%%%%%%%%%%%%%%%%%%%
${\cal A}_{1}=A_1$ &
$(\not \! P \gamma_5 C)_{\alpha \beta}$ &
$N_{\gamma}$ \\
%%%%%%%%%%%%%%%%%%%%%%%%%%
\end{tabular}
\label{eq:XY-twist3}
\end{equation}
Furthermore, when using Ioffe current,
there are the two twist-4 contributions
($ F^{(i)} \in \{ V_3, A_3 \}$)
calculable using \req{eq:LOcase1}
with  $N^{(i)}=1/2$,
and two twist-4 contributions
($ F^{(i)} \in \{ V_2, A_2 \}$)
corresponding to the more involved second case of \req{eq:cases}.
These contributions stem from
\begin{equation}
\begin{tabular}{l|l|l}
${\cal F}^{(i)}$ &
$X^{(i)}_{\alpha \beta}$ &
$Y^{(i)}_{\gamma}$ \\ \hline
%%%%%%%%%%%%%%%%%%%%%%%%%%
${\cal V}_{3}=1/2 \, V_3$ &
$M_N (\gamma_{\kappa} C)_{\alpha \beta}$ &
$(\gamma^{\kappa} \gamma_5 N)_{\gamma}$ \\
%%%%%%%%%%%%%%%%%%%%%%%%%%
${\cal A}_{3}=1/2 \, A_3$ &
$M_N (\gamma_{\kappa} \gamma_5 C)_{\alpha \beta}$ &
$(\gamma^{\kappa} N)_{\gamma}$ \\
%%%%%%%%%%%%%%%%%%%%%%%%%%
${\cal V}_{2}=1/(2 P \cdot x) (V_1-V_2-V_3)$ &
$M_N (\not \! P C)_{\alpha \beta}$ &
$(\not \! x \gamma_5 N)_{\gamma}$ \\
%%%%%%%%%%%%%%%%%%%%%%%%%%
${\cal A}_{2}=1/(2 P \cdot x) (-A_1+A_2-A_3)$ &
$M_N (\not \! P \gamma_5 C)_{\alpha \beta}$ &
$(\not \! x N)_{\gamma}$ \\
\end{tabular}
\label{eq:XY-twist4}
\end{equation}

Our main interest in this work are LO and NLO contributions
to twist-3 contributions and the generalization to other
first case contributions.
Hence, we do not discuss in detail the calculation
of second and third case LO contributions from \req{eq:cases},
nor higher-twist contributions, but rather refer to \cite{BraunLW06}
and references therein.

%%%%%%%%%%%%%%%%%%%%%%%%
\begin{figure}
\begin{center}
\includegraphics{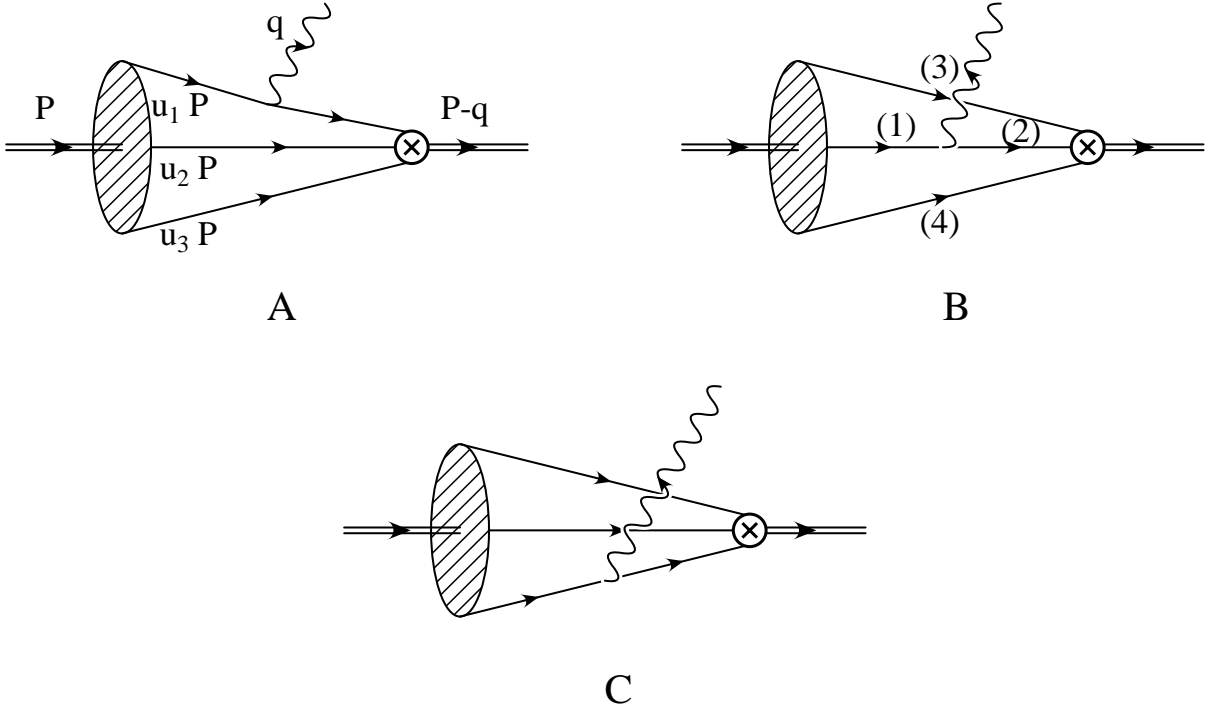}
\end{center}
\caption{LO diagrams contributing to the correlation function
\req{eq:correlator}. On diagram B the nomenclature for NLO diagrams
is sketched.}
\label{f:LO}
\end{figure}
%%%%%%%%%%%%%%%%%%%%%%%%

In order to simplify higher-order calculation
it is convenient to extract the Feynman rules 
and perform the calculation in momentum space.
The three terms contributing to \req{eq:LOgen}
correspond actually to three Feynman diagrams presented on 
Fig. \ref{f:LO}.

The Feynman rules for the contributions corresponding to
first case of \req{eq:cases}
are listed in Appendix \ref{sec:appFEYN}.
Using these rules it is trivial to write down the contributions
corresponding to the LO diagrams in Fig. \ref{f:LO}.
As it should, these contributions agree with
corresponding terms in Eq. \req{eq:LOcase1} and
\begin{eqnarray}
T_\mu^{{\rm LO}}&=&\sum \limits_i 
  \left( T_\mu^{A,(i)}+ T_\mu^{B,(i)}+ T_\mu^{C,(i)} \right)
\,.
\label{eq:LOcase1-sum}
\end{eqnarray}
\label{eq:LOcase1-mom}

Taking into account \req{eq:XT}
it is easy to see that
\begin{equation}
{\cal M}_{\mu}^{B,(i)}(u_1,u_2,u_3) =
\left\{
\begin{array}{l}
{\cal M}_{\mu}^{A,(i)}(u_2,u_1,u_3)
\quad \mbox{for} \quad 
F^{(i)} \in \{ V_1, V_3 \} \\
- {\cal M}_{\mu}^{A,(i)}(u_2,u_1,u_3)
\quad \mbox{for} \quad 
F^{(i)} \in \{ A_1, A_3 \} \\
\end{array}
\right.
 \, .
\label{eq:BtoA}
\end{equation}
It is also obvious 
that%
\footnote{
This result is expected since $A_1$ and $A_3$ are antisymmetric
in $u_1 \leftrightarrow u_2$, while 
${\cal M}^{C,(i)}$ (${\cal M}^{(5)C,(i)}$) 
contribution is obviously ``blind'', i.e., symmetric, to
$u_1 \leftrightarrow u_2$ exchange since the photon 
couples to $d$-quark 
(which carries the momentum fraction $u_3$ -- see Fig. \ref{f:LO}).
This property no longer holds at NLO order at which 
gluon can couple to $u$-quarks (see Fig. \ref{f:NLO}).} 
%\begin{equation}
${\cal M}^{C,A1}(\{u_k\}) 
= {\cal M}^{C,A3}(\{u_k\}) = 0
$.
%\, .
%\end{equation}

Hence for the first case of \req{eq:cases}
one can write 
\begin{eqnarray}
{\cal M}^{{\rm LO},(i)}(\{u_k\})
  & =& 
-\frac{1}{4}  N^{(i)} 
\nonumber \\ & &
\times \left\{
e_u  \; 
{\rm Tr} \left[ 
\left(\frac{u_1 \not \! P - \not \! q}{(u_1 P - q)^2}
\pm
\frac{u_2 \not \! P - \not \! q}{(u_2 P - q)^2} \right)
\gamma_{\mu} 
X^{(i)} (C \gamma_{\lambda})
\right] \; 
\gamma_5 \gamma^\lambda Y^{(i)}_{\gamma}
\right. \nonumber \\[0.3cm] && \left.
+
e_d \; 
{\rm Tr} \left[ X^{(i)} (C \gamma_\lambda) \right] \; 
\gamma_5 \gamma^\lambda
\frac{u_3 \not \! P - \not \! q}{(u_3 P - q)^2}
\gamma_{\mu} 
Y^{(i)}_{\gamma}
\right\}
\, ,
\end{eqnarray}
where $``+''$ sign in the first term corresponds to 
$F^{(i)} \in \{ V_1, V_3\}$ and $``-''$ sign to 
$F^{(i)} \in \{ A_1, A_3\}$.
Finally, note that, since we are interested in NLO calculation
in which dimensional regularization will be employed,
one has to use $D=4-2\epsilon$ dimensions also at LO.

%%%%%%%%%%%%%%%%%%%%%%%%%%%%%%%%%%%%%%%%%%%%%%%%%%%%%%%%%%%%%%%%%%%%%%%%%%%%%%%
\subsection{Twist-3 results}
%%%%%%%%%%%%%%%%%%%%%%%%%%%%%%%%%%%%%%%%%%%%%%%%%%%%%%%%%%%%%%%%%%%%%%%%%%%%%%%

%%%%%%%%%%%%%%%%%%%%%%%%%%%%%%%%%%%%%%%%%%%%%%%%%%%%%%%%
\begin{table}
\caption{LO twist-3 contributions corresponding to the diagrams
of Fig. \protect\ref{f:LO}.
The contributions ${\cal M}_\mu^{B,(i)}$ can be obtained using 
\protect\req{eq:BtoA}. 
}
\begin{tabular}{l|l}
\hline \hline
${\cal M}_\mu^{A,V1}$ & 
$\displaystyle
\frac{e_u}{Q^2+2 u_1 P\cdot q- u_1^2 M_N^2}
\left[ P_\mu \left( 2 u_1 M_N - \not \! q\right)
- q_\mu M_N
+ \gamma_\mu \left( P \cdot q -  u_1 M_N^2\right) 
\right]$ \\[0.5cm]
${\cal M}_\mu^{A,A1}$ & 
$\displaystyle
\frac{e_u}{Q^2+2 u_1 P\cdot q- u_1^2 M_N^2}
\left[ P_\mu \not \! q
- q_\mu M_N
- \gamma_\mu P \cdot q 
+ \gamma_\mu \not \! q M_N
\right]$ \\[0.5cm]
\hline
${\cal M}_\mu^{C,V1}$ & 
$\displaystyle
\frac{e_d}{Q^2+2 u_3 P\cdot q- u_3^2M_N^2}
\left[ 2 P_\mu \not \! q
- 2 q_\mu M_N
+ \gamma_\mu \left( -2 P \cdot q +  u_3 M_N^2\right) 
+ \gamma_\mu \not \! q M_N
\right]$ \\[0.5cm]
${\cal M}_\mu^{C,A1}$ & 0
 \\[0.5cm]
\hline \hline
\end{tabular}
\label{t:LO}
\end{table}
%%%%%%%%%%%%%%%%%%%%%%%%%%%%%%%%%%%%%%%%%%%%%%%%%%%%%%%%

In Table \ref{t:LO} we present the LO twist-3 contributions 
to diagrams displayed on Fig. \ref{f:LO}.
The complete LO contributions to amplitudes ${\cal M}_{\mu}^{{\rm LO},(i)}$
are given by
\begin{eqnarray}
{\cal M}_\mu^{{\rm LO},(i)}(u_1,u_2,u_3) &=&
{\cal M}_\mu^{A,(i)}(u_1,u_2,u_3) +
{\cal M}_\mu^{B,(i)}(u_1,u_2,u_3) +
{\cal M}_\mu^{C,(i)}(u_1,u_2,u_3) 
\nonumber \\[0.19cm] &=&
{\cal M}_\mu^{A,(i)}(u_1,u_2,u_3) 
 \pm {\cal M}_\mu^{A,(i)}(u_2,u_1,u_3) 
+ {\cal M}_\mu^{C,(i)}(u_1,u_2,u_3) 
\, ,
\nonumber \\
\end{eqnarray}
where $``+''$ sign in the second line corresponds to 
$V_1$ contributions and $``-''$ sign to 
$A_1$ contributions.
The correlation function is given by
\req{eq:LOcase1convol} and
taking into account the DA symmetry properties \req{eq:Fsym},
one obtains
\begin{eqnarray}
\lefteqn{\left({\cal M}_\mu^{A,(i)}(u_1,u_2,u_3) 
 \pm {\cal M}_\mu^{A,(i)}(u_2,u_1,u_3) \right) \otimes
 F^{(i)}(u_1,u_2,u_3) }
\nonumber \\ &\qquad =&
2 \, {\cal M}_\mu^{A,(i)}(u_1,u_2,u_3) \otimes F^{(i)}(u_1,u_2,u_3) 
\, ,
\end{eqnarray}
but it is advantageous for NLO calculation to leave 
full $u_1$ and $u_2$ dependence of the amplitude ${\cal M}$.

Our LO as well as NLO results for $T_{\mu}$ 
can be 
presented in terms of six invariant functions which multiply the
Lorentz structures 
$P_{\mu} M_N$, $P_{\mu}\not \! \! q$, 
$q_{\mu} M_N$, $q_{\mu}\not \! \! q$, 
$\gamma_{\mu}$, and $\gamma_{\mu}\not \! \! q M_N$.
Thus it is advantageous for future calculations to present the results
in a form
\begin{eqnarray}
{\cal M}_{\mu}^{X,(i)}(\{u_k\}) &=&
C^{X,(i)}_{P_{\mu} M_N}(\{u_k\}) \; P_{\mu} M_N
+ C^{X,(i)}_{P_{\mu} \not q}(\{u_k\}) \;P_{\mu} \not \! q 
\nonumber \\ & &
+ C^{X,(i)}_{q_{\mu} M_N}(\{u_k\}) \; q_{\mu} M_N 
+ C^{X,(i)}_{q_{\mu} \not  q}(\{u_k\}) \;q_{\mu} \not \! q 
\nonumber \\ & &
+ C^{X,(i)}_{\gamma_{\mu}}(\{u_k\}) \;\gamma_{\mu} 
+ C^{X,(i)}_{\gamma_{\mu} \not q M_N}(\{u_k\}) \;\gamma_{\mu} \not \! q M_N
\, .
\nonumber \\ & &
\label{eq:Mcoeff}
\end{eqnarray}
The corresponding LO twist-3 coefficients are given in Table \ref{t:coeffLO1}.
%%%%%%%%%%%%%%%%%%%%%%%%%%%%%%%%%%%%%%%%%%%%%%%%%%%%%%%%
\begin{table}
\caption{LO coefficients \protect\req{eq:Mcoeff} 
corresponding to $V_1$ and $A_1$ DAs.}
\begin{center}
\begin{tabular}{l|l}
\hline \hline
$C^{{\rm LO},V1}_{P_{\mu} M_N}$ 
&
$\displaystyle
\left(\frac{2 u_1 e_u}{Q^2+2 u_1 P\cdot q- u_1^2 M_N^2}
+ (u_1 \leftrightarrow u_2) \right)
$ 
\\[0.5cm]
$C^{{\rm LO},V1}_{P_{\mu} \not q}$ 
&
$\displaystyle
\left(\frac{- e_u}{Q^2+2 u_1 P\cdot q- u_1^2 M_N^2}
+ (u_1 \leftrightarrow u_2) \right)
+ \frac{2 e_d}{Q^2+2 u_3 P\cdot q- u_3^2 M_N^2}
$ 
\\[0.5cm]
$C^{{\rm LO},V1}_{q_{\mu} M_N}$ 
&
$\displaystyle
\left(\frac{- e_u}{Q^2+2 u_1 P\cdot q- u_1^2 M_N^2}
+ (u_1 \leftrightarrow u_2) \right)
+ \frac{-2 e_d}{Q^2+2 u_3 P\cdot q- u_3^2 M_N^2}
$ 
\\[0.5cm]
$C^{{\rm LO},V1}_{q_{\mu} \not q}$ 
&
$0$
\\[0.5cm]
$C^{{\rm LO},V1}_{\gamma_{\mu}}$ 
&
$\displaystyle
\left(\frac{e_u (P \cdot q - u_1 M_N^2)}{Q^2+2 u_1 P\cdot q- u_1^2 M_N^2}
+ (u_1 \leftrightarrow u_2) \right)
+ \frac{e_d(- 2 P \cdot q + u_3 M_N^2)}{Q^2+2 u_3 P\cdot q- u_3^2 M_N^2}
$ 
\\[0.5cm]
$C^{{\rm LO},V1}_{\gamma_{\mu} \not q M_N}$ 
&
$\displaystyle
 \frac{e_d}{Q^2+2 u_3 P\cdot q- u_3^2 M_N^2}
$ 
\\[0.5cm]
\hline
$C^{{\rm LO},A1}_{P_{\mu} M_N}$ 
&
$0$ 
\\[0.5cm]
$C^{{\rm LO},A1}_{P_{\mu} \not q}$ 
&
$\displaystyle
\left(\frac{e_u}{Q^2+2 u_1 P\cdot q- u_1^2 M_N^2}
- (u_1 \leftrightarrow u_2) \right)
$ 
\\[0.5cm]
$C^{{\rm LO},A1}_{q_{\mu} M_N}$ 
&
$\displaystyle
\left(\frac{-e_u}{Q^2+2 u_1 P\cdot q- u_1^2 M_N^2}
- (u_1 \leftrightarrow u_2) \right)
$ 
\\[0.5cm]
$C^{{\rm LO},A1}_{q_{\mu} \not q}$ 
&
$0$
\\[0.5cm]
$C^{{\rm LO},A1}_{\gamma_{\mu}}$ 
&
$\displaystyle
\left(\frac{-P\cdot q \, e_u}{Q^2+2 u_1 P\cdot q- u_1^2 M_N^2}
- (u_1 \leftrightarrow u_2) \right)
$ 
\\[0.5cm]
$C^{{\rm LO},A1}_{\gamma_{\mu} \not q M_N}$ 
&
$\displaystyle
\left(\frac{e_u}{Q^2+2 u_1 P\cdot q- u_1^2 M_N^2}
- (u_1 \leftrightarrow u_2) \right)
$ 
\\[0.5cm]
\hline \hline
\end{tabular}
\end{center}
\label{t:coeffLO1}
\end{table}
%%%%%%%%%%%%%%%%%%%%%%%%%%%%%%%%%%%%%%%%%%%%%%%%%%%%%%%%

Finally, we present the results for the invariant functions
${\mathcal A}$ and ${\mathcal B}$ defined in \req{eq:projections}.
Multiplying $T_{\mu}$ with $z^{\mu} \Lambda_+$
the invariant functions ${\mathcal A}$ and ${\mathcal B}$ 
are projected.
It is easy to see that these correspond to the
invariant functions multiplying 
$P_{\mu} M_N$ and $P_{\mu} \not \! q$, 
and thus
\begin{eqnarray}
{\mathcal A}^{\rm twist-3}(Q^2, (P-q)^2) &=&
 C^{{\rm LO},V1}_{P_{\mu} M_N}(\{u_k\}) \; \otimes \;  V_1(\{u_k\})
+ C^{{\rm LO},A1}_{P_{\mu} M_N}(\{u_k\}) \; \otimes \;  A_1(\{u_k\})
\, ,
\nonumber \\[0.19cm] & &
 \\[0.25cm] 
{\mathcal B}^{\rm twist-3}(Q^2, (P-q)^2) &=&
 C^{{\rm LO},V1}_{P_{\mu} \not \! q}(\{u_k\}) \; \otimes \;  V_1(\{u_k\})
+ C^{{\rm LO},A1}_{P_{\mu} \not \! q }(\{u_k\}) \; \otimes \;  A_1(\{u_k\})
\, .
\nonumber \\[0.19cm] & &
\end{eqnarray}
The results presented here are in agreement with the results 
from \cite{BraunLW06}.

%%%%%%%%%%%%%%%%%%%%%%%%%%%%%%%%%%%%%%%%%%%%%%%%%%%%%%%%%%%%%%%%%%%%%%%%%%%%%%%

\subsection{Twist-4 results}

%%%%%%%%%%%%%%%%%%%%%%%%%%%%%%%%%%%%%%%%%%%%%%%%%%%%%%%%%%%%%%%%%%%%%%%%%%%%%%%

The twist-4 contributions corresponding to $V_3$ and $A_3$ 
are obtained analogously to the twist-3 results discussed
in the preceding subsection.
The LO coefficients corresponding to \req{eq:Mcoeff} are given 
in Table \ref{t:coeffLO3}.
Note the dependence on $\epsilon=(4-D)/2$.

%%%%%%%%%%%%%%%%%%%%%%%%%%%%%%%%%%%%%%%%%%%%%%%%%%%%%%%%
\begin{table}
\caption{LO coefficients \protect\req{eq:Mcoeff}
corresponding to $V_3$ and $A_3$ DAs.}
\begin{center}
\begin{tabular}{l|l}
\hline \hline
$C^{{\rm LO},V3}_{P_{\mu} M_N}$ 
&
$\displaystyle
\left(\frac{-(3 - \epsilon) u_1 e_u}{Q^2+2 u_1 P\cdot q- u_1^2 M_N^2}
+ (u_1 \leftrightarrow u_2) \right)
+ \frac{-2 (1-\epsilon) u_3 e_d}{Q^2+2 u_3 P\cdot q- u_3^2 M_N^2}
$ 
\\[0.5cm]
$C^{{\rm LO},V3}_{P_{\mu} \not q}$ 
&
$
0 
$ 
\\[0.5cm]
$C^{{\rm LO},V3}_{q_{\mu} M_N}$ 
&
$\displaystyle
\left(\frac{ (3 - \epsilon) e_u}{Q^2+2 u_1 P\cdot q- u_1^2 M_N^2}
+ (u_1 \leftrightarrow u_2) \right)
+ \frac{2 (1-\eps) e_d}{Q^2+2 u_3 P\cdot q- u_3^2 M_N^2}
$ 
\\[0.5cm]
$C^{{\rm LO},V3}_{q_{\mu} \not q}$ 
&
$0$
\\[0.5cm]
$C^{{\rm LO},V3}_{\gamma_{\mu}}$ 
&
$\displaystyle
\left(\frac{u_1 M_N^2 e_u}{Q^2+2 u_1 P\cdot q- u_1^2 M_N^2}
+ (u_1 \leftrightarrow u_2) \right)
+ \frac{- \epsilon u_3 M_N^2 e_d}{Q^2+2 u_3 P\cdot q- u_3^2 M_N^2}
$ 
\\[0.5cm]
$C^{{\rm LO},V3}_{\gamma_{\mu} \not q M_N}$ 
&
$\displaystyle
\left(\frac{- e_u}{Q^2+2 u_1 P\cdot q- u_1^2 M_N^2}
+ (u_1 \leftrightarrow u_2) \right)
+ \frac{\epsilon e_d}{Q^2+2 u_3 P\cdot q- u_3^2 M_N^2}
$ 
\\[0.5cm]
\hline
$C^{{\rm LO},A3}_{P_{\mu} M_N}$ 
&
$ \displaystyle
\left(\frac{ u_1 (1 - \epsilon) e_u}{Q^2+2 u_1 P\cdot q- u_1^2 M_N^2}
- (u_1 \leftrightarrow u_2) \right)
$
\\[0.5cm]
$C^{{\rm LO},A3}_{P_{\mu} \not q}$ 
&
$
0
$ 
\\[0.5cm]
$C^{{\rm LO},A3}_{q_{\mu} M_N}$ 
&
$\displaystyle
\left(\frac{ -(1-\epsilon) e_u}{Q^2+2 u_1 P\cdot q- u_1^2 M_N^2}
- (u_1 \leftrightarrow u_2) \right)
$ 
\\[0.5cm]
$C^{{\rm LO},A3}_{q_{\mu} \not q}$ 
&
$0$
\\[0.5cm]
$C^{{\rm LO},A3}_{\gamma_{\mu}}$ 
&
$\displaystyle
\left(\frac{-(1-\epsilon) u_1 M_N^2 e_u}{Q^2+2 u_1 P\cdot q- u_1^2 M_N^2}
- (u_1 \leftrightarrow u_2) \right)
$ 
\\[0.5cm]
$C^{{\rm LO},A3}_{\gamma_{\mu} \not q M_N}$ 
&
$\displaystyle
\left(\frac{(1- \epsilon) e_u}{Q^2+2 u_1 P\cdot q- u_1^2 M_N^2}
- (u_1 \leftrightarrow u_2) \right)
$ 
\\[0.5cm]
\hline \hline
\end{tabular}
\end{center}
\label{t:coeffLO3}
\end{table}
%%%%%%%%%%%%%%%%%%%%%%%%%%%%%%%%%%%%%%%%%%%%%%%%%%%%%%%%

Furthermore, without going into much detail, 
we present also the contributions corresponding 
to ${\cal V}_2,{\cal A}_2$ -- see  \req{eq:XY-twist4}.
We therefore introduce the shorthand notation
\begin{eqnarray}
V_{123}=V_1-V_2-V_3,\nonumber\\
A_{123}=-A_1+A_2-A_3.
\label{eq:V123A123}
\end{eqnarray}

Furthermore, as in \cite{BraunLW06}, we use the definition
%\begin{eqnarray}
%\widetilde{F}(x_1)=\int_1^{x_1} dx'_1 \int_0^{1-x'_1} dx_2 F(x'_1,x_2,1-x'_1-x_2),\\
%\widetilde{F}(x_3)=\int_1^{x_3} dx'_3 \int_0^{1-x'_3} dx_1 F(x_1,1-x_1-x'_3,x'_3)
%\end{eqnarray}
\begin{eqnarray}
\widetilde{F}(u_l)&\equiv&\int_1^{u_l} dv_l \int_0^{1-v_l} 
du_m \, F(\{v_l,u_m,1-v_l-u_m\})
\, .
\label{eq:wtilde}
\end{eqnarray}
Here $F$ is the nucleon DA or the combination of nucleon DAs
that depends on three valence 
quark momentum fractions, and the integration over one momentum
fraction has already been performed using $\delta(1-v_l-u_m-u_k)$. 
Note that in this notation, which follows closely Ref. \cite{BraunLW06},
$\widetilde{F}(u_l)$ 
is not a simple function of $u_l$ and that the form of the function
itself depends on $u_l=u_1$, $u_2$ or $u_3$,
i.e., whether $u_l$ corresponds to the momentum fraction of  
the first u-quark, second u-quark or d-quark%
\footnote{Hence, the shorthand expression \req{eq:wtilde} encompasses
three functions
\begin{eqnarray*}
\widetilde{F}(u_1)
&\equiv&\int_1^{u_1} dv_1 \int_0^{1-v_1} du_2 \, F(v_1,u_2,1-v_1-u_2),
\nonumber\\
&=&\int_1^{u_1} dv_1 \int_0^{1-v_1} du_3 \, F(v_1,1-v_1-u_3,u_3),
\nonumber\\[0.3cm]
\widetilde{F}(u_2)
&\equiv&\int_1^{u_2} dv_2 \int_0^{1-v_2} du_1 \, F(u_1,v_2,1-u_1-v_2),
\nonumber\\
&=&\int_1^{u_2} dv_2 \int_0^{1-v_2} du_3 \, F(1-v_2-u_3,v_2,u_3),
\nonumber\\[0.3cm]
\widetilde{F}(u_3)
&\equiv&\int_1^{u_3} dv_3 \int_0^{1-v_3} du_1 \, F(u_1,1-u_1-v_3,v_3)
\nonumber\\
&=&\int_1^{u_3} dv_3 \int_0^{1-v_3} du_2 \, F(1-u_2-v_3,u_2,v_3)\, .
\label{tilde}
\end{eqnarray*}
Strictly speaking, it would be better that 
different names are introduced for these three functions
instead of using argument to determine the form of the function.
But for historical reasons and simplicity of the notation
we adopt this notation hoping that it will not lead to too
much confusion.}.
Note that 
$\widetilde{F}(u_1)=\pm \widetilde{F}(u_2)$
for
$F(u_1,u_2,u_3)=\pm F(u_2,u_1,u_3)$, respectively.

The function $\widetilde{F}(u_i)$ depends on only one momentum fraction
and enters the expression analogous to \req{eq:convolI},
but that expression contains the integration with amplitude ${\cal M}(u_i)$ 
only over one remaining momentum fraction $u_i$.
This property holds in LO (where the dependence on $u_1$, $u_2$ and $u_3$ 
is clearly separated) while in NLO the expressions will be more
involved.
For tabulated LO contributions see Table \ref{t:coeffLO2}.
%%%%%%%%%%%%%%%%%%%%%%%%%%%%%%%%%%%%%%%%%%%%%%%%%%%%%%%%
\begin{table}
\caption{LO coefficients analogous to \protect\req{eq:Mcoeff} and
corresponding to $\widetilde{V}_{123}(u_i)$ and $\widetilde{A}_{123}(u_i)$
(\ref{eq:V123A123}-\ref{eq:wtilde}): 
$\int_0^1 du_i C^{{\rm LO},F_{123}}_{\cdots} (u_i) \widetilde{F}_{123}(u_i)$\,
for $u_i \in \{ u_1,u_2,u_3\}$.}
\begin{center}
\begin{tabular}{l|l}
\hline \hline
\\[0.2cm]
$\{C^{{\rm LO},V_{123}}_{P_{\mu} M_N} (u_i) \}$ 
&
$\displaystyle
\left\{
\frac{ e_u
((2-\epsilon) Q^2 + 2 (1-\epsilon)u_1 P \cdot q -(1-\epsilon) M_N^2 u_1^2 )}%
{(Q^2+ 2  u_1 P\cdot q- u_1^2M_N^2)^2} 
\, ,
(u_1 \leftrightarrow u_2) 
\, ,
\right.
$
\\[0.5cm]
&
$\displaystyle
\left.
 -\frac{e_d (- 2\epsilon \, Q^2 +4(1-\epsilon) u_3 P \cdot q+ 2(\epsilon - 1)u_3^2 M_N^2)}%
{(Q^2+ 2  u_3 P\cdot q- u_3^2M_N^2)^2} 
\right\}
$ 
\\[0.5cm]
$\{ C^{{\rm LO},V_{123}}_{P_{\mu} \not q} (u_i) \}$ 
&
$
\displaystyle
\left\{
\frac{e_u u_1 M_N^2}{(Q^2+ 2  u_1 P\cdot q- u_1^2M_N^2)^2}
\, , 
(u_1 \leftrightarrow u_2)
\, , 
\frac{2 e_d u_3 M_N^2}{(Q^2+ 2  u_3 P\cdot q- u_3^2M_N^2)^2}
\right\}
$ 
\\[0.5cm]
$\{ C^{{\rm LO},V_{123}}_{q_{\mu} M_N} (u_i) \}$ 
&
$\displaystyle
\left\{
\frac{e_u (2 P \cdot q -u_1 M_N^2)}{(Q^2+ 2  u_1 P\cdot q- u_1^2M_N^2)^2}
\, ,
(u_1 \leftrightarrow u_2) 
\, ,
\frac{e_d \, (4 P \cdot q -2 u_3 M_N^2)}{(Q^2+ 2  u_3 P\cdot q- u_3^2M_N^2)^2}
\right\}
$ 
\\[0.5cm]
$\{ C^{{\rm LO},V_{123}}_{q_{\mu} \not q} (u_i) \}$ 
&
$\displaystyle
\left\{
\frac{-e_u M_N^2}{(Q^2+ 2  u_1 P\cdot q- u_1^2M_N^2)^2}
\, ,
(u_1 \leftrightarrow u_2)
\, ,
\frac{-2 e_d M_N^2 }{(Q^2+ 2  u_3 P\cdot q- u_3^2M_N^2)^2}
\right\}
$ 
\\[0.5cm]
$\{ C^{{\rm LO},V_{123}}_{\gamma_{\mu}} (u_i) \}$ 
&
$\displaystyle
\left\{
\frac{- e_u M_N^2 (Q^2+ u_1 P \cdot q )}{(Q^2+ 2  u_1 P\cdot q- u_1^2M_N^2)^2}
\, ,
(u_1 \leftrightarrow u_2) 
\, ,
\frac{- e_d \, \epsilon M_N^2}{Q^2 + 2  u_3 P\cdot q- u_3^2M_N^2}
\right\}
$ 
\\[0.5cm]
$\{ C^{{\rm LO},V_{123}}_{\gamma_{\mu} \not q M_N} (u_i) \}$ 
&
$\displaystyle
\left\{
\frac{e_u( -P \cdot q + u_1 M_N^2)}{(Q^2+ 2  u_1 P\cdot q- u_1^2M_N^2)^2}
\, ,
(u_1 \leftrightarrow u_2) 
\, ,
0
\right\}
$ 
\\[0.5cm]
\hline
\\[0.2cm]
$\{C^{{\rm LO},A_{123}}_{P_{\mu} M_N} (u_i) \}$ 
&
$ \displaystyle
\left\{
\frac{e_u \, (- \epsilon Q^2 + 2 (1-\epsilon) u_1 P \cdot q  -(1-\epsilon) u_1^2 M_N^2)}%
{(Q^2+ 2  u_1 P\cdot q- u_1^2M_N^2)^2}
\, ,
- (u_1 \leftrightarrow u_2) 
\, ,
0
\right\}
$
\\[0.5cm]
$\{C^{{\rm LO},A_{123}}_{P_{\mu} \not q} (u_i) \}$ 
&
$
\displaystyle
\left\{
(\frac{-e_u u_1 M_N^2}{(Q^2+ 2  u_1 P\cdot q- u_1^2M_N^2)^2}
\, ,
- (u_1 \leftrightarrow u_2)
\, ,
0
\right\}
$ 
\\[0.5cm]
$\{C^{{\rm LO},A_{123}}_{q_{\mu} M_N} (u_i) \}$ 
&
$\displaystyle
\left\{
\frac{e_u( -2 P\cdot q+ u_1 M_N^2)}{(Q^2+ 2  u_1 P\cdot q- u_1^2M_N^2)^2}
\, ,
- (u_1 \leftrightarrow u_2) 
\, ,
0
\right\}
$ 
\\[0.5cm]
$\{C^{{\rm LO},A_{123}}_{q_{\mu} \not q} (u_i) \}$ 
&
$\displaystyle
\left\{
\frac{e_u M_N^2}{(Q^2+ 2  u_1 P\cdot q- u_1^2M_N^2)^2}
\, ,
- (u_1 \leftrightarrow u_2) 
\, ,
0
\right\}
$
\\[0.5cm]
$\{C^{{\rm LO},A_{123}}_{\gamma_{\mu}} (u_i) \}$ 
&
$\displaystyle
\left\{
\frac{- e_u M_N^2 (-\epsilon Q^2 +(1-2 \epsilon) u_1 P \cdot q -(1-\epsilon) u_1^2 M_N^2 )}%
{(Q^2+ 2  u_1 P\cdot q- u_1^2M_N^2)^2}
\, ,
- (u_1 \leftrightarrow u_2) 
\, , 
0
\right\}
$ 
\\[0.5cm]
$\{C^{{\rm LO},A_{123}}_{\gamma_{\mu} \not q M_N} (u_i) \}$ 
&
$\displaystyle
\left\{
\frac{e_u(P \cdot q -u_1 M_N^2)}{(Q^2+ 2  u_1 P\cdot q- u_1^2M_N^2)^2}
\, ,
- (u_1 \leftrightarrow u_2) 
\, , 
0
\right\}
$ 
\\[0.5cm]
\hline \hline
\end{tabular}
\end{center}
\label{t:coeffLO2}
\end{table}
%%%%%%%%%%%%%%%%%%%%%%%%%%%%%%%%%%%%%%%%%%%%%%%%%%%%%%%%
The contributions to functions ${\cal A}$
and ${\cal B}$ are given by
\req{eq:ABF1F3} and \req{eq:ABF123}.

%%%%%%%%%%%%%%%%%%%%%%%%%%%%%%%%%%%%%%%%%%%%%%%%%%%%%%%%%%%%%%%%%%%%%%%%%%%%%%%
\subsection{Gauge invariance}
%%%%%%%%%%%%%%%%%%%%%%%%%%%%%%%%%%%%%%%%%%%%%%%%%%%%%%%%%%%%%%%%%%%%%%%%%%%%%%%
\label{sec:gaugeinv}

Due to the presence of nucleon interpolation current,
%
%\footnote{
%Starting with
%\begin{displaymath}
%i\! \int\! d^4 x \, e^{i q x}
%\left\langle 0 \right| T\left[\eta(0) 
%       \partial^\mu j_{\mu}(x) 
%\right] 
%\left| N(P) \right \rangle = 0
%\, ,
%\end{displaymath}
%and making use of 
%\begin{displaymath}
%\partial_\mu T [ J^\mu(x) J^\nu(y) ] =
% T [ (\partial_\mu J^\mu(x)) J^\nu(y) ] 
%+ \delta^4(x-y) (-Q_{J^\nu}) J^\nu 
%\, 
%\end{displaymath}
%one derives the condition of gauge invariance
%for the correlator function
%\protect\req{eq:correlator}.
%Here
%$Q_{J^\nu}$ is the difference of the charges of all fields 
%(and their derivatives)
%and of all field adjoints (and their derivatives) in $J^\nu$.
%For the electromagnetic current
%\begin{displaymath}
%Q_{j^\nu}=Q_q-Q_q=0
%\,,
%\end{displaymath}
%while
%$\partial_\mu j^\mu(x)=0 $.
%For nucleon interpolation current \req{eq:nucleon-curr}
%\begin{displaymath}
%Q_{\eta}=\frac{2}{3}+\frac{2}{3}-\frac{1}{3}=1
%\,.
%\end{displaymath}
%When one considers only the part $(2/3 \, \bar{u} \gamma_\mu u)$
%of the electromagnetic current 
%\req{eq:current} then
%\begin{displaymath}
%Q_{\eta}=\frac{2}{3}+\frac{2}{3}+0=\frac{4}{3}
%\,,
%\end{displaymath}
%while for the part $(-1/3 \, \bar{d} \gamma_\mu d)$
%\begin{displaymath}
%Q_{\eta}=0+0+\frac{-1}{3}=-\frac{1}{3}
%\,.
%\end{displaymath}
%},
the condition of gauge invariance takes for the correlator function
\req{eq:correlator} the form
\begin{equation}
q_\mu T^\mu = Q_{\eta} \, \left\langle 0 \right| \eta(0) \left| N(P) \right \rangle
\, ,
\label{eq:gaugecond}
\end{equation}
with $Q_\eta=1$ for the proton interpolation current
 and full electromagnetic current
\req{eq:current}, 
while 
when one considers $e_u$- and $e_d$-proportional parts of the 
electromagnetic current separately 
$Q_\eta=2 e_u=4/3$ and $Q_\eta=e_d=-1/3$,
respectively (in the neutron case, as usual, $e_u \leftrightarrow e_d$).
Furthermore for Ioffe current we have
\begin{equation}
\left\langle 0 \right| \eta(0) \left| N(P) \right \rangle
= \lambda_1 \, M_N \, N(P)
\, .
\label{eq:Ioffematr}
\end{equation}

In the preceding subsections we have presented the 
complete LO twist-3 and twist-4 contributions to the correlator
function, i.e., in contrast to Ref. \cite{BraunLW06},
not just the terms corresponding to the functions
of interest ${\cal A}$ and ${\cal B}$.
This enables us to check the gauge invariance.
By making use of the results given in Tables
\ref{t:coeffLO1}, \ref{t:coeffLO3} and \ref{t:coeffLO2}
one can easily see that for $M_N^2 \ne 0$
%for the results
%obtained using  $M_N^2 \ne 0$, or even $M_N^2=0$ but $M_N\ne0$,
the gauge invariance does not hold,
i.e., that \req{eq:gaugecond} is not satisfied,
for separate ${\cal F}^{(i)}$-proportional
terms nor for separate twists.

Let us consider the expansion in $M_N$ given as outlined
in Sec. \ref{sec:Mdependence}. 

In the $M_N=0$ approximation, taking into account
\req{eq:Ioffematr},
the gauge condition \req{eq:gaugecond} 
takes the simple form
\begin{equation}
q_\mu T^\mu = 0
\, .
\label{eq:gaugecond1}
\end{equation}
In this approximation only twist-3 contributions proportional to
$V_1$ and $A_1$ exist.
As can be easily seen from LO results listed in Table \ref{t:coeffLO1}, 
these contributions are separately gauge invariant, 
as well as, $e_u$ and $e_d$ parts separately.
The same condition applies and holds for NLO%
\footnote{ 
We will use this condition of gauge invariance to check the
NLO results and resolve the 
$\gamma_5$-ambiguity.} and higher-order contributions.

Furthermore, we have checked the gauge invariance of the 
LO results in the $M_N^2=0$ but $M_N\ne0$
approximation to which $V_1$, $V_3$, $V_2$, $A_1$, $A_3$, and $A_2$ 
DAs contribute.
This approximation corresponds to the first two terms in Eq. \req{eq:Mexpansion}. 
Using the LO results given in Tables \ref{t:coeffLO1}, \ref{t:coeffLO3}
and \ref{t:coeffLO2} and taking $M_N^2=0$,
one can show that 
\begin{equation}
q_\mu T^\mu 
= Q_\eta \, \lambda_1 \, M_N \, N(P)
\, 
\label{eq:gaugecondIoffe}
\end{equation}
is satisfied when the sum of all contributing terms 
is taken into account (i.e., both twist-3 and twist-4 contributions)
and the asymptotic forms of twist-3 DAs% 
\footnote{If one, as natural, demands gauge invariance of $e_u$ and $e_d$
terms separately then $V_{1}^{d}=1/3$ and $A_{1}^{u}=0$ is enforced,
while for the sum of $e_u$ and $e_d$ terms $A_{1}^{u}=0$
is sufficient.} 
are used.
There are no, at least at this order
of expansion in $M_N$, conditions on twist-4 DAs.

Hence, gauge invariance can be satisfied  order by order in the
expansion in $M_N$ \req{eq:Mexpansion} with possibly some additional
conditions on the form of DAs.
For the check of higher order terms in $M_N$ one should calculate 
the complete LO higher-twist contributions to the correlation function.

%%%%%%%%%%%%%%%%%%%%%%%%%%%%%%%%%%%%%%%%%%%%%%%%%%%%%%%%%%%%%%%%%%%%%%%%%%%%%%%
\section{NLO contributions}
\label{sec:NLO}
\setcounter{equation}{0}
%%%%%%%%%%%%%%%%%%%%%%%%%%%%%%%%%%%%%%%%%%%%%%%%%%%%%%%%%%%%%%%%%%%%%%%%%%%%%%%

We are finally ready to address in this section the NLO contributions
to the correlator function $T_\mu$ \req{eq:correlator}.
We will present the results obtained in $M_N=0$ approximation
(corresponding to the first term in \req{eq:Mexpansion}),
discuss the problems encountered in $M_N \ne 0$ but $M_N^2=0$ approximation 
(corresponding to the first two terms in \req{eq:Mexpansion}),
and outline the obstacles present in general $M_N^2\ne 0$ calculation.

\subsection{Topological structure}
%%%%%%%%%%%%%%%%%%%%%%%%%%%%%%%%%

\begin{figure}
\begin{center}
\includegraphics{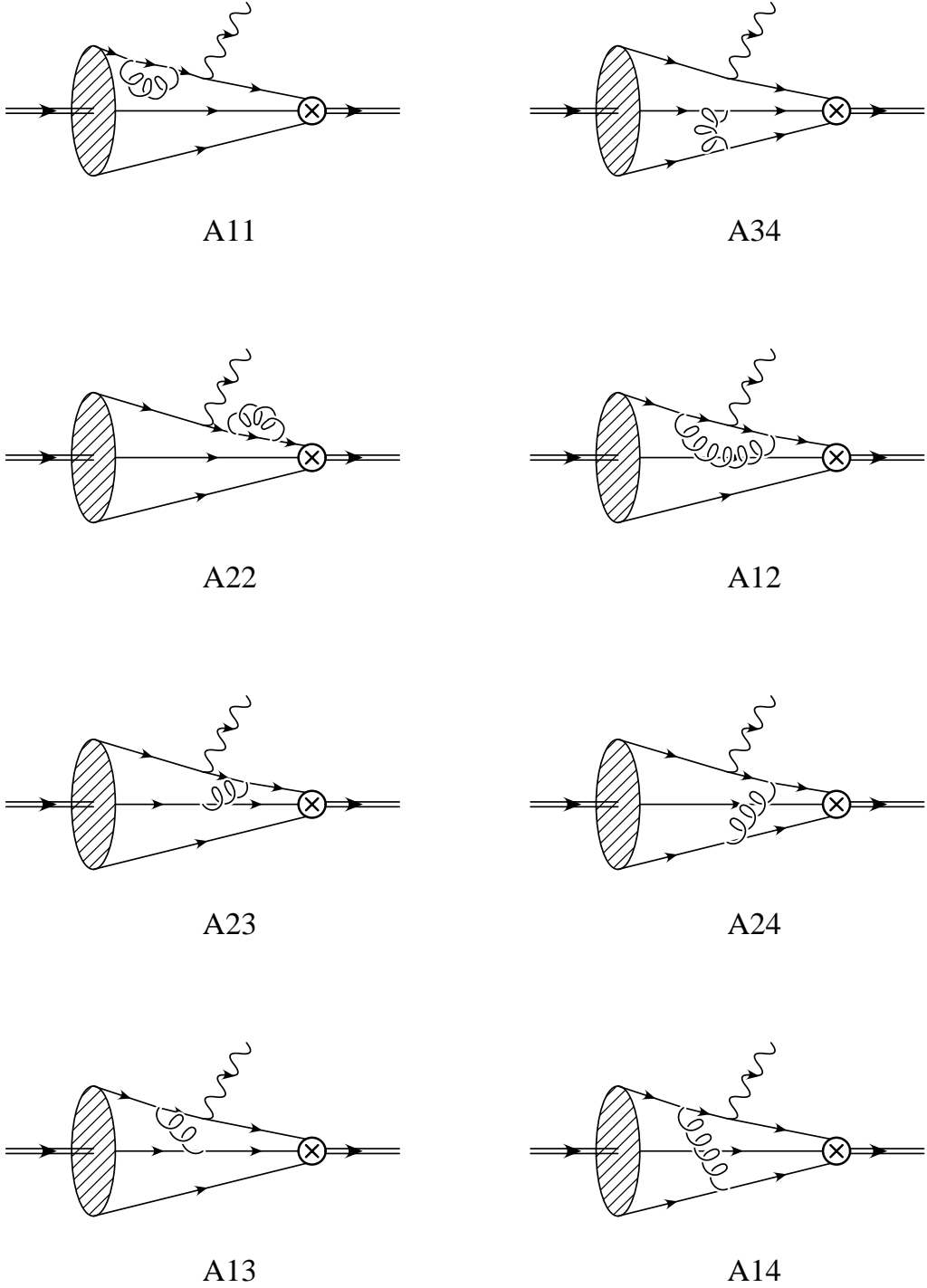}
\end{center}
\caption{Typical NLO diagrams. The contributions of ``self-energy`` diagrams
$A33$ and $A44$ is equal to the contribution of diagram $A11$. 
Diagrams $Bij$ and $Cij$ are similar to the ones presented above, 
while the nomenclature is sketched on Fig. \protect\ref{f:LO}.}
\label{f:NLO}
\end{figure}

Three LO diagrams displayed in Fig. \ref{f:LO} lead to $30$ NLO diagrams.
The nomenclature we use can be deduced from Fig. \ref{f:LO} (diagram B):
at NLO the gluon is attached in all possible ways.
Typical NLO diagrams are presented in Fig. \ref{f:NLO}.

The contributions of $Bij$ diagrams one can
obtain from the contribution of $Aij$ diagrams by 
$u_1 \leftrightarrow u_2$ exchange analogous to \req{eq:BtoA}.
The similar relation connects diagrams $C24$ and $C23$,
as well as, $C14$ and $C13$.
Taking all this into account, there are 8 more complicated
diagrams to calculate ($A12$, $A23$, $A24$, $A13$, $A14$, $C12$, $C24$, $C14$)
and the rest are either proportional to LO ($X11$, $X33$, $X44$, $X22$, $X34$)
or obtainable from above mentioned symmetries.

%%%%%%%%%%%%%%%%%%%%%%%%
%%%%%%%%%%%%%%%%%%%%%%%%

Our calculation is performed in Feynman gauge.
The colour factors were calculated using usual 
$SU(N_C=3)$-algebra relations.
In the first group of diagrams where gluon couples to the same quark line
the colour factor is $C_F=(N_C^2-1)/(2 N_C)=4/3$
while in the second in which gluon  connects two different lines
colour factors equal $(-C_B)=-(N_C+1)(2 N_C)=-2/3$.
As mentioned in App. \ref{sec:appFEYN}, since we are describing 
nucleon as colour singlet state of three quarks ($\varepsilon^{abc}$),
the $N_C=3$ choice is enforced.

\subsection{Using dimensional regularization and resolving $\gamma_5$ ambiguity}
\label{sec:Dgamma5}
%%%%%%%%%%%%%%%%%%%%%%%%%%%%%%%%%%%%%%%%%%%%%%%%%%%%%%%%%%%%%%%%%%%%%%%%%%%

In the following NLO calculation 
we take 
\begin{equation}
P^2=M_N^2=0
\label{eq:M2approx}
\end{equation}
%(approximation equivalent to considering first two terms in \req{eq:Mexpansion})
and consider diagrams with  massless on-shell external legs.
Hence, apart from UV divergences the IR divergences of collinear
type appear (there are no ``true'' IR divergences).
Both divergences are regularized using dimensional regularization 
in
\[ D=4-2 \epsilon \]
dimensions.

We introduce the abbreviations% 
\begin{subequations}
\begin{eqnarray}
     \Gamma_{UV}(\eps)&=&\Gamma(\eps)
\frac{\Gamma(1-\eps) \Gamma(1-\eps)}{\Gamma(1- 2 \eps)}
          (4 \pi )^{\eps} 
= \frac{1}{\eps} - \gamma + \ln (4 \pi) + O(\eps)
\, ,
\label{eq:Gamma0UV} \\[0.2cm]
     \Gamma_{IR}(\eps)&=&\Gamma(1+\eps)
\frac{\Gamma(-\eps) \Gamma(1-\eps)}{\Gamma(1- 2 \eps)}
          (4 \pi )^{\eps}
= \frac{1}{- \eps} + \gamma - \ln (4 \pi) + O(\eps)
        \, .
\label{eq:Gamma0IR}
\end{eqnarray}
\label{eq:Gamma0}
\end{subequations}
The first $\Gamma$ function on the right-hand side of
Eqs. \req{eq:Gamma0} originates
from the loop momentum integration, while
the integration over Feynman parameters
produces $\Gamma$s collected in a fraction.
Consequently, the singularity contained in $\Gamma(\eps)$
appearing in \req{eq:Gamma0UV} is of UV origin, while the
singularity contained in
$\Gamma(-\eps)$ appearing in \req{eq:Gamma0IR}
is of infrared (IR), i.e., collinear, origin%
\footnote{The UV divergent integrals are finite 
in $D<4$ dimensions, while IR ones in $D>4$ dimensions.
Since we regularize both in $D=4-2 \eps$ dimensions,
$\Gamma(\eps)$ represents a signature of UV divergence, and 
$\Gamma(-\eps)$ of the IR one.}.
From
$
  \Gamma(z) \Gamma(1-z) = \pi/\sin \pi z
$
one can see that to all orders in $\eps$
\begin{equation}
\Gamma_{UV}(\eps)=-\Gamma_{IR}(\eps)
         \, .
\label{eq:G0uvG0ir}
\end{equation}
Nevertheless, we find it useful to keep track of the
origin of the UV and collinear singularities
(for details, see also Refs. \cite{MelicNP01,Smirnov04}).

In dimensional regularization the ``trivial'' self-energy diagrams 
($X11$, $X33$, $X44$), as well as, ``trivial'' 3-point integral ($X34$) 
vanish if one allows that UV and IR divergences to cancel.
%\footnote{
%Namely, these contributions  simplify to a one point integral
%$I_{1}^{[n]}$ 
%\begin{displaymath}
%I_{1}^{[n]}(\eps)=
%  \mu^{2\epsilon} \int \frac{d^D l}{(2 \pi)^D}
%  \frac{1}{[l^2 + i \eta]^n}
%%\label{eq:1point}
%\end{displaymath}
%which in dimensional regularization equates to 0.
%Nevertheless, in the spirit of distinguishing the UV and IR
%singularities (as shown in \cite{MelicNP01} and discussed
%in, for example, \cite{Smirnov04}, Sec. 2.6),
%$I_{1}^{[n]}$ takes the form
%\begin{displaymath}
%I_{1}^{[n]}(\eps)= \left\{
%\begin{array}{l}
%\displaystyle
%\frac{i}{(4 \pi)^2} 
%  \left(\frac{-q^2-i\eta}{\mu^2} \right)^{-\eps}
% (\Gamma_{UV}(\eps)+\Gamma_{IR}(\eps))
%\;
%    \frac{2-\eps}{2(1-2\eps)} 
% \qquad\mbox{for $n=2$}
%\\
%0\qquad \mbox{for $n \ne 2$}
%\end{array}
%\right.
%%\label{eq:I1,2}
%\end{displaymath}
%where $q^2$ is any scale which satisfies $q^2\ne0$.
%Note that when \req{eq:G0uvG0ir} is taken into account 
%$n=2$ case simplifies to 0 as expected.}.
However, as mentioned above, we adopt here an approach of
consistent tracking of UV and collinear singularities,
and their separate removal by renormalization and factorization,
respectively.

Additionally, 
when calculating
the contributions to $T_{\mu}$ corresponding to $A_i$ distribution
amplitudes
we encounter $\gamma_5$-ambiguity 
-- see App. \ref{sec:gamma5-app} for details. 
In these cases the general Lorentz decomposition \req{eq:gLdecomp}
and the choice of nucleon interpolating current \req{eq:nucleon-curr} 
lead to the appearance of the traces with one $\gamma_5$ matrix.
At NLO these traces contain contracting $\gamma$ matrices and, as such,
trace $\gamma_5$-ambiguity. Moreover, after the trace operation is
performed one is left with one or more Levi-Civita tensors 
which get contracted with additional $\gamma$ matrices.
Hence, the ambiguity related to the use of Chisholm identity is also present.

We choose to use the naive-$\gamma_5$ scheme \cite{ChanowitzFH79}. 
We could choose to use HV scheme \cite{tHooftV72,BreitenlohnerM77}
but then we would have to know
or somehow calculate the terms which  remove ``spurious`` anomalies
violating Ward identities. Moreover we would have to use HV scheme also
for the calculation of otherwise nonproblematic contributions 
corresponding to $V_i$.

We remember that the choice of general decomposition
\req{eq:gLdecomp} is not unique and that using Fierz transformations
one could get the representation in which 
there is no trace and as such no $\gamma_5$ ambiguity
(no trace ambiguity and no appearance of Levi-Civita tensor).
Hence, the intermediate appearance of the problems with $\gamma_5$ are caused 
by our choice of the Lorentz decomposition of the nucleon matrix element 
and by the choice of the interpolating current. 
One can, for example, use the
Lorentz decomposition of the form
$X_{\gamma \beta}^{(i)} Y_{\alpha}^{(i)}$
(see App. A in Ref. \cite{BraunFMS00} for useful relations)
which when used with \req{eq:nucleon-curr}
does not lead to the appearance of the trace.
But the price to pay when using this decomposition are
much larger expressions,
$\eps$ proportional terms at LO even for $V_1$ and $A_1$ etc..

Nevertheless, that possibility lead us to the correct
way to handle contractions using Chisholm identity: 
''follow'' the fermion line (as usual, that means to go opposite the fermion line) 
and always perform the contraction of the Levi-Civita
with the ''last'' $\gamma$ matrix (with an open index) on the $d$ line.
The generalized recipe follows that one should  write also the traces 
as a part of an expression obtained following the fermion lines
-- remember that the essence of trace ambiguity is loosing
the cyclicity of the trace -- see App. \ref{sec:gamma5-app}. 

When calculating NLO contributions in $M_N=0$ approximation,
we have used the gauge invariance (\ref{eq:gaugecond})
as a check and a help to resolve $\gamma_5$-ambiguity.
After adopting this simple recipe 
-- {\it to write all parts of expression ''following'' the fermion lines:
opposite d-line, along u-line 
($C \cdots C^{-1}$ present - see App. \ref{sec:appFEYN}), 
opposite u-line, and to perform all evaluations obeying that order } -- we obtain the gauge invariant NLO results as one should.

%%%%%%%%%%%%%%%%%%%%%%%%%%%%%%%%%%%%%%%%%%%%%%%%%%%%%%%%%%%%%%%%%%%%%%%%%%%%%%%
\subsection{Twist-3 and $M_N=0$}
\label{sec:tw3NLO}
%%%%%%%%%%%%%%%%%%%%%%%%%%%%%%%%%%%%%%%%%%%%%%%%%%%%%%%%%%%%%%%%%%%%%%%%%%%%%%%
\label{sec:firstcaseMN0}

Let us first investigate the approximation in which we 
neglect nucleon mass completely and take consistently
throughout the calculation $M_N=0$.
As can be seen from \req{eq:Mdep}, the general Lorentz decomposition
of nucleon matrix element of three quark operator
\req{eq:gLdecomp} has for $M_N=0$ only three terms: the ones proportional to 
${\cal V}_1$, ${\cal A}_1$, and ${\cal T}_1$.
The tensor contributions vanish for our choice of interpolating nucleon current,
and we are left with two contributions
convoluted with $V_1$ and $A_1$:
\begin{equation}
\left. T_{\mu} \right|_{M_N=0}
={\cal M}_{\mu}^{V1}(u_1,u_2,u_3) \otimes V_1(u_1,u_2,u_3) 
 + {\cal M}_{\mu}^{A1}(u_1,u_2,u_3) \otimes A_1(u_1,u_2,u_3)
\, .
\label{eq:Tm0}
\end{equation}

As explained in, for example, Ref. \cite{BraunFMS00},
since $V_1$ and $A_1$ have different symmetry properties
(see \req{eq:Fsym}), they can be combined together
to define a single independent twist-3 nucleon distribution
amplitude:
\begin{equation}
\Phi_3(x_1,x_2,x_3) = V_1(x_1,x_2,x_3)- A_1(x_1,x_2,x_3)
\, .
\label{eq:Phi}
\end{equation}
Now, by taking into account that the convolution of symmetric
and antisymmetric functions gives 0, as well as, relation \req{eq:Phi}
and symmetry properties \req{eq:Fsym},
one can write
 \req{eq:Tm0}
as 
\begin{eqnarray}
\left. T_{\mu} \right|_{M_N=0}
&=&\frac{1}{2} \left( {\cal M}_{\mu}^{V1}(u_1,u_2,u_3) 
                   + {\cal M}_{\mu}^{V1}(u_2,u_1,u_3) 
               \right) \otimes V_1(u_1,u_2,u_3) 
\nonumber \\ & &
               + \frac{1}{2}\left({\cal M}_{\mu}^{A1}(u_1,u_2,u_3) 
                   - {\cal M}_{\mu}^{A1}(u_2,u_1,u_3) 
               \right) \otimes A_1(u_1,u_2,u_3) 
\nonumber \\ 
&=&\frac{1}{2} \left( {\cal M}_{\mu}^{V1}(u_1,u_2,u_3) 
                   + {\cal M}_{\mu}^{V1}(u_2,u_1,u_3)
                   - {\cal M}_{\mu}^{A1}(u_1,u_2,u_3) 
                   + {\cal M}_{\mu}^{A1}(u_1,u_2,u_3) 
               \right)
\nonumber \\ & &
\otimes \, \Phi_3(u_1,u_2,u_3) 
\nonumber \\ 
&=& {\cal M}_{\mu}^{\Phi}(u_1,u_2,u_3) \otimes  \Phi_3(u_1,u_2,u_3) 
\, .
\label{eq:Tm0-Phi}
\end{eqnarray}
From \req{eq:Tm0-Phi} the definition of ${\cal M}_{\mu}^{\Phi}$ is obvious.

\subsubsection{Renormalization and factorization of collinear singularities:
$\Phi_3$}

We start here with the explanation of renormalization procedure
for \req{eq:Tm0-Phi} in which $T_{\mu}$ is given in terms
of convolution of only two functions
${\cal M}_{\mu}^{\Phi}(u_1,u_2,u_3)$ and  $\Phi_3(u_1,u_2,u_3)$.
We follow closely Ref. \cite{MelicNP01}.

The amplitude ${\cal M}_{\mu}^{\Phi}(u_1,u_2,u_3)$
is of the general form
\begin{subequations}
\label{eq:Mexp1}
\begin{equation}
{\cal M} =  Z_{\rm curr} \;
     \left[ {\cal M}^{{\rm LO}} + \frac{\alpha_s}{4 \pi}  {\cal M}^{{\rm NLO}} 
           + \cdots \right]
\, ,
\end{equation}
where
\begin{eqnarray}
{\cal M}^{{\rm LO}}&=& a_0 + \eps a_1 + {\cal O}(\eps^2)
\, ,
 \\[0.19cm]
{\cal M}^{{\rm NLO}}&=& \left\{
\Gamma_{UV}(\eps) \left[ b_0^{UV} + \eps b_1^{UV} + {\cal O}(\eps^2) \right]
+
\Gamma_{IR}(\eps) \left[ b_0^{IR} + \eps b_1^{IR} + {\cal O}(\eps^2) \right]
\right\} \left( \frac{\mu^2}{Q^2} \right)^\eps
\, ,
\nonumber \\ & &
\end{eqnarray}
\end{subequations}
and ${\cal M}^{{\rm LO}}$ and ${\cal M}^{{\rm NLO}}$ are
calculated from LO and NLO diagrams from Figs. \ref{f:LO} and \ref{f:NLO},
respectively.

The bare coupling constant $\alpha_s$ can be defined in terms
of the running coupling constant $\alpha_s(\mu_R^2)$
as%
\footnote{
In this as in the rest of the presentation we
prefer to retain all terms in the expansion over $\epsilon$.
}
\begin{equation}
\alpha_s=\alpha_s(\mu_R^2) 
   \left(1 - \frac{\alpha_s(\mu_R^2)}{4 \pi} \beta_0 \frac{1}{\eps} \right)
   \left( \frac{\mu_R^2}{\mu^2} \right)^{\eps} 
   \left[ \eps \Gamma_{UV}(\eps) \right]^{-1}
\, ,
\end{equation}
and to the order we are calculating this essentially means that
the bare coupling is replaced by the renormalized one and 
no singularities are removed as yet.

The expansion of the amplitude ${\cal M}$ takes the form
\begin{subequations}
\begin{equation}
{\cal M} = Z_{\rm curr} \; \left\{ {\cal M}^{{\rm LO}} 
           + \frac{\alpha_s(\mu_R^2)}{4 \pi}  {\hat {\cal M}}^{{\rm NLO}} 
           + \cdots \right\}
\, ,
\end{equation}
where
\begin{eqnarray}
{\cal M}^{{\rm LO}}&=& a_0 + \eps a_1 + {\cal O}(\eps^2)
\, ,
\\[0.19cm]
{\hat {\cal M}}^{{\rm NLO}}&=& \left\{
\frac{1}{\eps} \left[ b_0^{UV} + \eps b_1^{UV} + {\cal O}(\eps^2) \right]
+
\frac{1}{-\eps} \left[ b_0^{IR} + \eps b_1^{IR} + {\cal O}(\eps^2) \right]
\right\} \left( \frac{\mu_R^2}{Q^2} \right)^\eps
\, .
\qquad
\end{eqnarray}
\end{subequations}

In our case the coefficients of the $1/\eps$ poles of UV origin, i.e.,
$b_0^{UV}$, are removed by renormalization 
of the nucleon interpolating current (Ioffe current in this calculation).
For that purpose $Z_{\rm curr}$ has been introduced and it is of the form
\begin{equation}
Z_{\rm curr} = 
  1 - \alpha_s(\mu_{R,1}^2) \frac{C_{\rm curr}^{(1)}}{\eps} + {\cal O}(\alpha_s^2) 
\, ,
\end{equation}
where $\mu_{R,1}^2$ is a scale at which the nucleon interpolation
current is renormalized.
The change of the scale of the renormalization constant is given by
\begin{equation}
\alpha_s(\mu^2)=\left( \frac{\mu_R^2}{\mu^2} \right)^{\eps}
    \alpha_s(\mu_R^2) \left[1 + {\cal O}(\alpha_s) \right]
\end{equation}
Hence, ${\cal M}$ takes the form
\begin{subequations}
\label{eq:Mexp2}
\begin{equation}
{\cal M} = {\cal M}^{{\rm LO}} 
           + \frac{\alpha_s(\mu_R^2)}{4 \pi}  {\widehat {\cal M}}^{{\rm NLO}} 
           + \cdots 
\, ,
\end{equation}
where
\begin{eqnarray}
{\cal M}^{{\rm LO}}&=& a_0 + \eps a_1 + {\cal O}(\eps^2)
\, ,
\end{eqnarray}
and
\begin{eqnarray}
\lefteqn{{\widehat {\cal M}}^{{\rm NLO}}}
\nonumber \\[0.19cm]
&=& \left\{
\frac{1}{\eps} \left[  
    \left( b_0^{UV} 
 - C_{\rm curr}^{(1)} \, a_0 \left( \frac{Q^2}{\mu_{R,1}^2} \right)^\eps \right)
 + \eps \left( b_1^{UV} 
- C_{\rm curr}^{(1)} \, a_1 
   \left( \frac{Q^2}{\mu_{R,1}^2} \right)^\eps\right) + {\cal O}(\eps^2) \right]
\right. \nonumber \\[0.19cm] & & \left.
+
\frac{1}{-\eps} \left[ b_0^{IR} + \eps b_1^{IR} + {\cal O}(\eps^2) \right]
\right\} \left( \frac{\mu_R^2}{Q^2} \right)^\eps
\, .
\end{eqnarray}
\end{subequations}
For 
\begin{equation}
    b_0^{UV}
 - C_{\rm curr}^{(1)} \, a_0 =0
\label{eq:UVcond}
\end{equation}
the $1/\eps$ poles of UV origin vanish, and as only signature
of their existence the logarithms
$C_{\rm curr}^{(1)} \, a_0 \ln (\mu_{R,1}^2/Q^2)$ 
will remain in the end result.
Notice that we have shown here that in principle coupling constant
renormalization and the renormalization of the current can be 
performed at different scales, $\mu_R^2$ and $\mu_{R,1}^2$.
Generally, we write the amplitude ${\cal M}$
as an expansion in $\alpha_s(\mu_R^2)$ and independent
of $\mu_R^2$. The truncation of this series in actual calculation 
will introduce the dependence of the results on $\mu_R^2$
(frequently discussed in the literature). 

The remaining collinear singularities get cancelled by the renormalization
constant of the nucleon distribution amplitude.
Namely,
\begin{eqnarray}
T_{\mu}\left|_{M_N=0}\right.
&=& {\cal M}_{\mu}^{\Phi}(u_1,u_2,u_3) 
\otimes \,  \Phi_3(u_1,u_2,u_3)
\nonumber \\
&=& {\cal M}_{\mu}^{\Phi}(u_1,u_2,u_3) 
\otimes \, Z_{\Phi}(u_1,u_2,u_3;x_1,x_2,x_3;\mu_F^2)
\otimes \, \Phi_3(x_1,x_2,x_3;\mu_F^2)
\nonumber \\
&=& {\cal M}_{\mu}^{\Phi}(x_1,x_2,x_3;\mu_F^2) 
\otimes \, \Phi_3(x_1,x_2,x_3;\mu_F^2)
\, .
\end{eqnarray}
The renormalization constant $Z_{\Phi}$ is of the form
\begin{equation}
Z_{\Phi}(\{u_k\},\{x_k\};\mu_F^2) = 
  \delta(u_1-x_1) \delta(u_2-x_2) 
 - \alpha_s(\mu_F^2) \frac{V_{\Phi}^{(1)}(\{u_k\},\{x_k\})}{-\eps} 
 + {\cal O}(\alpha_s^2) 
\, ,
\end{equation}
where $V_{\Phi}^{(1)}$ is a leading term of
the kernel of the evolution equation for twist-3 DA:
\begin{equation}
\mu^2 \frac{\partial}{\partial \mu^2} \Phi_3(\{u_k\}; \mu^2) =
  V_{\Phi}(\{u_k\},\{x_k\}; \mu^2) \, \otimes\Phi_3(\{x_k\}; \mu^2)
\, ,
\label{eq:Phi-evoleq}
\end{equation}
and
\begin{equation}
V_{\Phi}(\{u_k\};\{x_k\}; \mu^2) =
\frac{\alpha_s(\mu^2)}{4 \pi} \, V_{\Phi}^{(1)}(\{ u_k\};\{x_k\})
+ {\cal O}(\alpha_s)
\, .
\end{equation}
The kernel $V_{\Phi}^{(1)}$ was given in Ref. \cite{LepageB80}
and confirmed, for example, by the calculation of anomalous dimensions
in Ref. \cite{Nyeo92}.
Here we present it in a convenient form
\begin{eqnarray}
\lefteqn{V^{(1)}_{\Phi}(\{u_k\},\{x_k\})} 
\nonumber \\ 
&=& 
- \frac{3}{2} C_F \delta(u_1-x_1) \delta(u_2-x_2)
\nonumber \\ & &
+ C_F \left\{
\delta(u_3-x_3) \left[
\frac{u_1}{x_1} \frac{1}{x_1-u_1} \theta(x_1-u_1) 
+ \frac{u_2}{x_2} \frac{1}{x_2-u_2} \theta(x_2-u_2) 
\right]
+ \left(
\begin{array}{l}
3 \leftrightarrow 1 \\
3 \leftrightarrow 2 
\end{array}
\right)
\right\}_+
\nonumber \\ & &
+ 2 C_B \left\{
\delta(u_3-x_3) \left[
\frac{u_1}{x_1}  \theta(x_1-u_1) 
+ \frac{u_2}{x_2} \theta(x_2-u_2) 
\right] \frac{\delta_{h_1 \bar{h}_2}}{u_1+u_2}
+ \left(
\begin{array}{l}
3 \leftrightarrow 1 \\
3 \leftrightarrow 2 
\end{array}
\right)
\right\}
\, ,
\label{eq:Vphi}
\end{eqnarray}
where $h_i=-\bar{h}_i$ is a helicity of quark $i$,
while 
\begin{equation}
\left \{ F(\{u_k\},\{x_k\}) \right \}_+ = F(\{u_k\},\{x_k\}) 
- \delta(u_1-x_1) \delta(u_2-x_2) \int {\cal D}z F(\{z_k \},\{x_k\})
\, .
\end{equation}
We stress that one should take $C_F=2 C_B=4/3$.
Furthermore, in \req{eq:Vphi} one has to take
\begin{equation}
 \delta_{h_1 \bar{h}_2}=1 \, , \quad
 \delta_{h_2 \bar{h}_3}=1 \, , \quad
 \delta_{h_1 \bar{h}_3}=0 \, ,
\end{equation}
i.e., the helicities of the quarks correspond to 
\begin{equation}
u_{\uparrow (\downarrow)} u_{\downarrow (\uparrow)} d_{\uparrow(\downarrow)}
\, ,
\end{equation}
which is in agreement with App.C from Ref. \cite{BraunLW06}.

For the finite amplitude ${\cal M}$
one then gets
\begin{subequations}
\begin{equation}
{\cal M}(\{x_k\}; \mu_F^2) = 
      \left\{ {\cal M}^{{\rm LO}}(\{x_k\})
    + \frac{\alpha_s(\mu_R^2)}{4 \pi}  
        {\cal M}^{{\rm NLO}} (\{x_k\}; \mu_F^2,\mu_{R,1}^2)
    + \cdots \right\} 
\, ,
\end{equation}
where
\begin{eqnarray}
{\cal M}^{{\rm LO}}(\{x_k\})&=& a_0 + \eps a_1 + {\cal O}(\eps^2)
\, ,
\end{eqnarray}
\begin{eqnarray}
\lefteqn{{\cal M}^{{\rm NLO}} (\{x_k\}; \mu_F^2,\mu_{R,1}^2)}
\nonumber \\ 
&=& \left\{
\frac{1}{\eps} \left[  
    \left( b_0^{UV}
 - C_{\rm curr}^{(1)}  a_0 \left( \frac{Q^2}{\mu_{R,1}^2} \right)^\eps \right)
 + \eps \left( b_1^{UV} 
  - C_{\rm curr}^{(1)} a_1 \left( \frac{Q^2}{\mu_{R,1}^2} \right)^\eps \right) + {\cal O}(\eps^2) \right]
\right. \nonumber \\ & & \left.
+
\frac{1}{-\eps} \left[ 
\left(b_0^{IR} 
    - a_0 \, \otimes V^{(1)}\left( \frac{Q^2 }{\mu_F^2} \right)^\eps \right)
+ \eps \left( b_1^{IR} 
  - a_1 \otimes V^{(1)} \left( \frac{Q^2}{\mu_F^2} \right)^\eps \right)+ {\cal O}(\eps^2) \right]
\right\} 
\nonumber \\ & & \times
\left( \frac{\mu_R^2}{Q^2} \right)^\eps
\, ,
%\nonumber \\
\end{eqnarray}
\end{subequations}
where $\mu_F$ is a (usual) factorization scale
and the collinear singularities cancel for
\begin{equation}
b_0^{IR} - a_0 \, \otimes V^{(1)} = 0
\, .
\label{eq:IRcond}
\end{equation}

With all singularities cancelled,
we can now take the limit $\eps \rightarrow 0$ and finally obtain
\begin{subequations}
\label{eq:finiteM-Phi}
\begin{equation}
{\cal M}(\{x_k\}; \mu_F^2) = 
      \left\{ {\cal M}^{{\rm LO}}(\{x_k\})
    + \frac{\alpha_s(\mu_R^2)}{4 \pi}  
        {\cal M}^{{\rm NLO}} (\{x_k\}; \mu_F^2, \mu_{R,1}^2)
    + \cdots \right\} 
\, ,
\end{equation}
with
\begin{eqnarray}
{\cal M}^{{\rm LO}}(\{x_k\})&=& a_0 
\, ,
 \\[0.19cm]
{\cal M}^{{\rm NLO}} (\{x_k\}; \mu_F^2, \mu_{R,1}^2)&=& 
\left( b_1^{UV} - C_{\rm curr}^{(1)} a_1 \right)
-\left( b_1^{IR} - a_1 \otimes V^{(1)}\right)
\nonumber \\[0.19cm] & &
+ C_{\rm curr}^{(1)} a_0 \ln(\mu_{R,1}^2/Q^2) 
- a_0 \otimes V^{(1)} \ln(\mu_F^2/Q^2) 
\, .
\quad
\label{eq:finiteM-NLO}
\end{eqnarray}
\end{subequations}
For Ioffe current
\begin{equation}
C_{\rm curr}^{(1)} = C_{\rm Ioffe}^{(1)} =2
\, .
\end{equation}
Furthermore, one can check in Table \ref{t:LO} that
for our twist-3 results there are no $\eps$ proportional 
LO contribution (in contrast to, for example, twist-4 contributions)
%\begin{equation}
and 
$
a_1=0
$.
%\, .
%\end{equation}
Taking this into account along with the conditions for
cancelling UV \req{eq:UVcond}
and collinear singularities \req{eq:IRcond}, simplifies
the NLO result \req{eq:finiteM-NLO} to
\begin{equation}
{\cal M}^{{\rm NLO}} (\{x_k\}; \mu_F^2, \mu_{R,1}^2)= 
 b_1^{UV} -  b_1^{IR} 
+ b_0^{UV} \ln(\mu_{R,1}^2/Q^2) 
-  b_0^{IR} \ln(\mu_F^2/Q^2) 
\, .
\label{eq:FiniteM-NLO}
\end{equation}

\subsubsection{Renormalization and factorization of collinear singularities:
$V_1$ and $A_1$}

Although \req{eq:Tm0-Phi} and the analysis given in the preceding subsection
are sufficient for obtaining twist-3 results,
we now turn to renormalization procedure for $T_{\mu}$ expressed
in terms of $V_1$ and $A_1$. i.e, as in \req{eq:Tm0}:
\begin{displaymath}
\left. T_{\mu} \right|_{M_N=0}
={\cal M}_{\mu}^{V1}(u_1,u_2,u_3) \otimes V_1(u_1,u_2,u_3) 
 + {\cal M}_{\mu}^{A1}(u_1,u_2,u_3) \otimes A_1(u_1,u_2,u_3)
\, .
\end{displaymath}
The crucial difference in comparison to the preceding subsection
is that $T_{\mu}$ is no longer expressed as just one convolution 
but rather as a sum of convolutions. As we shall see, there is a mixing
between these terms. The procedure is similar to the one used
in Ref. \cite{KrollPK02} and, 
although the finite results are the same as the ones in preceding 
subsection, the experience that we gain in this subsection
should be very useful for the $M_N \ne 0$ case.

It is instructive to write \req{eq:Tm0} in a matrix form:
\begin{equation}
\left. T_{\mu} \right|_{M_N=0}
=\left( 
{\cal M}_{\mu}^{V1}(u_1,u_2,u_3), {\cal M}_{\mu}^{A1}(u_1,u_2,u_3) 
 \right ) \otimes 
\left(
\begin{array}{l}
V_1(u_1,u_2,u_3)\\
A_1(u_1,u_2,u_3)
\end{array}
\right)
\, .
\label{eq:Tmu0-matrix}
\end{equation}
Both ${\cal M}_{\mu}^{V1}$ and ${\cal M}_{\mu}^{A1}$
can be expanded as in \req{eq:Mexp1}
and the UV renormalization of these expressions
proceeds the same way as 
explained in the previous subsection.
One ends up with the UV-finite expressions of the form
\req{eq:Mexp2}:
\begin{eqnarray}
{\cal M}_{\mu}^{V1}(\{u_k\}) &=&  {\cal M}^{{\rm LO},V1}_{\mu}(\{u_k\}) 
+ \frac{\alpha_s(\mu_R^2)}{4 \pi}  {\widehat {\cal M}}^{{\rm NLO},V1}_{\mu}(\{u_k\})
+ \cdots 
\, ,
\nonumber \\
{\cal M}_{\mu}^{A1}(\{u_k\})&=& {\cal M}^{{\rm LO},A1}_{\mu}(\{u_k\})  
+ \frac{\alpha_s(\mu_R^2)}{4 \pi}  {\widehat {\cal M}}^{{\rm NLO},A1}_{\mu}(\{u_k\}) 
+ \cdots 
\, ,
\end{eqnarray}
and the conditions \req{eq:UVcond} have to be satisfied, i.e.,
\begin{eqnarray}
    b_0^{UV,V1}
 - C_{\rm curr}^{(1)} \, a_0^{V1} &=&0
\, ,
\nonumber \\[0.19cm]
    b_0^{UV,A1}
 - C_{\rm curr}^{(1)} \, a_0^{A1} &=&0
\, .
\label{eq:UVcond2}
\end{eqnarray}

In order to cancel remaining collinear singularities one has to know
the evolution kernels, i.e., renormalization constants, for $V_1$ and $A_1$ distributions
amplitudes. One can probably derive it by additional one-loop calculation,
or, as we will do here, make use of our knowledge of $V_{\Phi}$.
Knowing the symmetry properties of $V_1$ and $A_1$ distribution
amplitudes, we write $V_{\Phi}$ as
\begin{subequations}
\begin{eqnarray}
V_{\Phi}(\{u_k\};\{x_k\};\mu^2)&=&
V_{V1,V1}(\{u_k\};\{x_k\};\mu^2) 
+V_{V1,A1}(\{u_k\};\{x_k\};\mu^2) 
\nonumber \\[0.19cm] & &
+ \, V_{A1,V1}(\{u_k\};\{x_k\};\mu^2) 
+V_{A1,A1}(\{u_k\};\{x_k\};\mu^2) 
\, , \quad
\label{eq:VVAA}
\end{eqnarray}
where
\begin{eqnarray}
V_{V1,V1}(\{u_k\};\{x_k\};\mu^2) &=& 
\frac{1}{4} \left( V_{\Phi}(u_1,u_2,u_3;x_1,x_2,x_3;\mu^2) 
+V_{\Phi}(u_1,u_2,u_3;x_2,x_1,x_3;\mu^2) 
\right. \nonumber \\[0.19cm] & & \left.
+V_{\Phi}(u_2,u_1,u_3;x_1,x_2,x_3;\mu^2) 
+V_{\Phi}(u_2,u_1,u_3;x_2,x_1,x_3;\mu^2) \right)
\, , 
\nonumber \\[0.19cm] 
V_{V1,A1}(\{u_k\};\{x_k\};\mu^2) &=& 
\frac{1}{4} \left( V_{\Phi}(u_1,u_2,u_3;x_1,x_2,x_3;\mu^2) 
-V_{\Phi}(u_1,u_2,u_3;x_2,x_1,x_3;\mu^2) 
\right. \nonumber \\[0.19cm] & & \left.
+V_{\Phi}(u_2,u_1,u_3;x_1,x_2,x_3;\mu^2) 
-V_{\Phi}(u_2,u_1,u_3;x_2,x_1,x_3;\mu^2) \right)
\, , 
\nonumber \\[0.19cm] 
V_{A1,V1}(\{u_k\};\{x_k\};\mu^2) &=& 
\frac{1}{4} \left( V_{\Phi}(u_1,u_2,u_3;x_1,x_2,x_3;\mu^2) 
+V_{\Phi}(u_1,u_2,u_3;x_2,x_1,x_3;\mu^2) 
\right. \nonumber \\[0.19cm] & & \left.
-V_{\Phi}(u_2,u_1,u_3;x_1,x_2,x_3;\mu^2) 
-V_{\Phi}(u_2,u_1,u_3;x_2,x_1,x_3;\mu^2) \right)
\, , 
\nonumber \\[0.19cm] 
V_{A1,A1}(\{u_k\};\{x_k\};\mu^2) &=& 
\frac{1}{4} \left( V_{\Phi}(u_1,u_2,u_3;x_1,x_2,x_3;\mu^2) 
-V_{\Phi}(u_1,u_2,u_3;x_2,x_1,x_3;\mu^2) 
\right. \nonumber \\[0.19cm] & & \left.
-V_{\Phi}(u_2,u_1,u_3;x_1,x_2,x_3;\mu^2) 
+V_{\Phi}(u_2,u_1,u_3;x_2,x_1,x_3;\mu^2) \right)
\, . 
\nonumber \\
\end{eqnarray}
\end{subequations}
Obviously,
\begin{equation}
\begin{tabular}{ll}
$V_{V1,V1}(\{u_k\};\{x_k\};\mu^2)$ &
$\left| \quad \mbox{ symmetric in } 
  u_1 \leftrightarrow u_2 \mbox{ and symmetric in } 
  x_1 \leftrightarrow x_2 \right. $ \\[0.19cm]
$V_{V1,A1}(\{u_k\};\{x_k\};\mu^2)$ &
$\left| \quad \mbox{ symmetric in } 
  u_1 \leftrightarrow u_2 \mbox{ and antisymmetric in } 
  x_1 \leftrightarrow x_2 \right. $ \\[0.19cm]
$V_{A1,V1}(\{u_k\};\{x_k\};\mu^2)$ &
$\left| \quad \mbox{ antisymmetric in } 
  u_1 \leftrightarrow u_2 \mbox{ and symmetric in } 
  x_1 \leftrightarrow x_2 \right. $ \\[0.19cm]
$V_{A1,A1}(\{u_k\};\{x_k\};\mu^2)$ &
$\left| \quad \mbox{ antisymmetric in } 
  u_1 \leftrightarrow u_2 \mbox{ and antisymmetric in } 
  x_1 \leftrightarrow x_2 \right. $ 
\end{tabular}
\end{equation}

We can now substitute \req{eq:Phi} and \req{eq:VVAA}
in the evolution equation \req{eq:Phi-evoleq}
and taking into account the symmetry properties
with respect to $x_k$ one gets
\begin{eqnarray}
\mu^2 \frac{\partial}{\partial \mu^2}
\left[ V_1 - A_1 \right] (\{u_k\}; \mu^2)
&=& 
\left[V_{V1,V1}+V_{V1,A1} +V_{A1,V1}+V_{A1,A1}\right](\{u_k\};\{x_k\};\mu^2) 
\nonumber \\[0.19cm] & & 
   \otimes \left[ V_1 - A_1 \right] (\{x_k\};\mu^2) 
\nonumber \\[0.5cm] 
&=& 
\left[V_{V1,V1}+V_{A1,V1} \right](\{u_k\};\{x_k\};\mu^2) \otimes V_1(\{x_k\};\mu^2) 
\nonumber \\[0.5cm] & & 
-\left[V_{V1,A1}+V_{A1,A1} \right](\{u_k\};\{x_k\};\mu^2) \otimes A_1(\{x_k\};\mu^2) 
\, .
\nonumber \\
\end{eqnarray}
Furthermore, the symmetry properties with respect to $u_k$
allows us to write the evolution equation
 in a matrix form as
\begin{eqnarray}
\mu^2 \frac{\partial}{\partial \mu^2}
\left(
\begin{array}{l}
V_1\\
A_1
\end{array}
\right) (\{u_k\};\mu^2) &=&
\left(
\begin{array}{ll}
 V_{V1,V1} & - V_{V1,A1} \\
 - V_{A1,V1} & V_{A1,A1} 
\end{array}
\right) (\{u_k\};\{x_k\};\mu^2) \otimes 
\left(
\begin{array}{l}
V_1\\
A_1
\end{array}
\right) (\{x_k\};\mu^2) 
\, .
\nonumber \\
\end{eqnarray}

The DAs $V_1$ and $A_1$ obviously mix under renormalization
and we can write
\begin{equation}
\left(
\begin{array}{l}
V_1\\
A_1
\end{array}
\right) (\{u_k\}) 
=
{\mathbf Z}(\{u_k\};\{x_k\};\mu^2)
\otimes
\left(
\begin{array}{l}
V_1\\
A_1
\end{array}
\right) (\{x_k\};\mu^2) 
\, ,
\label{eq:ZV1A1}
\end{equation}
where
\begin{equation}
{\mathbf Z}(\{u_k\};\{x_k\};\mu^2)
={\mathbf 1} -
\frac{\alpha_s(\mu^2)}{4 \pi} \frac{1}{-\eps} {\mathbf V}^{(1)}(\{u_k\};\{x_k\})
\, ,
\end{equation}
and
\begin{equation}
{\mathbf V}^{(1)}(\{u_k\};\{x_k\}) =
\left(
\begin{array}{ll}
 V_{V1,V1}^{(1)} & - V_{V1,A1}^{(1)} \\
 - V_{A1,V1}^{(1)} & V_{A1,A1}^{(1)} 
\end{array}
\right) (\{u_k\};\{x_k\}) 
\, .
\end{equation}

By substituting \req{eq:ZV1A1} in \req{eq:Tmu0-matrix},
we get 
\begin{eqnarray}
T_{\mu}\left|_{M_N=0} \right. &=&
\left( 
{\cal M}_{\mu}^{V1}, {\cal M}_{\mu}^{A1} 
 \right )(\{u_k\}) \, \otimes \, 
{\mathbf Z}(\{u_k\},\{x_k\}; \mu^2)
\, \otimes \,
\left(
\begin{array}{l}
V_1\\
A_1
\end{array}
\right)(\{x_k\}; \mu^2)
\nonumber \\[0.19cm]
&=&
\left( 
{\cal M}_{\mu}^{V1}, {\cal M}_{\mu}^{A1} 
 \right )(\{x_k\},\mu^2) \, 
\, \otimes \,
\left(
\begin{array}{l}
V_1\\
A_1
\end{array}
\right)(\{x_k\}; \mu^2)
\, .
\label{eq:Tmu0-matrix-2}
\end{eqnarray}
The condition for cancelling collinear singularities
\req{eq:IRcond}
now takes the more involved form
\begin{eqnarray}
b_0^{IR,V1} 
- a_0^{V1} \, \otimes V^{(1)}_{V1,V1} 
+ a_0^{A1} \, \otimes V^{(1)}_{A1,V1}&=& 0
\, ,
\nonumber \\[0.19cm]
b_0^{IR,A1} 
+ a_0^{V1} \, \otimes V^{(1)}_{V1,A1} 
- a_0^{A1} \, \otimes V^{(1)}_{A1,A1}&=& 0
\, .
\label{eq:IRcond2}
\end{eqnarray}

Finally, we take $\eps \rightarrow 0$ limit
and obtain
\begin{eqnarray}
{\cal M}^{V1}_{\mu}(\{x_k\}; \mu_F^2) &=& 
      \left\{ {\cal M}^{{\rm LO},V1}_{\mu}(\{x_k\})
    + \frac{\alpha_s(\mu_R^2)}{4 \pi}  
    {\cal M}^{{\rm NLO},V1}_{\mu} (\{x_k\}; \mu_F^2, \mu_{R,1}^2)
    + \cdots \right\} 
\, ,
\nonumber \\[0.19cm]
{\cal M}^{A1}_{\mu}(\{x_k\}; \mu_F^2) &=& 
      \left\{ {\cal M}^{{\rm LO},A1}_{\mu}(\{x_k\})
    + \frac{\alpha_s(\mu_R^2)}{4 \pi}  
    {\cal M}^{{\rm NLO},A1}_{\mu} (\{x_k\}; \mu_F^2, \mu_{R,1}^2)
    + \cdots \right\} 
\, ,
\nonumber \\[0.19cm]
\label{eq:finiteM-V1A1}
\end{eqnarray}
with
\begin{eqnarray}
{\cal M}^{{\rm LO},V1}_{\mu}(\{x_k\})&=& a_0^{V1} 
\, ,
\nonumber \\[0.19cm] 
{\cal M}^{{\rm LO},A1}_{\mu}(\{x_k\})&=& a_0^{A1} 
\, .
\end{eqnarray}
In our calculation
%\begin{equation}
$
a_1^{V1} = a_1^{A1} = 0
$
%\, .
%\end{equation}
and taking this into account along with the conditions of cancellation
of UV \req{eq:UVcond2} and collinear singularities \req{eq:IRcond2},
one finally gets
\begin{eqnarray}
{\cal M}^{{\rm NLO},V1}_{\mu} (\{x_k\}; \mu_F^2, \mu_{R,1}^2)&=& 
b_1^{UV,V1} - b_1^{IR,V1} 
+ b_0^{UV,V1} \ln(\mu_{R,1}^2/Q^2) 
- b_0^{IR,V1} \ln(\mu_F^2/Q^2) 
\, ,
\nonumber \\[0.5cm] 
{\cal M}^{{\rm NLO},A1}_{\mu} (\{x_k\}; \mu_F^2, \mu_{R,1}^2)&=& 
b_1^{UV,A1} - b_1^{IR,A1} 
+ b_0^{UV,A1} \ln(\mu_{R,1}^2/Q^2) 
- b_0^{IR,A1} \ln(\mu_F^2/Q^2) 
\, ,
\nonumber \\[0.5cm] 
\label{eq:FiniteM-V1A1-NLO}
\end{eqnarray}
and when one uses the actual values for $b_i$s
one can see the agreement
with
\req{eq:FiniteM-NLO}.

\subsubsection{Results}

The cancellation of singularities for the
$M_N=0$ case has been checked and shown for two equivalent 
representations: 
one corresponding to $\Phi_3$ and the other
corresponding to $V_1$ and $A_1$ DAs.
In the latter case the mixing appears.

In App. \ref{sec:appBfunM0} we list our finite NLO results
contributing to the function of interest ${\cal B}$.
The function ${\cal A}$ cannot be accessed in $M_N=0$
approximation -- see \req{eq:projections}.
For completeness sake we list both the results corresponding
to $V_1$ and $A_1$ distribution amplitudes, as well as, to $\Phi_3$.
The latter results are shorter and actually used in
further numerical calculation.

%%%%%%%%%%%%%%%%%%%%%%%%%%%%%%%%%%%%%%%%%%%%%%%%%%%%%%%%%%%%%%%%%%%%%%%%%%%%%%%
\subsection{Away from  $M_N = 0$ approximation}
\label{sec:awzayMN0}
%%%%%%%%%%%%%%%%%%%%%%%%%%%%%%%%%%%%%%%%%%%%%%%%%%%%%%%%%%%%%%%%%%%%%%%%%%%%%%%

Next we consider the approximation $M_N \ne 0$ while $M_N^2= 0$,
i.e., the approximation corresponding to the first two terms of the 
expansion \req{eq:Mexpansion}.

We have calculated the NLO contributions corresponding to
$V_1$ and $A_1$.
The UV singularities get cancelled as in the previous
subsection. In contrast, the collinear singularities
cannot be cancelled considering only the contributions
corresponding to $V_1$ and $A_1$.
For example, both $C^{{\rm LO},V1}_{P\mu M_N}$ and
$C^{{\rm LO},A1}_{P\mu M_N}$ parts proportional to $e_d$ are $0$,
while NLO counterparts are different from $0$ and contain 
collinear singularities. Obviously, mixing between $V_1$ and $A_1$ alone 
cannot cancel these terms even if we knew the new kernels
$\bar{V}_{V1,V1}$,$\bar{V}_{V1,A1}$ etc. corresponding 
to the $M_N \ne 0$ case.

Hence, it follows that the mixing with twist-4 DAs $V_3$, $A_3$, and even maybe 
$V_2$ and $A_2$, should be taken into account in this $M_N^2=0$
but $M_N \ne 0$ approximation.
This is similar to the observation from Sec. \ref{sec:gaugeinv}
that the gauge invariant results are obtained not twist-by-twist
but order-by-order in $M_N$.
The problem which we then encounter is that we do not know
these corresponding new kernels and they play a role
not only in cancelling the singularities but also
change the finite parts ($a_1$, i.e., $\eps$ proportional
LO parts, are not $0$ for $V_3$ and $A_3$ terms). 

For example, for mixing just between $V_1$, $A_1$, $V_3$ and
$A_3$ we encounter the unknown kernel of the form
\begin{equation}
\left(
\begin{array}{llll}
\bar{V}_{V1,V1} & \bar{V}_{V1,A1} & \bar{V}_{V1,V3} & \bar{V}_{V1,A3} \\
\bar{V}_{A1,V1} & \bar{V}_{A1,A1} & \bar{V}_{A1,V3} & \bar{V}_{A1,A3} \\
\bar{V}_{V3,V1} & \bar{V}_{V3,A1} & \bar{V}_{V3,V3} & \bar{V}_{V3,A3} \\
\bar{V}_{A3,V1} & \bar{V}_{A3,A1} & \bar{V}_{A3,V3} & \bar{V}_{A3,A3} 
\end{array}
\right).
\end{equation}

By substituting the conditions for cancellation
of UV and collinear singularities, 
for the finite NLO contributions one gets
\begin{eqnarray}
{\cal M}^{{\rm NLO},V1}_{\mu} (\{x_k\}; \mu_F^2)
 &=& b_1^{UV,V1} - b_1^{IR,V1}
       + b_0^{UV,V1} \ln(\mu_{R,1}^2/Q^2) 
       - b_0^{IR,V1} \ln(\mu_F^2/Q^2) 
\nonumber \\[0.19cm] & &
+\left(a_1^{V3} \otimes \bar{V}^{(1)}_{V3,V1}
     + a_1^{A3} \otimes \bar{V}^{(1)}_{A3,V1}\right)
\, ,
\nonumber \\[0.5cm] 
{\cal M}^{{\rm NLO},A1}_{\mu} (\{x_k\}; \mu_F^2)
 &=& b_1^{UV,A1} - b_1^{IR,A1}
       + b_0^{UV,A1} \ln(\mu_{R,1}^2/Q^2) 
       - b_0^{IR,A1} \ln(\mu_F^2/Q^2) 
\nonumber \\[0.19cm] & &
+\left(a_1^{V3} \otimes \bar{V}^{(1)}_{V3,A1}
     + a_1^{A3} \otimes \bar{V}^{(1)}_{A3,A1}\right)
\, ,
\nonumber \\[0.5cm] 
{\cal M}^{{\rm NLO},V3}_{\mu} (\{x_k\}; \mu_F^2)
 &=& b_1^{UV,V3} - b_1^{IR,V3}
       + b_0^{UV,V3} \ln(\mu_{R,1}^2/Q^2) 
       - b_0^{IR,V3} \ln(\mu_F^2/Q^2) 
\nonumber \\[0.19cm] & &
+\left(a_1^{V3} \otimes \bar{V}^{(1)}_{V3,V3}
     + a_1^{A3} \otimes \bar{V}^{(1)}_{A3,V3}\right)
- C_{\rm curr}^{(1)} \, a_1^{V3} 
\, ,
\nonumber \\[0.5cm] 
{\cal M}^{{\rm NLO},A3}_{\mu} (\{x_k\}; \mu_F^2)
 &=& b_1^{UV,A3} - b_1^{IR,A3}
       + b_0^{UV,A3} \ln(\mu_{R,1}^2/Q^2) 
       - b_0^{IR,A3} \ln(\mu_F^2/Q^2) 
\nonumber \\[0.19cm] & &
+\left(a_1^{V3} \otimes \bar{V}^{(1)}_{V3,A3}
     + a_1^{A3} \otimes \bar{V}^{(1)}_{A3,A3}\right)
- C_{\rm curr}^{(1)} \, a_1^{A3} 
\, .
\label{eq:FiniteM-V1A1V3A3-NLO}
\end{eqnarray}
But we do not know $\bar{V}^{(1)}_{V3,X}$ and $\bar{V}^{(1)}_{A3,X}$
and hence we do not know how to calculate the finite terms
\begin{equation}
a_1^{V3} \otimes \bar{V}^{(1)}_{V3,X}
+ a_1^{A3} \otimes \bar{V}^{(1)}_{A3,X}
\, ?!
\end{equation}

%%%%%%%%%%%%%%%%%%%%%%%%%%%%%%%%%%%%%%%%%%%%%%%%%%%%%%%%%%%%%%%%%%%%%%%%%%%%%%%
\subsubsection{Open problems in higher-twist calculations}
%%%%%%%%%%%%%%%%%%%%%%%%%%%%%%%%%%%%%%%%%%%%%%%%%%%%%%%%%%%%%%%%%%%%%%%%%%%%%%%

If we take $M_N^2 \ne 0$, there are no collinear divergences
whose cancellation we should take care of.
%Still gauge invariance (see Sec. \ref{sec:gaugeinv})
%will could pose restrictions on NLO results.
But there are additional open problems one should solve
before attacking higher-order higher-twist calculations. 

For example,
there is an open problem in the calculation of the NLO contributions
to second and third case defined in \req{eq:cases}.
Let us for a moment go back to coordinate space.
When calculating NLO contributions one encounters
the matrix elements of the form
\begin{equation}
\left \langle 0 \right |
\varepsilon^{abc} u_\alpha^a(x) 
               u_\beta^b(z_2) 
               d_\gamma^c(0) 
\left | N(P,\lambda) \right \rangle
\, ,
\end{equation}
and
\begin{equation}
\left \langle 0 \right |
\varepsilon^{abc} u_\alpha^a(x) 
               u_\beta^b(z_2) 
               d_\gamma^c(z_1) 
\left | N(P,\lambda) \right \rangle
\, ,
\end{equation}
where $z_1$ and $z_2$ come from the gluon coordinates.
In contrast at LO only matrix elements of the form 
\begin{equation}
\left \langle 0 \right |
\varepsilon^{abc} u_\alpha^a(x) 
               u_\beta^b(0) 
               d_\gamma^c(0) 
\left | N(P,\lambda) \right \rangle
\end{equation}
appear.
For first case of \req{eq:cases} these additional coordinates,
not apparently proportional either to $x$-coordinate 
(photon coordinate) nor $0$, do not pose a problem
but in the second (and third) case it seems not to be clear 
how to identify the $a_k$ coefficients and $x$ in \req{eq:gLdecomp}.

These and similar problems are postponed for future investigations.

%%%%%%%%%%%%%%%%%%%%%%%%%%%%%%%%%%%%%%%%%%%%%%%%%%%%%%%%%%%%%%%%%%%%%%%%%%%%%%%
\section{Light-cone sum rules}
\label{sec:LCSR}
\setcounter{equation}{0}
%%%%%%%%%%%%%%%%%%%%%%%%%%%%%%%%%%%%%%%%%%%%%%%%%%%%%%%%%%%%%%%%%%%%%%%%%%%%%%%

%%%%%%%%%%%%%%%%%%%%%%%%%%%%%%%%%%%%%%%%%%%%%%%%%%%%%%%
\subsection{LCSR for $M_N=0$ case}
%%%%%%%%%%%%%%%%%%%%%%%%%%%%%%%%%%%%%%%%%%%%%%%%%%%%%%%

In $M_N=0$ approximation we cannot asses the function 
${\cal A}$ and thus the form factor $F_1$.
The form factor $F_2(Q^2)$ is given in terms of ${\cal B}(Q^2,P'^2)$
by the sum rule \req{eq:LCSRb}
\begin{eqnarray}
 F_2(Q^2) &=& 
\frac{1}{\lambda_1 \pi }
\int_0^{s_0} ds \; e^{-s/M_B^2} \; {\rm Im} {\cal B}_{M_N=0}(Q^2,s)
\, . 
\label{eq:LCSR-MN0}
\end{eqnarray}

The function ${\cal B}$ calculated for $M_N=0$ case and to NLO 
is given in App. \ref{sec:appBfunM0}. 
Note that the convenient dimensionless quantity
\begin{equation}\label{W}
W = \frac{P'^2 + Q^2}{Q^2}
\end{equation}
has been introduced and that 
the function ${\cal B}$ has been expressed through this
new variable, ${\cal B}(Q^2,P'^2) \to {\cal B'}(Q^2,W)$.
For simplicity sake, the same name for the function ${\cal B}$
has been retained.
Thus in Eq. \req{eq:LCSR-MN0} the change of variables
$s \to w=(s+Q^2)/Q^2$ has to be performed.

Furthermore, note that $i \eta$-terms ($\eta >0$ and $\eta\ll$)
originating from the Feynman rules for quark and gluon
propagators
were explicitly kept in resulting logarithm terms
throughout perturbative calculation. 
The analogous terms in denominators 
can be easily recovered % from the original calculation 
resulting in $W \to W + i \eta$.
Hence, as it turns out, in our calculation the sign infront
$W$ and $i \eta$ is the same both in denominators and logarithm terms%
\footnote{In practice, one often suppresses $i \eta$ terms 
during calculation and recovers them when the analytical
continuation is needed. This approach can in some cases 
(when more complicated functions appear) be non-trivial
and can even lead to mistakes. Actually, it is much simpler just to keep track
of $i \eta$ terms from Feynman diagrams to resulting higher order
expressions. In our work we adopt that approach.}
.
The imaginary part can now be determined using the expressions 
listed in App. \ref{sec:appIm}.

Let us see this in more detail.
As can be seen from App. \ref{sec:appBfunM0},
the function ${\cal B}_{M_N=0}$ can be expressed in a
form of convolution 
\begin{eqnarray}
{\cal B}_{M_N=0}(Q^2,W)= 
\frac{1}{Q^2}
\sum_i \, T_{{\cal B},F^{(i)},M_N=0}(\{x_k\},W; \mu_F^2/Q^2) \otimes
                          F^{(i)}(\{x_k\};\mu_F^2)
\end{eqnarray}
with $F^{(i)}$ denoting, as before, nucleon DAs
($F^{(i)} \in \{ V_1, A_1\}$, or $F^{(i)}=\Phi_3$ when the sum has only one term 
-- see Eqs. \req{eq:convolB-V1A1} and \req{eq:convolB-Phi}).\\
The imaginary parts of $T_{{\cal B},F^{(i)},M_N=0}(\{x_k\},W;\mu_F^2/Q^2)$
determine the imaginary parts of \\ ${\cal B}_{M_N=0}(Q^2,W)$.
Furthermore, the hard-scattering part 
$T_{{\cal B},F^{(i)},M_N=0}(\{x_k\},W;\mu_F^2/Q^2)$
can be conveniently expressed as a sum of the terms of 
general form
\begin{equation}
g(\{x_k\},W) \; \; f(\{x_k\};\mu_F^2/Q^2,\mu_{R,1}^2/Q^2,\mu_R^2)
\, ,
\end{equation}
where only $g$-functions possess poles leading to imaginary parts.
The imaginary part is then determined from the imaginary part
of the function $g(\{x_k\},W)$ and, when from four integrations
present in \req{eq:LCSR-MN0} two are performed,
one obtains the following rule for separate terms contributing
to $T_{{\cal B}}$ and leading to separate terms contributing to $F_2$:
\begin{eqnarray}
 F_2(Q^2) &:& 
\frac{1}{\lambda_1 \pi } \int {\cal D}x
\int_1^{(s_0+Q^2)/Q^2} dw \; e^{-(w-1) Q^2/M_B^2} \; 
\nonumber \\[0.2cm]&& \times \;
\; {\rm Im} \,g(\{x_k\},w) \, 
\nonumber \\[0.2cm]&& \times \;
f(\{x_k\};\mu_F^2/Q^2,\mu_{R,1}^2/Q^2,\mu_R^2) \, F^{(i)}(\{x_k\};\mu_F^2) 
\nonumber \\[0.5cm]&\to&
\frac{1}{\lambda_1} 
\int_{Q^2/(Q^2+s_0)}^{1}  dx_i \int_{0}^{1-x_i} dx_j \;  
\; e^{-(1-x_i) Q^2/(x_i M_B^2)}
\nonumber \\[0.2cm]&& \times \;
\; \widetilde{g}(\{x_i,x_j,1-x_i-x_j\},Q^2,s_0,M_B^2)
\nonumber \\[0.2cm]&& \times \;
f(\{x_i,x_j,1-x_i-x_j\};\mu_F^2/Q^2,\mu_{R,1}^2/Q^2,\mu_R^2) 
\nonumber \\[0.2cm]&& \times \;
\; \;   F^{(i)} (\{x_i,x_j,1-x_i-x_j\};\mu_F^2)
\, .
\nonumber \\
\label{eq:substitution}
\end{eqnarray}
The selected $g$ functions (see Eq. \req{eq:g-functions}) that appear in our
LO and NLO calculation of ${\cal B}$ are listed
in Table \ref{tab:sub1} along with corresponding $\widetilde{g}$ functions 
which contribute to $F_2$ as shown in \req{eq:substitution}.
This table along with Eq. \req{eq:substitution}
thus provides us with necessary substitution rules
for calculation of $F_2$ from perturbatively calculated results
for ${\cal B}$ summarized in App. \ref{sec:appBfunM0}.

The resulting nucleon form factor $F_2$ calculated to NLO 
takes the form
\begin{equation}
F_2(Q^2, s_0, M_B^2; \mu_F^2, \mu_R^2, \mu_{R,1}^2)
 =
F_2^{\rm LO}(Q^2, s_0, M_B^2; \mu_F^2)
 + 
F_2^{\rm NLO}(Q^2, s_0, M_B^2; \mu_F^2, \mu_R^2, \mu_{R,1}^2)
\, ,
\label{eq:F2toNLO}
\end{equation}
where
\begin{eqnarray}
F_2^{\rm NLO}(Q^2, s_0, M_B^2; \mu_F^2, \mu_R^2, \mu_{R,1}^2)
 &=& 
\frac{\alpha_s(\mu_R^2)}{\pi}
\left[
F_2^{\rm NLO, fin}(Q^2, s_0, M_B^2; \mu_F^2)
\right. \nonumber \\ && \left.
+
F_2^{\rm NLO, UV}(Q^2, s_0, M_B^2; \mu_F^2)
\; \ln \frac{\mu_{R,1}^2}{Q^2}
\right. \nonumber \\ && \left.
+
F_2^{\rm NLO, IR}(Q^2, s_0, M_B^2; \mu_F^2)
\; \ln \frac{\mu_{F}^2}{Q^2}
\right]
\, ,
\label{eq:F2exp}
\end{eqnarray}
where, as defined before,
$Q^2$ is the virtuality of the photon probe,
$\mu_R^2$ is the coupling constant renormalization scale,
$\mu_{R,1}^2$ is the renormalization scale at which the renormalization
of the nucleon interpolation current has been performed
(often taken the same as $\mu_R^2$ but in principle an independent scale),
$\mu_F^2$ is the factorization scale at which the collinear singularities
corresponding to the nucleon DA factorize,
$s_0$ is the scale corresponding to the continuum subtraction in LCSRs,
and $M_B^2$ is a Borel mass, which can be regarded as a matching scale 
of hadronic and partonic part of the calculation.

%%%%%%%%%%%%%%%%%%%%%%%%%%%%%%%%%%%%%%%%%%%%%%%%%%%%%%%%%%%%%%%%%%%%%%%%
\begin{table}
\caption{ Table of the substitution rules corresponding to
Eq. \protect\req{eq:substitution} for $g_n$ functions
($n=0,\ldots,12$)
defined in Eq. \protect\req{eq:g-functions}.}
\begin{tabular}{l|l}
\hline
\\[0.1cm] $g_n(x_i , x_j,W)$ & 
$\widetilde{g}_n(x_i , x_j,Q^2,s_0,M_B^2)$
\\[0.5cm]
\hline\hline
\\[0.1cm]
$\displaystyle
\frac{1}{(x_i W-1 +i \eta)}$ & 
$\displaystyle
-\frac{1}{x_i} ~ $
\\[0.5cm]
\hline
\\[0.1cm]
$\displaystyle
- \frac{\ln(1-x_i W -i \eta)}{(1-x_i W -i \eta)}$ & 
$\displaystyle
-\frac{1}{x_i} \ln\left(x_i\frac{s_0 +Q^2}{Q^2}-1\right) ~ 
$
\\[0.5cm]
&$\displaystyle 
+ 
\frac{1}{x_i} \int_1^{x_i \frac{s_0 +Q^2}{Q^2}} dt~ \frac{1}{(1-t)} ~ 
 \left[ \exp\left(\frac{(1-t)Q^2}{x_i M_B^2} \right)
        -1 \right]
$ 
 \\[0.5cm]
\hline 
\\[0.1cm]
$\displaystyle
-\frac{\ln^2(1-x_i W -i \eta)}{(1-x_i W -i \eta)}$ & 
$\displaystyle
-\frac{1}{x_i}\left[ \ln^2\left(x_i\frac{s_0 +Q^2}{Q^2}-1\right) ~ 
- \frac{\pi^2}{3}\right]
$
\\[0.5cm]
&$\displaystyle 
+\frac{2}{x_i} \int_1^{x_i \frac{s_0 +Q^2}{Q^2}}dt ~ 
\frac{\ln(t-1)}{(1-t)} ~ 
  \left[ \exp\left(\frac{(1-t)Q^2}{x_i M_B^2} \right)
        -1 \right]
$ 
\\[0.5cm]
\hline
\\[0.1cm]
$\displaystyle 
\frac{\ln(1-x_i W -i \eta)}{(W +i \eta)}$ & 
$\displaystyle
-\int_1^{x_i\frac{s_0 +Q^2}{Q^2}}dt ~ \frac{1}{t} ~ 
   \exp\left(\frac{(1-t)Q^2}{x_i M_B^2} \right)
$ \\[0.5cm]
\hline
\\[0.1cm]
$\displaystyle
\frac{\ln(1-x_i W -i \eta)}{(W +i \eta)^2}$ & 
$\displaystyle
-x_i \int_1^{x_i\frac{s_0 +Q^2}{Q^2}}dt ~ \frac{1}{t^2} ~  
   \exp\left(\frac{(1-t)Q^2}{x_i M_B^2} \right)
$
 \\[0.5cm]
\hline 
\\[0.1cm]
$\displaystyle
\frac{\ln^2(1-x_i W -i \eta)}{(W +i \eta)}$ & 
$\displaystyle
-2 \int_1^{x_i\frac{s_0 +Q^2}{Q^2}}dt ~ \frac{\ln(t-1)}{t} 
   \exp\left(\frac{(1-t)Q^2}{x_i M_B^2} \right)
$
\\[0.5cm]
\hline 
\\[0.1cm]
$\displaystyle
\frac{\ln^2(1-x_i W -i \eta)}{(W +i \eta)^2}$ & 
$\displaystyle
-2 x_i \int_1^{x_i\frac{s_0 +Q^2}{Q^2}}dt ~ \frac{\ln(t-1)}{t^2} 
   \exp\left(\frac{(1-t)Q^2}{x_i M_B^2} \right)
$
\\[0.5cm]
\hline 

\end{tabular}
\label{tab:sub1}
\end{table}

\begin{table}
\begin{tabular}{l|l}
$\displaystyle
-\frac{\ln(1-(x_i + x_j)W - i \eta)}{(1-x_i W -i \eta)}$ 
& 
$\displaystyle
-\frac{1}{x_i}\left[\ln\left(x_i\frac{s_0+Q^2}{Q^2}-1\right)+\ln\left(\frac{x_i+x_j}{x_i}\right)\right]
$
\\[0.5cm]
& 
$\displaystyle
+\frac{1}{x_i}\int_{\frac{x_i}{x_i+x_j}}^{x_i\frac{s_0+Q^2}{Q^2}}dt ~ \frac{1}{(1-t)}\left[ \exp\left(\frac{(1-t)Q^2}{x_i M_B^2} \right)
        -1 \right]$
\\[0.5cm]
\hline 
\\[0.1cm]
$\displaystyle
-\frac{\ln^2(1-(x_i + x_j)W - i \eta)}{(1-x_i W -i \eta)}$ 
&
$\displaystyle
-\frac{1}{x_i} \left[ \left(\ln\left(x_i\frac{s_0+Q^2}{Q^2}-1\right)+\ln\left(\frac{x_j+x_i}{x_i}\right)\right)^2 \right.
$
\\[0.5cm]
& 
$\displaystyle
\qquad
\left.
-\frac{\pi^2}{3}-\pi^2\left(1-\delta(x_j)\right)
+ 2 \,\mbox{Li}_2\left(\frac{x_j}{(x_i+x_j)(1-x_i\frac{s_0+Q^2}{Q^2})}\right)
\right]
$
\\[0.5cm]
&
$\displaystyle
+\frac{2}{x_i}\int_{\frac{x_i}{x_i+x_j}}^{x_i\frac{(s_0+Q^2)}{Q^2}}dt ~ \frac{\ln\left(t(1+\frac{x_j}{x_i})-1\right)}{(1-t)}\left[ \exp\left(\frac{(1-t)Q^2}{x_i M_B^2} \right)
        -1 \right]
$
\\[0.5cm]
\hline 
\\[0.1cm]
$\displaystyle
\frac{\ln(1-(x_i + x_j)W - i \eta)}{(W+i \eta)}$ & 
$\displaystyle
-\int_\frac{x_i}{x_i+x_j}^{x_i\frac{s_0 +Q^2}{Q^2}}dt ~  \frac{1}{t} 
   \exp\left(\frac{(1-t)Q^2}{x_i M_B^2} \right)
$
\\[0.5cm]
\hline 
\\[0.1cm]
$\displaystyle
\frac{\ln(1-(x_i + x_j)W - i \eta)}{(W+i \eta)^2}$ & 
$\displaystyle
-x_i\int_\frac{x_i}{x_i+x_j}^{x_i\frac{s_0 +Q^2}{Q^2}}dt ~ 
\frac{1}{t^2} 
   \exp\left(\frac{(1-t)Q^2}{x_i M_B^2} \right)
$
\\[0.5cm]
\hline 
\\[0.1cm]
$\displaystyle
\frac{\ln^2(1-(x_i + x_j)W - i \eta)}{(W+i \eta)}$ & 
$\displaystyle
-2 \int_\frac{x_i}{x_i+x_j}^{x_i\frac{s_0 +Q^2}{Q^2}}dt ~ \frac{\ln\left(t(1+\frac{x_j}{x_i})-1\right)}{t} 
   \exp\left(\frac{(1-t)Q^2}{x_i M_B^2} \right)
$
\\[0.5cm]
\hline 
\\[0.1cm]
$\displaystyle
\frac{\ln^2(1-(x_i + x_j)W - i \eta)}{(W+i \eta)^2}$ & 
$\displaystyle
-2 x_i \int_\frac{x_i}{x_i+x_j}^{x_i\frac{s_0 +Q^2}{Q^2}}dt ~ \frac{\ln\left(t(1+\frac{x_j}{x_i})-1\right)}{t^2} 
   \exp\left(\frac{(1-t)Q^2}{x_i M_B^2} \right)
$
\\[0.5cm]
\hline 
\end{tabular}
Note: As expected, the case $x_j=0$ corresponds 
to the results from previous page.
\label{tab:sub2}
\end{table}
%%%%%%%%%%%%%%%%%%%%%%%%%%%%%%%%%%%%%%%%%%%%%%%%%%%%%%%%%%%%%%%%%%%%%%%%
%%%%%%%%%%%%%%%%%%%%%%%%%%%%%%%%%%%%%%%%%%%%%%%%%%%%%%%
\subsection{Nucleon DAs}
\label{sec:DA}
%%%%%%%%%%%%%%%%%%%%%%%%%%%%%%%%%%%%%%%%%%%%%%%%%%%%%%%

We refer to App. B of Ref. \cite{BraunLW06} for 
detailed account of nucleon distribution amplitudes 
and list here only the selected expressions.

For $M_N=0$ case only the twist-3 DAs are relevant:
\begin{eqnarray}
V_1(\{x_k\},\mu^2) &=&
120 \, x_1 \, x_2 \, x_3 
\left[ 
\phi_3^0(\mu^2)
+
\phi_3^+(\mu^2)
(1 -3 x_3) 
\right]
\, ,
\nonumber \\[0.2cm]
A_1(\{x_k\},\mu^2) &=&
120 \, x_1 \, x_2 \, x_3 
(x_2-x_1) 
\phi_3^-(\mu^2)
\, ,
\end{eqnarray}
or, equivalently%
\footnote{
As explained in, for example, Ref. \cite{BraunFMS00},
$V_1$ and $A_1$ have different symmetry properties
and they can be combined together
to define a single independent twist-3 nucleon distribution
amplitude $\Phi_3$.}
\begin{eqnarray}
\Phi_3(\{x_k\},\mu^2) &=&
V_1(\{x_k\},\mu^2) - 
A_1(\{x_k\},\mu^2) 
\nonumber \\
&=&
120 \, x_1 \, x_2 \, x_3 
\left[ 
\phi_3^0(\mu^2)
+
\phi_3^+(\mu^2)
(1 -3 x_3) 
- \phi_3^-(\mu^2)
(x_2-x_1) 
\right]
\, .
\qquad
\label{eq:PhiV1A1}
\end{eqnarray}
Here
\begin{eqnarray}
\phi_3^0 &=& f_N
\, ,
\nonumber \\[0.2cm]
\phi_3^+ &=& \frac{7}{2} f_N (1 - 3 V_1^d)
\, ,
\nonumber \\[0.2cm]
\phi_3^- &=& \frac{21}{2} f_N A_1^u
\, .
\end{eqnarray}

The normalization
obtained using QCD sum rules
amounts to \cite{BraunLW06}
\begin{eqnarray}
f_N(\mu_0^2=1 {\rm GeV}^2) &=&
(5.0 \pm 0.5) \cdot 10^{-3} {\rm GeV}^2
\, .
\end{eqnarray}
Actually the normalization
of twist-3, -4 and -5 DAs is determined by three
dimensionful parameters $f_N$, $\lambda_1$ and $\lambda_2$
that are well known from the QCD sum rule literature
and correspond to nucleon couplings
to the existing three different three-quark local
operators.
The numerical values
of the other two normalization constants, obtained by QCD sum rules
\cite{BraunLW06}, 
are
\begin{eqnarray}
\lambda_1(\mu_0^2=1 {\rm GeV}^2) &=&
-(2.7 \pm 0.9) \cdot 10^{-2} {\rm GeV}^2
\, ,
\end{eqnarray}
and 
$\lambda_2(\mu_0^2=1 {\rm GeV}^2)=(5.4 \pm 1.9)\cdot 10^{-2} {\rm GeV}^2$.
For the evolution of these parameters we refer to, for example,
\cite{BraunFMS00}.

The shape of the twist-3 DAs, i.e., the deviation from the
asymptotic form, is determined by dimensionless parameters
$V_1^d$ and $A_1^u$, while three more parameters 
($f_1^d$, $f_1^u$ and $f_2^d$) appear in twist-4 and at higher twists.
The values of these parameters and thus the shape of DAs are 
controversial.
The older prediction from QCD sum rules, which is sometimes referred to as the
Chernyak-Zhitnitsky-like (CZ-like) model \cite{ChernyakOZ87}
amounts to
\begin{eqnarray}
V_1^d &=& 0.23 \pm 0.03
\, ,
\nonumber \\
A_1^u &=& 0.38 \pm 0.15
\, ,
\label{eq:V1A1qcd}
\end{eqnarray}
while 
$f_1^d=0.40\pm0.05$, $f_1^u=0.07\pm0.05$ and $f_2^d=0.22\pm0.05$.
In Ref. \cite{BraunLW06} (Eq. (42,\cite{BraunLW06})) the following values (we refer to them as BLW parameters)
were introduced 
\begin{eqnarray}
V_1^d &=& 0.30 
\, ,
\nonumber \\
A_1^u &=& 0.13 
\, ,
\label{eq:V1A1blw}
\end{eqnarray}
while 
$f_1^d=0.33$, $f_1^u=0.09$ and $f_2^d=0.25$.
Finally, for the asymptotic DAs
\begin{eqnarray}
V_1^d &=& 1/3
\, ,
\nonumber \\
A_1^u &=& 0 
\, ,
\label{eq:V1A1asy}
\end{eqnarray}
and $f_1^d=3/10$, $f_1^u=1/10$ and $f_2^d=4/15$.
Thus, for asymptotic twist-3 DAs
\begin{equation}
\phi_3^0 = f_N \, ,
\qquad \qquad
\phi_3^+ = \phi_3^- = 0
\, ,
\end{equation}
leading to
\begin{equation}
\Phi_3(\{x_k\},\mu^2) =
V_1(\{x_k\},\mu^2) = 
120 \, x_1 \, x_2 \, x_3 \, f_N(\mu^2)
\,.
\label{eq:Phiasy}
\end{equation}

%%%%%%%%%%%%%%%%%%%%%%%%%%%%%%%%%%%%%%%%%%%%%%%%%%%%%%%
%%%%%%%%%%%%%%%%%%%%%%%%%%%%%%%%%%%%%%%%%%%%%%%%%%%%%%%
\section{Numerical results}

\setcounter{equation}{0}
\label{sec:num}
%%%%%%%%%%%%%%%%%%%%%%%%%%%%%%%%%%%%%%%%%%%%%%%%%%%%%%%
%%%%%%%%%%%%%%%%%%%%%%%%%%%%%%%%%%%%%%%%%%%%%%%%%%%%%%%

In order to obtain numerical predictions for the proton
form factor $F_2$ calculated at twist-3 to NLO order
\req{eq:F2toNLO}, we use
the results (\ref{eq:finresPhiLO}-\ref{eq:finresPhiNLO}),
twist-3 DA $\Phi_3$ defined in Sec. \ref{sec:DA}, 
recipe \req{eq:substitution}
and Table \ref{tab:sub1}.
The additional necessary numerical values listed in
preceding subsections and taken from Ref. \cite{BraunLW06},
are
\begin{eqnarray}
s_0&=&(1.5 \mbox{GeV})^2 \, ,
\nonumber \\
M_B^2&=& 2 \mbox{GeV}^2 \, ,
\label{eq:s0MB2}
\end{eqnarray}
and
\begin{equation}
\lambda_1(\mu_0^2=1 {\rm GeV}^2)=-2.7 \cdot 10^{-2} \, \rm GeV^{2}
\, .
\label{eq:lambda1def}
\end{equation}
We will check the sensitivity of the results on the
variation of $M_B^2$.
Alternatively, one can use the modified value
of $\lambda_1$ calculated using
the expression (B10,\cite{BraunFMS00}) from Ref. \cite{BraunFMS00} 
where $E_3$ is corrected by a factor $(1+71/12 \alpha_s/\pi)$
from Eq. (24) in Ref. \cite{SadovnikovaDR05}
%\begin{eqnarray}
%\lambda_1^{corr.}(2 \rm GeV^2)=-3.3 \cdot 10^{-2} \, \rm GeV^{2}
%\, ,
%\label{eq:lambda1corr}
%\end{eqnarray}
%i.e.
\begin{eqnarray}
\lambda_1^{corr.}(1-10 \rm GeV^2)=((-3.4) - (-3.2)) \cdot 10^{-2} \, \rm GeV^{2}
\, .
\label{eq:lambda1corrA}
\end{eqnarray}

The one-loop expression for $\alpha_s$ reads
\begin{equation}
\alpha_s(\mu^2)=\frac{4 \pi}{\beta_0 \ln \mu^2/\Lambda^2}
\, ,
\end{equation}
where
$\beta_0=11-2/3 n_f$, with $n_f=3$ being the number of flavours.
For $\Lambda$ we take the value $\Lambda=0.2$ GeV.
In this work we do not take into account the DA evolution
i.e. we neglect the evolution (see, for example, \cite{BraunFMS00})
of $f_N(\mu^2)$ and of $\lambda_1(\mu^2)$. 
One takes that these effects are small, 
especially since in the most part of the numerical
analysis we fix the relevant scale to 2 GeV$^2$.

The LO twist-3 and twist-4 results calculated
for $M_N\ne0$ are presented in App. \ref{sec:LOMN},
while the higher-twist results and $x^2$ corrections we
take from Ref. \cite{BraunLW06}.
The parameters for the DAs are given in Sec. \ref{sec:DA},
while for the exact form of the higher-twist DAs we again refer
to Ref. \cite{BraunLW06}.

It is convenient to normalize the results 
for $F_2$ and $G_M$ to dipole form factor,
i.e., $G_M/(\mu_p G_D)$ where
\begin{equation}
G_D=\frac{1}{(1+Q^2/0.71 {\rm GeV}^2)^2}
\, .
\label{eq:GD}
\end{equation}
%%%%%%%%%%%%55%%%%%%%%%%%%%%%%%%%%%%%%%%%%%%%%%%%%%%%%%%%
\subsection{$M_N=0$ case: $F_2$ at twist-3 to NLO order}
%%%%%%%%%%%%%%%%%%%%%%%%%%%%%%%%%%%%%%%%%%%%%%%%%%%%%%%%%%

\begin{figure}
 \centering
\begin{tabular}{cc}
 \includegraphics[scale=0.7]{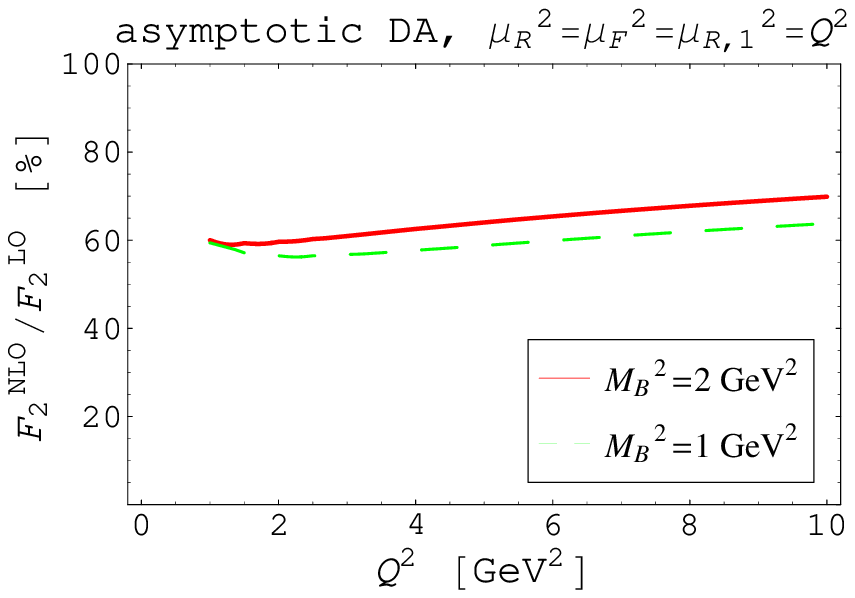} 
&
 \includegraphics[scale=0.7]{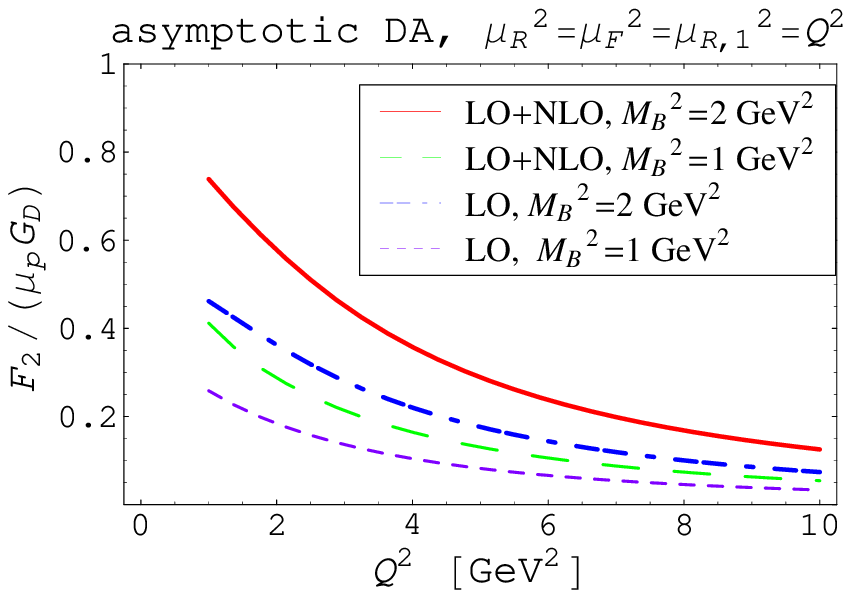} \\[0.2cm]
   \small (a) &
   \small (b)
\end{tabular}
\caption{NLO calculation of twist-3 contribution
to the proton form factor $F_2$ obtained for $M_N=0$ 
(\protect\ref{eq:F2toNLO}-\protect\ref{eq:F2exp}), 
using asymptotic DA \req{eq:Phiasy} and 
$\mu_R^2=\mu_F^2=\mu_{R,1}^2=Q^2$: 
a) The ratio  $F_2^{\rm NLO,fin}/F_2^{\rm LO}$ 
(i.e., the ratio of NLO correction to the LO prediction) 
evaluated for the default  choice (\protect\ref{eq:s0MB2})
$M_B^2=2$ GeV$^2$ (solid line), 
and for the test choice $M_B^2=1$ GeV$^2$ (dashed line).
b) The twist-3 prediction to $F_2$ normalized to $\mu_p G_D$:
NLO (thick solid line) and LO (thick dot-dashed line) for
$M_B^2=2$ GeV$^2$,
NLO (thin long dashed line) and LO (thin short dashed line)
for $M_B^2=1$ GeV$^2$.
}
 \label{f:F2NLOcAsy2}
\end{figure}

In Fig. \ref{f:F2NLOcAsy2}a
we present the relative size of the NLO correction \req{eq:F2exp}
taken at $\mu_R^2=\mu_F^2=\mu_{R,1}^2=Q^2$ compared to LO prediction,
i.e., the ratio $F_2^{\rm NLO}/F_2^{LO}$ in dependence on $Q^2$
and calculated for the asymptotic twist-3 DA \req{eq:Phiasy}.
Note that for that choice of scales only the
$F_2^{\rm NLO,fin}$ part from \req{eq:F2exp} contributes.
In Fig. \ref{f:F2NLOcAsy2}b
the complete NLO prediction $F_2$ \req{eq:F2toNLO}
normalized to $\mu_p G_D$ is displayed. 
In Fig. \ref{f:F2NLOcAsy2} we test the sensitivity of 
the results on the choice of $M_B^2$ by displaying the results
for the default choice \req{eq:s0MB2}, as well as,
for $M_B^2=1$ GeV$^2$.

From Fig. \ref{f:F2NLOcAsy2}a one can see that, 
for the asymptotic DA and $\mu_R^2=\mu_F^2=\mu_{R,1}^2=Q^2$
the ratio $F_2^{\rm NLO}/F_2^{LO}$,
and thus the NLO correction in comparison to the LO
prediction is quite large, being 60-70\% for $1<Q^2<10$ GeV$^2$,
and it increases with $Q^2$.
By decreasing $M_B^2$ to 1 GeV$^2$ the ratio drops at higher $Q^2$
only slightly, i.e., to 65\%.

The sensitivity of the LO (dot-dashed and dashed lines)
and NLO (solid and long-dashed lines) predictions
for $F_2$ on $M_B^2$ is illustrated 
in Fig. \ref{f:F2NLOcAsy2}b.
One can see that this effect is large
(comparable to the change from LO to NLO).

In the following we adopt the default choice \req{eq:s0MB2}
employed in Ref. \cite{BraunLW06}.

%%%%%%%%%%%%%%%%%%%%%%%%%%%%%%%%%%%%%%%%%%%%%%%%%%%%%%%%%%%%%%%%%%%%%%
%%%%%%%%%%%%%%%%%%%%%%%%%%%%%%%%%%%%%%%%%%%%%%%%%%%%%%%%%%%%%%%%%%%%%%

\begin{figure}
 \centering
\begin{tabular}{cc}
 \includegraphics[scale=0.7]{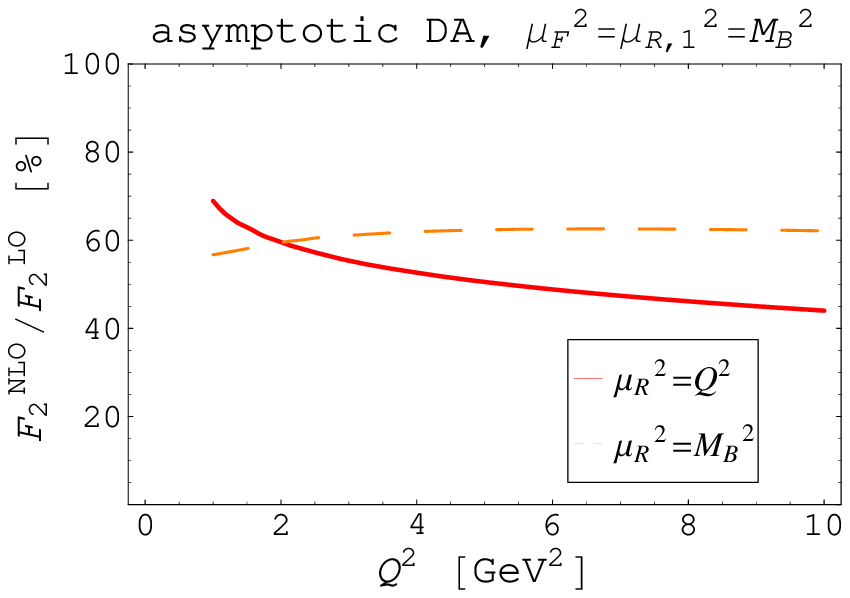} 
&
 \includegraphics[scale=0.7]{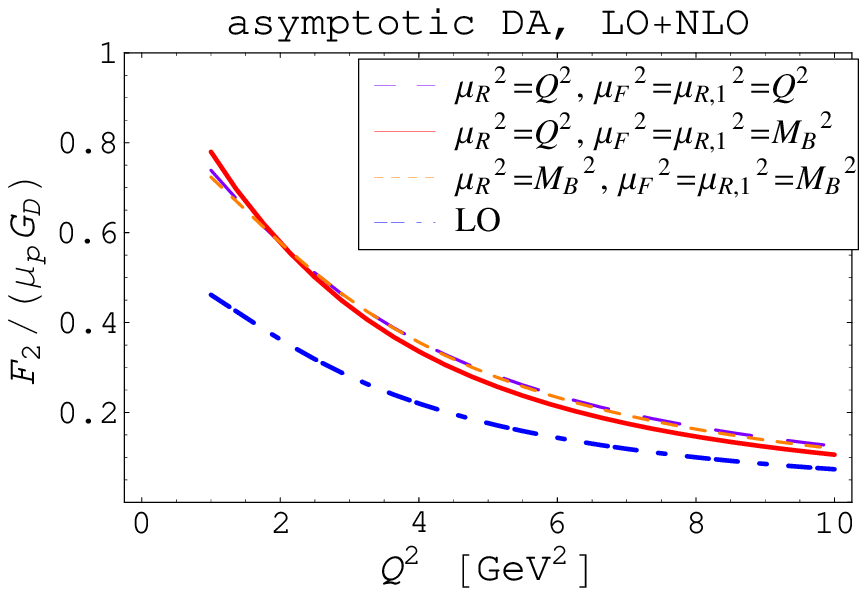} \\[0.2cm]
   \small (a) &
   \small (b)
\end{tabular}
\caption{NLO calculation of twist-3 contribution
to the proton form factor $F_2$ obtained for $M_N=0$ 
(\protect\ref{eq:F2toNLO}-\protect\ref{eq:F2exp}) 
and using asymptotic DA \req{eq:Phiasy}: 
a) The ratio  $F_2^{\rm NLO}/F_2^{\rm LO}$ 
(i.e., the ratio of NLO correction calculated 
at to the LO prediction) 
calculated using  $\mu_F^2=\mu_{R,1}^2=M_B^2$, 
while $\mu_{R}^2=Q^2$ (solid line), 
and  $\mu_{R}^2=M_B^2$ (dashed line).
b) The twist-3 prediction to $F_2$ normalized to $\mu_p G_D$:
NLO for $\mu_{R}^2=Q^2$ and $\mu_F^2=\mu_{R,1}^2=Q^2$ (thin long dashed line),
NLO for $\mu_{R}^2=Q^2$ and $\mu_F^2=\mu_{R,1}^2=M_B^2$ (thick solid line),
NLO for $\mu_{R}^2=M_B^2$ and $\mu_F^2=\mu_{R,1}^2=M_B^2$ (thin short dashed line),
and LO (thick dot-dashed line).
}
 \label{f:F2NLOAsy1}
\end{figure}

In Fig. \ref{f:F2NLOAsy1} the change of scales $\mu_R^2$,
$\mu_F^2$ and $\mu_{R,1}^2$ is investigated.
As can be seen by comparing Figs. \ref{f:F2NLOcAsy2}a
and \ref{f:F2NLOAsy1}a, by taking $\mu_F^2=\mu_{R,1}^2=M_B^2$
and retaining $\mu_R^2=Q^2$,
the ratio of NLO correction to the LO prediction
is lowered to 69\% - 44\% for $1<Q^2<10$ GeV$^2$
and it decreases with $Q^2$.
Thus, one can see that the $F^{\rm NLO,IR}_2$ 
and $F^{\rm NLO,UV}_2$ terms from Eq. \req{eq:F2exp}
decrease the size of NLO correction.
By changing $\mu_R^2$ to $M_B^2$ the ratio gets bigger
since then there is no suppression due to the running of the
$\alpha_s$.
From Fig. \ref{f:F2NLOAsy1}b one can see that 
the change of scales does not influence much the
value of the complete $F_2$ NLO prediction.
In further calculation, if not specified differently,
%we take as a reasonable choice
%$\mu_R^2=Q^2$ and $\mu_F^2=\mu_{R,1}^2=M_B^2$.
we, following \cite{BraunKM99}, take 
$\mu_R^2=\mu_F^2=\mu_{R,1}^2=M_B^2$.

%%%%%%%%%%%%%%%%%%%%%%%%%%%%%%%%%%%%%%%%%%%%%%%%%%%%%%%%%%%%%%%%%%%%%%
%%%%%%%%%%%%%%%%%%%%%%%%%%%%%%%%%%%%%%%%%%%%%%%%%%%%%%%%%%%%%%%%%%%%%%

\begin{figure}
 \centering
\begin{tabular}{cc}
 \includegraphics[scale=0.7]{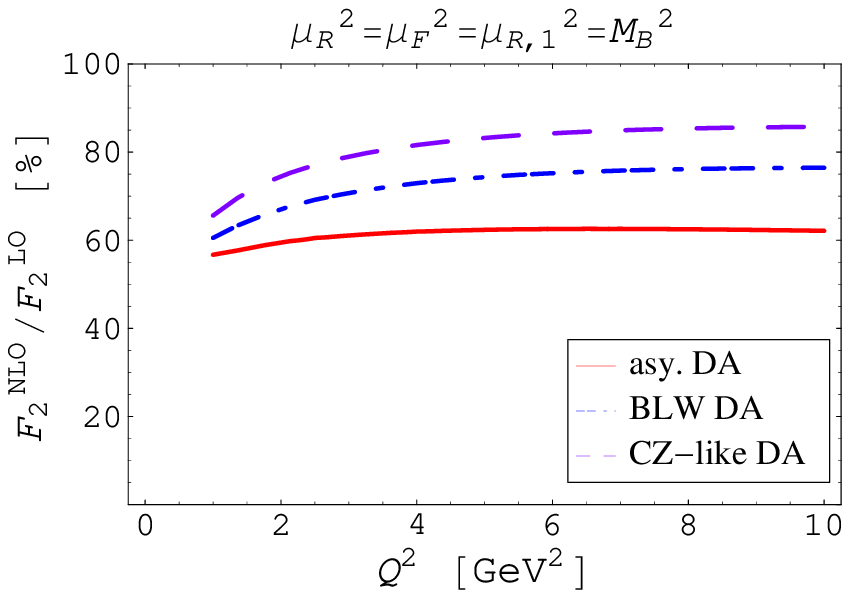} 
&
 \includegraphics[scale=0.7]{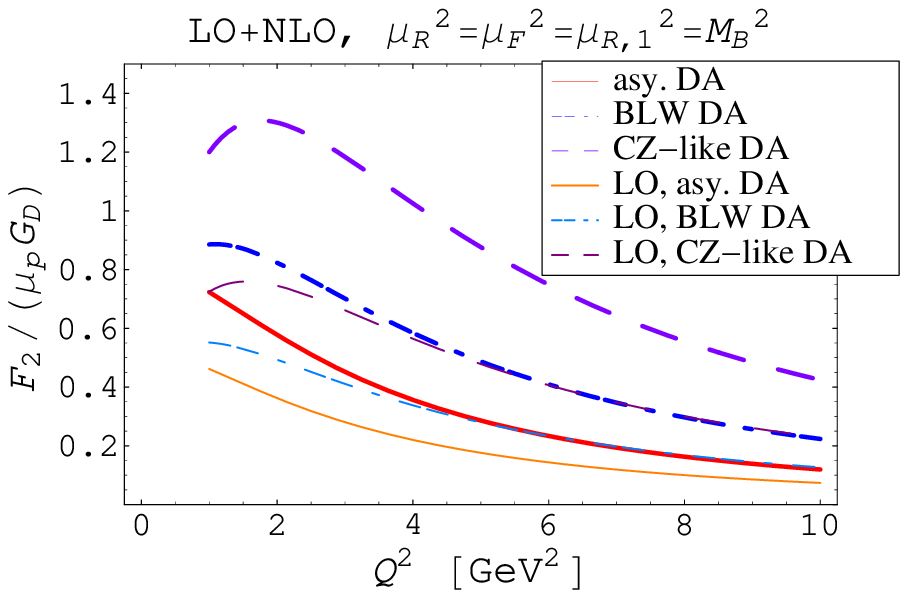} \\[0.2cm]
   \small (a) &
   \small (b)
\end{tabular}
\caption{NLO calculation of twist-3 contribution
to the proton form factor $F_2$ obtained for $M_N=0$ 
(\protect\ref{eq:F2toNLO}-\protect\ref{eq:F2exp}) 
%and using $\mu_F^2=\mu_{R,1}^2=M_B^2$ and $\mu_{R}^2=Q^2$: 
using $\mu_{R}^2=\mu_F^2=\mu_{R,1}^2=M_B^2$. 
The displayed results correspond to
asymptotic  DAs (solid line),
BLW DAs (dash-dotted line),
and CZ-like DAs (dashed line).
a) The ratio  $F_2^{\rm NLO}/F_2^{\rm LO}$, 
i.e., the ratio of the NLO correction to the LO prediction. 
b) The twist-3 prediction to $F_2$ normalized to $\mu_p G_D$.
Thick lines: NLO prediction. Thin lines: LO prediction.
}
 \label{f:F2NLO1}
\end{figure}

Finally, in Fig. \ref{f:F2NLO1} we investigate the size of 
the NLO correction and of the complete NLO prediction 
in dependence on the choice of DA \req{eq:PhiV1A1}.
We employ  the asymptotic (\ref{eq:V1A1asy}) DA (solid line),
BLW (\ref{eq:V1A1blw}) DA (dash-dotted line),
and the CZ-like DA (\ref{eq:V1A1qcd}) (dashed line).
In Fig. \ref{f:F2NLO1}b the LO predictions are denoted by thin lines
and NLO predictions by thick lines.
From Fig. \ref{f:F2NLO1}a one can see that for $1<Q^2<10$ GeV$^2$
the ratio of NLO corrections to LO predictions amounts to
%69\% - 44\% for the asymptotic DA,  
%74\% - 54\% for the BLW DA, and 
%80\% - 61\% for the CZ-like DA.  
57\% - 62\% for the asymptotic DA,  
61\% - 76\% for the BLW DA, and 
66\% - 86\% for the CZ-like DA.  
Both LO predictions and NLO corrections are larger
for the two DAs whose form differ from the asymptotic one.

In conclusion, we state that the NLO corrections to
$F_2$ calculated at twist-3 taking $M_N=0$ are large,
with $F_2^{\rm NLO}/F_2^{\rm LO}$ amounting to cca $60\%$,
but varying for different DAs 
and depending on the choice of renormalization and factorization scales.
The sensitivity
of both LO and NLO corrections 
%on the choice $s_0$ and $M_B$
%is large and in the following we take 
on the choice of $M_B$
is large and in the following we take 
the value  \req{eq:s0MB2} from Ref \cite{BraunLW06}.
In contrast to the dependence of $F_2^{\rm NLO}/F_2^{\rm LO}$,
the dependence of the complete NLO prediction of $F_2$
on the choice of renormalization and factorization scales
is small and, if not stated otherwise, we take in the following
%$\mu_R^2=Q^2$ and $\mu_F^2=\mu_{R,1}^2=M_B^2$. 
$\mu_R^2=\mu_F^2=\mu_{R,1}^2=M_B^2$. 
The results for different DAs differ significantly.

%%%%%%%%%%%%%%%%%%%%%%%%%%%%%%%%%%%%%%%%%%%%%%%%%%%%%%%%%%%%%%%%%%%%%%
%%%%%%%%%%%%%%%%%%%%%%%%%%%%%%%%%%%%%%%%%%%%%%%%%%%%%%%%%%%%%%%%%%%%%%

%%%%%%%%%%%%55%%%%%%%%%%%%%%%%%%%%%%%%%%%%%%%%%%%%%%%%%%%
\subsection{Various contributions to $F_1$ and $F_2$}
%%%%%%%%%%%%%%%%%%%%%%%%%%%%%%%%%%%%%%%%%%%%%%%%%%%%%%%%%%

In preceding section we have discussed the size of the NLO
corrections to $F_2$ calculated at twist-3 and taking $M_N=0$,
and we have compared it to the corresponding LO predictions.
In this section we want to analyze how large are these corrections
in comparison to other contributions: mass effects, higher-twists
and $x^2$ corrections calculated at LO.
These effects are calculated on the basis of App. \ref{sec:LOMN},
and Ref. \cite{BraunLW06}.

\subsubsection{$F_2$}

The effect of including $M_N^2$ terms in LO twist-3 contribution
to $F_2$ corresponds to the  change of the LO contribution 
obtained using the asymptotic DA
for  $(-3)$-$23$ \% when $1\le Q^2 \le 10$ GeV$^2$.
The change increases with $Q^2$ and one obtains
similar numbers for all three DAs.

Let us now discuss higher-twist effects starting with twist-4.
We note that for $M_N^2=0$  the twist-4 contribution to $F_2$ is 0  
(the whole contribution \req{eq:calBV123}, \req{eq:calBA123} 
is proportional to $M_N^2$).
The ratio of LO twist-4 and twist-3 contributions to $F_2$ 
is in the range (-5) to (-35) \% for asymptotic DAs, 
2 to (-0.7) \% for BLW DAs,
and 13 to 24\% for CZ-like DAs,
when  $1\le Q^2 \le 10$ GeV$^2$.
This behaviour is obviously very different for different DAs.
In the case of the asymptotic DAs and the CZ-like DAs
the absolute value of the ratio grows with $Q^2$,
while for BLW the ratio decreases, changes sign and then absolute value increases.
The role of the LO twist-4 contributions is thus small in the case
of the results obtained using the BLW DAs, and more pronounced
in the case of the other two investigated DAs (the BLW results
seem to fall in some kind of minima).
We have noticed the large sensitivity of these results on the choice 
of the parameters $f_{1}^{d}$ and $A_{1}^{u}$.

Finally, we summarize our findings about the
size of various contributions to $F_2$ in
Tables \ref{t:F2contrA}
and \ref{t:F2contrB}.
One can see that
twist-3 contributions are dominant and positive.
The $x^2$-contributions are negative and 
the ratio of $x^2$-contributions
and LO twist-3 contributions does not change much for various DAs.
The twist-5 contributions are more pronounced than twist-4
contributions and for both the ratio to twist-3 contribution
is very sensitive to the shape of DAs.
The twist-3 NLO corrections are positive and cca 60\%.
%for $\mu_R^2=\mu_F^2=\mu_{R,1}^2=Q^2$
%while in the range of cca 70-40\% 
%for $\mu_R^2=Q^2$ and $\mu_F^2=\mu_{R,1}^2=M_B^2$.

Hence, the twist-3 NLO corrections are indeed sizable
and important.

%%%%%%%%%%%%%%%%%%%%%%%%%%%%%%%%%%%%%%%%%%%%%%%%%%%%%%%%%%%%%%%
\begin{table}
\caption{The size of various contributions to the proton form factor
$F_2$ 
(normalized to
LO twist-3 contribution calculated for $M_N^2 \ne 0$)
in the range $1\le Q^2 \le 10$ GeV$^2$.}
\begin{tabular}{l|c|ccc}
\hline
 DAs & 
$\displaystyle\frac{F_2^{\rm LO, tw-3}|_{M_N^2=0}}{F_2^{\rm LO, tw-3}}$ &
$\displaystyle\frac{F_2^{\rm LO, tw-4}}{F_2^{\rm LO, tw-3}}$ &
$\displaystyle\frac{F_2^{\rm LO, tw-5}}{F_2^{\rm LO, tw-3}}$ &
$\displaystyle\frac{F_2^{{\rm LO},x^2-{corr.}}}{F_2^{\rm LO, tw-3}}$ 
\\ \hline
asy. &
103 - 81\%& 
(-5) - (-35)\%& 
34 - 91\%& 
(-35) - (-43)\% 
 \\[0.2cm]
BLW &
98 - 81\%& 
2 - (-0.7)\%& 
14 - 21\%& 
(-33) - (-28)\% 
 \\[0.2cm] 
CZ-like &
92 - 81\%& 
13 - 24 \%& 
(-13) - (-29)\%& 
(-31) - (-18)\% 
\\[0.2cm] \hline
\end{tabular}
\label{t:F2contrA}
\end{table}
%%%%%%%%%%%%%%%%%%%%%%%%%%%%%%%%%%%%%%%%%%%%%%%%%%%%%%%%%%%%%%%
%%%%%%%%%%%%%%%%%%%%%%%%%%%%%%%%%%%%%%%%%%%%%%%%%%%%%%%%%%%%%%%
\begin{table}
\caption{The size of NLO twist-3 corrections to the proton form factor $F_2$ 
(normalized to
LO twist-3 contribution calculated for $M_N^2 \ne 0$)
in the range $1\le Q^2 \le 10$ GeV$^2$ and for three choices of scales:
a) $\mu_R^2=\mu_F^2=\mu_{R,1}^2=Q^2$,
b) $\mu_R^2=Q^2, \mu_F^2=\mu_{R,1}^2=M_B^2$,
c) $\mu_R^2=\mu_F^2=\mu_{R,1}^2=M_B^2$.}
\begin{tabular}{l|ccc}
\hline
 DAs & 
$\displaystyle\frac{F_2^{\rm NLO, tw-3}|_{M_N^2=0, a)}}{F_2^{\rm LO, tw-3}}$  &
$\displaystyle\frac{F_2^{\rm NLO, tw-3}|_{M_N^2=0, b)}}{F_2^{\rm LO, tw-3}}$ &
$\displaystyle\frac{F_2^{\rm NLO, tw-3}|_{M_N^2=0, c)}}{F_2^{\rm LO, tw-3}}$\\ \hline
asy. &
62 - 57\%& 
71 - 36\%& 
58 - 51\% \\[0.2cm]
BLW &
62 - 68\%& 
72 - 44\%& 
59 - 62\% \\[0.2cm] 
CZ-like &
63 - 76\%& 
74 - 49\%& 
61 - 69\% \\[0.2cm] \hline
\end{tabular}
\label{t:F2contrB}
\end{table}
%%%%%%%%%%%%%%%%%%%%%%%%%%%%%%%%%%%%%%%%%%%%%%%%%%%%%%%%%%%%%%%

\subsubsection{$F_1$}
 
In the next section we will proceed to the comparison of our results
to the experimental data and for that we need the $F_1$ contribution
as well (we have at our disposal the experimental data for $G_M$
and $G_E$).
Here we thus analyze the LO contributions to $F_1$.

The effect of including $M_N^2$ terms in LO twist-3 contribution
to $F_1$ corresponds to the  change of the LO contribution 
obtained using the asymptotic DA
for  $6$-$24$ \% when $1\le Q^2 \le 10$ GeV$^2$.
The change decreases and then increases with $Q^2$ 
(minimum at $Q^2 \approx 2$ GeV$^2$ and one obtains
similar numbers for all three DAs.
But in the case of $F_1$, twist-3 contribution is negative and
small in comparison with twist-4 contribution. 

In contrast to $F_2$, twist-4 contribution to $F_1$ is not proportional to 
$M_N^2$ (see App. \ref{sec:LOMN}).
The effect of including $M_N^2$ terms in LO twist-4 to $F_2$ corresponds to the 
change of the contribution 
obtained using the asymptotic DAs
for  $12$-$30$ \% when $1\le Q^2 \le 10$ GeV$^2$.
The numbers are similar for both asymptotic and BLW DAs, 
but smaller and negative for CZ-like DAs.
The ratio of twist-3 and twist-4 LO contributions to $F_1$ 
is (-19)-(-7)\% for the asymptotic DAs, 
(-25)-(-12)\% for the BLW DAs, and  (-46)-(-105)\%  
for the CZ-like DAs.
Hence, apart from the results for CZ-like DAs 
at higher $Q^2$, the twist-4 contributions is larger 
than the twist-3 contribution.

Finally, to summarize our findings about the
size of various contributions to $F_1$ in
Table \ref{t:F1contr},
we state that the
twist-4 contributions are dominant and positive.

%%%%%%%%%%%%%%%%%%%%%%%%%%%%%%%%%%%%%%%%%%%%%%%%%%%%%%%%%%%%%%%
\begin{table}
\caption{The size of various contributions to the proton form factor $F_1$ 
(normalized to
LO twist-4 contribution calculated for $M_N^2 \ne 0$)
in the range $1\le Q^2 \le 10$ GeV$^2$.}
\begin{tabular}{l|cccc}
\hline
 DAs & 
$\displaystyle\frac{F_1^{\rm LO, tw-3}}{F_1^{\rm LO, tw-4}}$ &
$\displaystyle\frac{F_1^{\rm LO, tw-5}}{F_1^{\rm LO, tw-4}}$ &
$\displaystyle\frac{F_1^{\rm LO, tw-6}}{F_1^{\rm LO, tw-4}}$ &
$\displaystyle\frac{F_1^{{\rm LO},x^2-{corr.}}}{F_1^{\rm LO, tw-4}}$ 
\\ \hline
asy. &
(-19) - (-7)\%& 
(-4) - (-5)\%& 
3 - 2\%& 
5 - 2\% 
 \\[0.2cm]
BLW &
(-25) - (-12)\%& 
(-2)\%& 
3 - 2\%& 
6 - 3\% 
 \\[0.2cm] 
CZ-like  &
(-46) - (-105)\%& 
5 - 37\%& 
5 - 14\%& 
10 - 22\% 
\\[0.2cm] \hline
\end{tabular}
\label{t:F1contr}
\end{table}
%%%%%%%%%%%%%%%%%%%%%%%%%%%%%%%%%%%%%%%%%%%%%%%%%%%%%%%%%%%%%%%
%\clearpage 

%%%%%%%%%%%%55%%%%%%%%%%%%%%%%%%%%%%%%%%%%%%%%%%%%%%%%%%%
\subsection{Comparison to experimental data}
%%%%%%%%%%%%%%%%%%%%%%%%%%%%%%%%%%%%%%%%%%%%%%%%%%%%%%%%%%

Finally, let us compare our findings to experimental data.

\begin{figure}
 \centering
\begin{tabular}{cc}
 \includegraphics[scale=0.7]{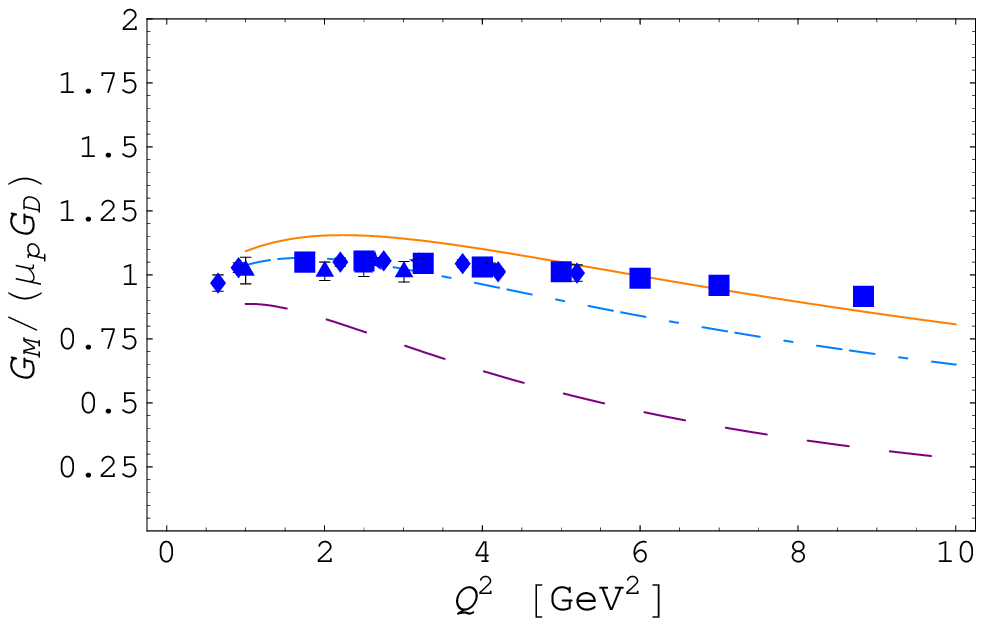} 
&
 \includegraphics[scale=0.7]{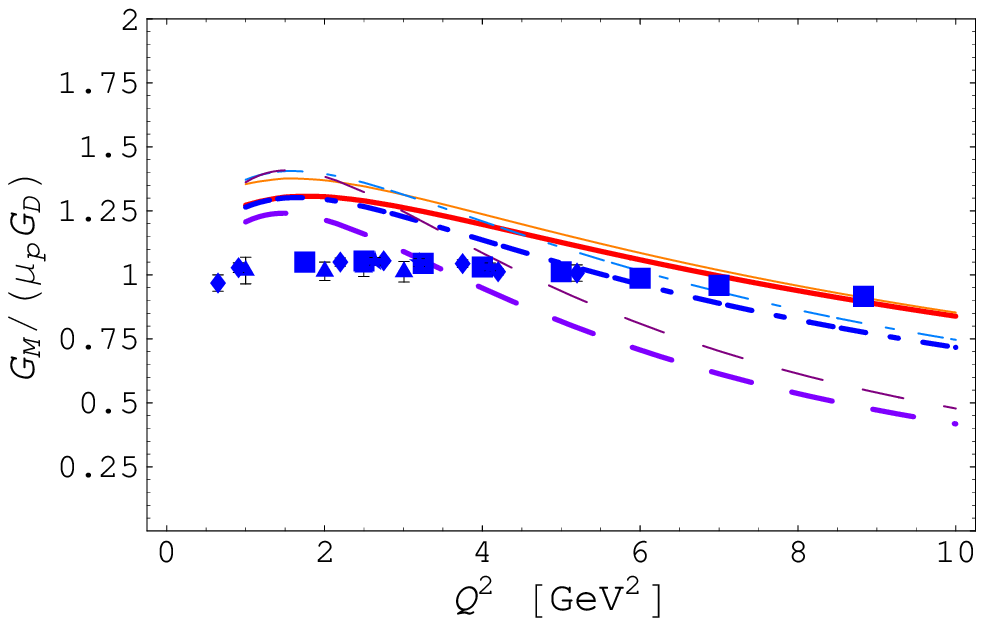} \\[0.2cm]
   \small (a) &
   \small (b)
\end{tabular}
\caption{
LCSR prediction for the proton form factor $G_M$ normalized to $\mu_p G_D$.
The displayed results correspond to
asymptotic  DAs (solid line),
BLW DAs (dash-dotted line),
and CZ-like DAs (dashed line) --
see Sec. \protect\ref{sec:DA} for the DA parameters.
a) LO prediction on the basis of the results from
Ref. \protect\cite{BraunLW06} (i.e., the higher-twist contributions
(up to twist-6) and $x^2$ correction included).
b) 
LO prediction on the basis of the results from Ref. \protect\cite{BraunLW06}
plus NLO correction for twist-3 ($M_N=0$ case) calculated in this
work (choice of scales as in Fig. \protect\ref{f:F2NLO1}). Thin lines: 
 the default value of $\lambda_1$ (\protect\ref{eq:lambda1def}).
Thick lines:
 the corrected value of $\lambda_1$ (\protect\ref{eq:lambda1corrA}).
Experimental data obtained using Rosenbluth separation:
$\blacktriangle$ SLAC  1994 \cite{Walkeretal94},
$\blacksquare$ SLAC 1994 \cite{Andivahisetal94},
$\blacklozenge$ JLab 2004 \cite{Christyetal04},
$\bigstar$ JLab 2005 \cite{Qattanetal04}.
}
 \label{f:GMmuGD}
\end{figure}

\begin{figure}
 \centering
\begin{tabular}{cc}
 \includegraphics[scale=0.7]{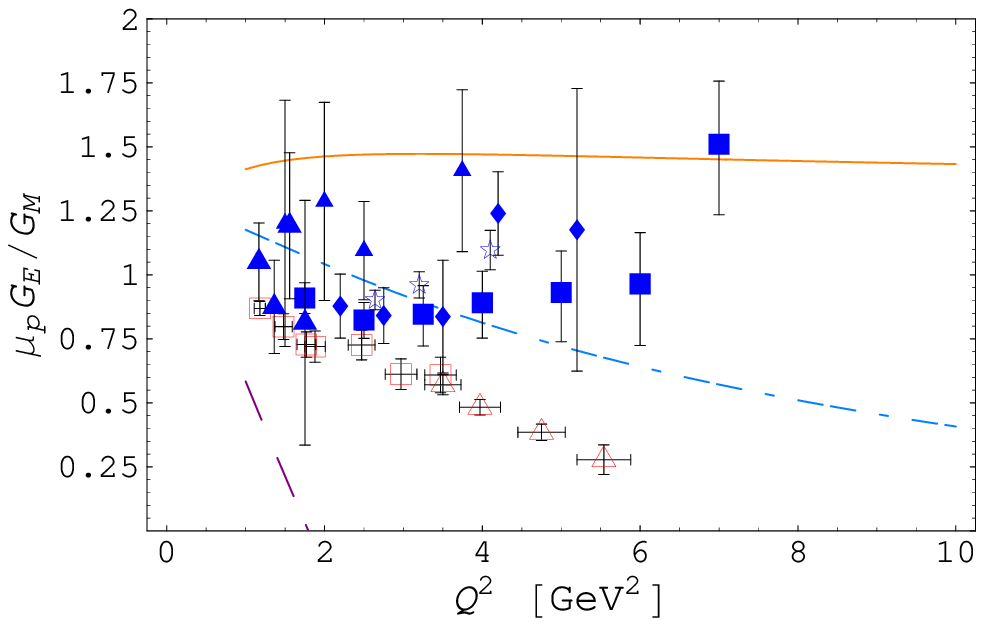} 
&
 \includegraphics[scale=0.7]{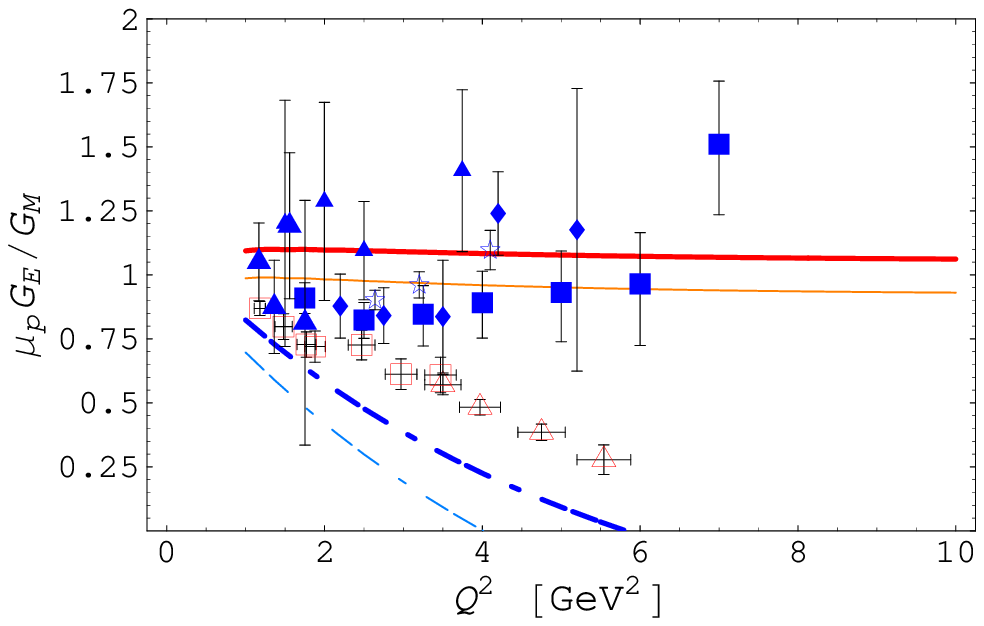} \\[0.2cm]
   \small (a) &
   \small (b)
\end{tabular}
\caption{
LCSR prediction for the proton form factors ratio $\mu_p G_E/G_M$.
The displayed results correspond to
asymptotic  DAs (solid line),
and BLW DAs (dash-dotted line) --
see Sec. \protect\ref{sec:DA} for the DA parameters.
a) LO prediction on the basis of the results from
Ref. \protect\cite{BraunLW06} (i.e., the higher-twist contributions
up to twist-6 and $x^2$ correction included).
b) 
LO prediction on the basis of the results from Ref. \protect\cite{BraunLW06}
plus NLO correction (to $F_2$) for twist-3 ($M_N=0$ case) calculated in this
work (choice of scales as in Fig. \protect\ref{f:F2NLO1}). Thin lines: 
 the default value of $\lambda_1$ (\protect\ref{eq:lambda1def}).
Thick lines:
 the corrected value of $\lambda_1$ (\protect\ref{eq:lambda1corrA}).
Experimental data obtained using Rosenbluth separation:
$\blacksquare$ SLAC 1994 \cite{Andivahisetal94},
$\blacklozenge$ JLab 2004 \cite{Christyetal04},
$\bigstar$ JLab 2005 \cite{Qattanetal04},
$\blacktriangle$ SLAC 1970 (small) \cite{Littetal70}
                  and Bonn 1971 (big) \cite{Bergeretal71} data
                  revised in \cite{Arrington03}.
Experimental data obtained via polarization transfer:
$\triangle$ JLab 2001 \cite{Gayouetal01},
$\Box$ JLab 1999 \cite{Jonesetal99}.
}
 \label{f:muGEGM}
\end{figure}

\begin{figure}
 \centering
\begin{tabular}{cc}
 \includegraphics[scale=0.7]{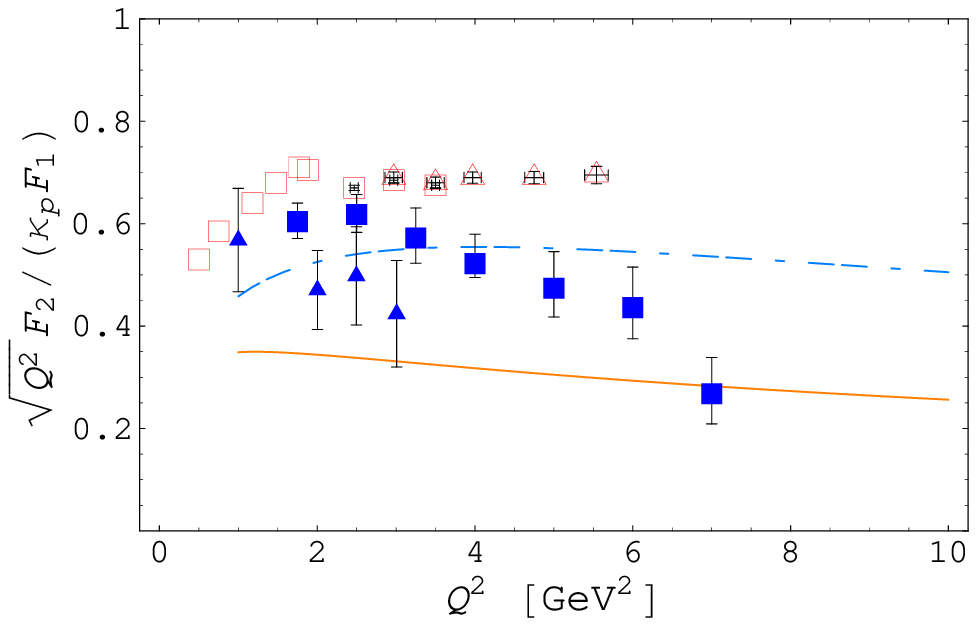} 
&
 \includegraphics[scale=0.7]{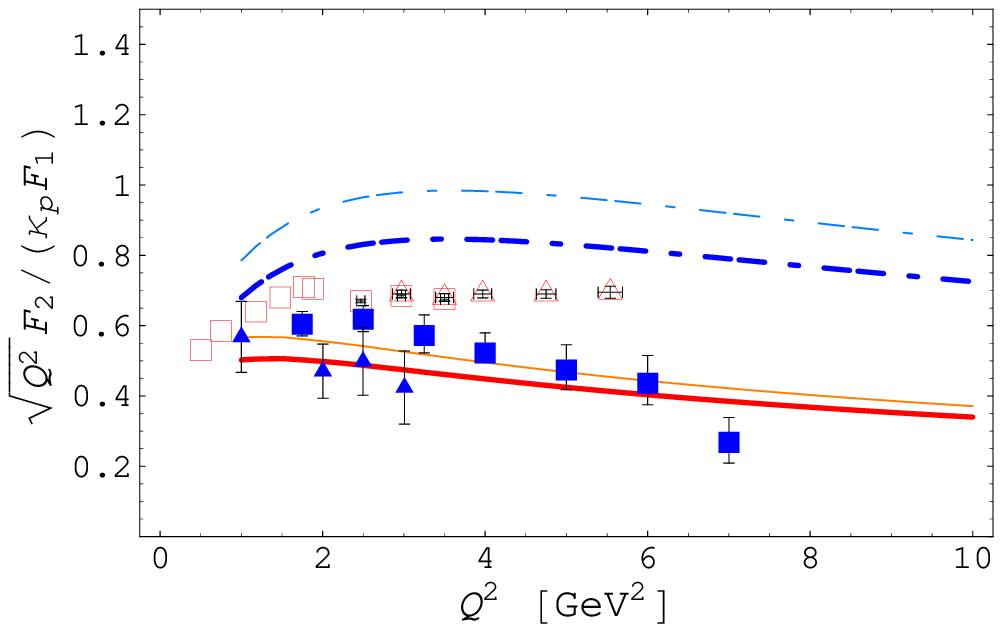} \\[0.2cm]
   \small (a) &
   \small (b)
\end{tabular}
\caption{
LCSR prediction for the proton form factors ratio 
$\sqrt{Q^2} F_2/(\kappa_p F_1)$.
The displayed results correspond to
asymptotic  DAs (solid line),
and BLW DAs (dash-dotted line) --
see Sec. \protect\ref{sec:DA} for the DA parameters.
a) LO prediction on the basis of the results from
Ref. \protect\cite{BraunLW06} (i.e., the higher-twist contributions
up to twist-6 and $x^2$ correction included).
b) 
LO prediction on the basis of the results from Ref. \protect\cite{BraunLW06}
plus NLO correction (to $F_2$) for twist-3 ($M_N=0$ case) calculated in this
work (choice of scales as in Fig. \protect\ref{f:F2NLO1}). Thin lines: 
 the default value of $\lambda_1$ (\protect\ref{eq:lambda1def}).
Thick lines:
 the corrected value of $\lambda_1$ (\protect\ref{eq:lambda1corrA}).
Experimental data obtained using Rosenbluth separation:
$\blacksquare$ SLAC 1994 \cite{Andivahisetal94},
$\blacktriangle$ SLAC  1994 \cite{Walkeretal94}.
Experimental data obtained via polarization transfer:
$\triangle$ and  $\Box$ as in Ref. \protect\cite{BraunLW06}, Fig. 15 (M. Jones, private communication).
}
 \label{f:F2F1}
\end{figure}

In Figs. \ref{f:GMmuGD}, \ref{f:muGEGM} and \ref{f:F2F1}
we display the results for
$G_M/(\mu_p G_D)$, $\mu_p G_E/G_M$ and $\sqrt{Q^2} F_2/(\kappa_p F_1)$
obtained using the
asymptotic  DAs (solid line),
the BLW DAs (dash-dotted line),
and the CZ-like DAs (dashed line). 
The DA parameters are given in Sec. \protect\ref{sec:DA}.

For comparison, in 
Figs. \ref{f:GMmuGD}a, \ref{f:muGEGM}a and \ref{f:F2F1}a
we present the LO predictions obtained on the basis of the results from
Ref. \protect\cite{BraunLW06}, where the higher-twist contributions
(up to twist-6) and $x^2$ correction were included
and the value of $\lambda_1$ (\protect\ref{eq:lambda1def}) was used.
The NLO predictions, 
i.e. the LO predictions obtained on the basis of the results from Ref. 
\protect\cite{BraunLW06} plus NLO corrections for twist-3 ($M_N=0$ case) 
calculated in this work 
(with $s_0$ and $M_B$ as from Eq. \req{eq:s0MB2}, 
%while $\mu_R^2=Q^2$ and $\mu_F^2=\mu_{R,1}^2=M_B^2$), 
while $\mu_R^2=\mu_F^2=\mu_{R,1}^2=M_B^2$), 
are displayed in Figs. \ref{f:GMmuGD}b, \ref{f:muGEGM}b and \ref{f:F2F1}b.
Here we investigate also the change of the results
with the choice of $\lambda_1$.
The NLO results obtained using
 the default value of $\lambda_1$ (\protect\ref{eq:lambda1def})
are, as in Fig. \ref{f:GMmuGD}a, \ref{f:muGEGM}a and \ref{f:F2F1}a, 
denoted by thin lines, while thick lines denote the NLO results
obtained employing the corrected value of $\lambda_1$ 
(\protect\ref{eq:lambda1corrA}).

%%%%%%%%%%%%%%%%%%%%%%%%%%%%%%%%%%%%%%%%%%%%%%%%%%%%%%%%%%%%%%%%%%%%%%
%%%%%%%%%%%%%%%%%%%%%%%%%%%%%%%%%%%%%%%%%%%%%%%%%%%%%%%%%%%%%%%%%%%%%%

In Fig. \ref{f:GMmuGD} we display the
LCSR prediction for $G_M$ normalized to $\mu_p G_D$.
The displayed experimental data were obtained using Rosenbluth separation:
$\blacktriangle$ SLAC  1994 \cite{Walkeretal94},
$\blacksquare$ SLAC 1994 \cite{Andivahisetal94},
$\blacklozenge$ JLab 2004 \cite{Christyetal04},
$\bigstar$ JLab 2005 \cite{Qattanetal04}.
The LO results \ref{f:GMmuGD}a favour the asymptotic and
BLW DAs.
The inclusion of NLO corrections (compare thin lines
in Figs. \ref{f:GMmuGD}a and \ref{f:GMmuGD}b) 
raises the predictions.
The change of $\lambda_1$ to the corrected value
(\protect\ref{eq:lambda1corrA}) lowers the NLO results (thick lines)
slightly.
The NLO results seem to overshoot the data at lower $Q^2$,
while at higher $Q^2$ again the asymptotic and BLW results
seem to be closer to the data than the results obtained
using the CZ-like DAs.

%%%%%%%%%%%%%%%%%%%%%%%%%%%%%%%%%%%%%%%%%%%%%%%%%%%%%%%%%%%%%%%%%%%%%%
%%%%%%%%%%%%%%%%%%%%%%%%%%%%%%%%%%%%%%%%%%%%%%%%%%%%%%%%%%%%%%%%%%%%%%

In Fig. \ref{f:muGEGM} we present the
LCSR prediction for $\mu_p G_E/G_M$.
We use the experimental data obtained using Rosenbluth separation
(
$\blacksquare$ SLAC 1994 \cite{Andivahisetal94},
$\blacklozenge$ JLab 2004 \cite{Christyetal04},
$\bigstar$ JLab 2005 \cite{Qattanetal04},
$\blacktriangle$ SLAC 1970 (small) \cite{Littetal70}
                  and Bonn 1971 (big) \cite{Bergeretal71} data
                  revised in \cite{Arrington03})
as well as more reliable experimental data obtained via polarization transfer
($\triangle$ JLab 2001 \cite{Gayouetal01},
$\Box$ JLab 1999 \cite{Jonesetal99}).
The LO results displayed in Fig. \ref{f:muGEGM}a
show that while the results obtained using the
CZ-like DAs are quite low and well beyond the data,
and the results obtained the asymptotic DAs are on the 
high edge of the data, the BLW results seem to be in better agreement
with the data, but one cannot really make some conclusive statements.
The inclusion of NLO corrections (compare thin lines
in Figs. \ref{f:muGEGM}a and \ref{f:muGEGM}b) 
lowers the predictions significantly, while
the change of $\lambda_1$ to the corrected value
(\protect\ref{eq:lambda1corrA}) raises the NLO results (thick lines)
slightly.
Note that the results obtained using the CZ-like DAs are ruled out
and left out.
One can see now the NLO results, especially the results obtained using
the corrected value for $\lambda_1$, are in quite good agreement
with the data. 
The results obtained using the asymptotic DAs
seem to describe well the experimental data obtained using
the Rosenbluth separation, while the NLO results obtained using the
BLW DAs seem to follow the slope of the preferred experimental data 
obtained via polarization transfer.

%%%%%%%%%%%%%%%%%%%%%%%%%%%%%%%%%%%%%%%%%%%%%%%%%%%%%%%%%%%%%%%%%%%%%%
%%%%%%%%%%%%%%%%%%%%%%%%%%%%%%%%%%%%%%%%%%%%%%%%%%%%%%%%%%%%%%%%%%%%%%

In Fig. \ref{f:F2F1} we present the
LCSR prediction for $\sqrt{Q^2} F_2/(\kappa_p F_1)$.
We display the experimental data obtained using Rosenbluth separation
($\blacksquare$ SLAC 1994 \cite{Andivahisetal94},
$\blacktriangle$ SLAC  1994 \cite{Walkeretal94})
and preferred experimental data obtained via polarization transfer
($\triangle$ and  $\Box$ as in Ref. \protect\cite{BraunLW06}, 
Fig. 15 (M. Jones, private communication)).
The LO results are displayed in Fig. \ref{f:F2F1}a
and while the results obtained using the asymptotic DAs are on the
lower edge of the data, the BLW results seem to fall
close to the data (at least for lower $Q^2$).
The inclusion of NLO corrections (compare thin lines
in Figs. \ref{f:F2F1}a and \ref{f:F2F1}b) 
raises the predictions significantly, while
the change of $\lambda_1$ to the corrected value
(\protect\ref{eq:lambda1corrA}) lowers the NLO results (thick lines)
slightly.
As in the case of the $\mu G_E/G_M$ results displayed in Fig. \ref{f:muGEGM},
one can see that the NLO results, especially the results obtained using
the corrected value for $\lambda_1$, are in good agreement
with the data. Again, the results obtained using the asymptotic DAs
seem to describe well the experimental data obtained using
the Rosenbluth separation, while the NLO results obtained using the
BLW DAs seem to follow the slope of the experimental data 
obtained via polarization transfer.

In conclusion, the inclusion of NLO corrections calculated at twist-3 for
$M_N = 0$ introduces significant changes in
the LCSR predictions for
$G_M/(\mu_p G_D)$, $\mu_p G_E/G_M$ and $\sqrt{Q^2} F_2/(\kappa_p F_1)$.
It seems that NLO corrections, as well as the use of the corrected value
for $\lambda_1$
(\protect\ref{eq:lambda1corrA}), bring the predictions for
$\mu_p G_E/G_M$ and $\sqrt{Q^2} F_2/(\kappa_p F_1)$
in better agreement with the experimental data.
For these quantities, the results obtained using the asymptotic DAs
seem to describe well the experimental data obtained using
the Rosenbluth separation, while the NLO results obtained using the
BLW DAs seem to follow the slope of the experimental data 
obtained via polarization transfer.

%%%%%%%%%%%%%%%%%%%%%%%%%%%%%%%%%%%%%%%%%%%%%%%%%%%%%%%%%%%%%%%%%%%%%%%%%%%%%%%
%%%%%%%%%%%%%%%%%%%%%%%%%%%%%%%%%%%%%%%%%%%%%%%%%%%%%%%%%%%%%%%%%%%%%%%%%%%%%%%
\section{Summary and conclusions}
\label{sec:concl}
%%%%%%%%%%%%%%%%%%%%%%%%%%%%%%%%%%%%%%%%%%%%%%%%%%%%%%%%%%%%%%%%%%%%%%%%%%%%%%%

In this work the first attempt 
has been made
to asses
the size of NLO corrections to nucleon form factors.

In LCSR approach dealing with nucleons is much more
demanding than dealing with mesons, even at LO.
For one, the number of contributing terms is rather
large, the expressions are more involved, and the presence of 
three, instead of two, partons with corresponding momenta
makes the calculation more complicated.
All this is present at NLO also, with additional
difficulties of one-loop calculation and larger number of
contributing Feynman diagrams.

In order to calculate the NLO corrections, 
we have started with the simple $M_N^2=0$ (but $M_N \ne 0$) approximation,
i.e.,  the approximation corresponding
to the first two terms in the expansion in nucleon mass \req{eq:Mexpansion}. 
In that approximation only the leading twist, twist-3,
and next-to-leading twist, twist-4, contributions appear.
But to our surprise it turned out that the collinear divergences
appearing in one-loop calculation do not cancel on the
level of separate twist and that actually mixing appears which,
without knowing the corresponding kernels, disables us in 
determing the finite contributions
(see Sec. \ref{sec:awzayMN0}).

Hence, 
we have strengthen our approximation
and considered $M_N=0$ approximation 
(corresponds to the first term in \req{eq:Mexpansion})  
in which only twist-3 contributes
and the evolution kernels are known. 
We have shown the explicit cancellation of 
collinear, as well as, UV singularities
(see  Sec. \ref{sec:tw3NLO}).
The finite twist-3 NLO contributions to the correlation
function are thus obtained in $M_N=0$ approximation
and relevant invariant functions are listed in App. \ref{sec:appBfunM0}.
 
We note that 
the observation of mixing of twist-3 and twist-4 NLO contributions
is in nature similar to the observation given in Sec. \ref{sec:gaugeinv}
that the gauge invariant results are obtained not twist-by-twist
but order-by-order in $M_N$.
The gauge condition is for $M_N=0$ case
satisfied both in LO and NLO order. 
For $M_N^2=0$ and $M_N \ne 0$
we have shown to LO that gauge condition
is satisfied only when the sum of all contributing terms 
is taken into account, i.e., both twist-3 and twist-4 contributions.
The additional condition is that the asymptotic forms of twist-3 DAs 
are used (no conditions, at least at this order, on twist-4 DAs).
Hence, gauge invariance can be satisfied  order by order in the
expansion in $M_N$ \req{eq:Mexpansion} with possibly some additional
conditions on the form of DAs.
%For the check of higher order terms in $M_N$ one should calculate 
%the complete LO higher-twist contributions to the correlation function.

To repeat, by switching-on the nucleon mass, which is, of course necessary
in order to determine higher-twists, we are 
at  $M_N^2=0$ and $M_N \ne 0$ stuck with
mixing of the contributions corresponding to different twists.
This ''mixing'' can be seen even at LO through the check of gauge
invariance with respect to photon.  
If we take $M_N^2 \ne 0$, there are no collinear divergences
whose cancellation we should take care of and no apparent ''mixing''.
But the NLO expressions are more involved and there are additional 
open problems 
one should solve before attacking higher-order higher-twist calculations. 
For example,
there is an open technical problem elaborated in Sec. \ref{sec:awzayMN0} 
and connected to the calculation of the NLO contributions 
to second and third case defined in \req{eq:cases}.
The calculation of NLO corrections for $M_N \ne 0$ case
and thus NLO corrections to higher twists we postpone for some other time. 

In Sec. \ref{sec:num} we have presented and analyzed
our numerical results based on the calculation of NLO corrections
in $M_N=0$ approximation, i.e., twist-3 NLO corrections
in that approximation. Using Ioffe current in this approximation
we are able to calculate only the corrections to $F_2$
nucleon form factor. To make a full analysis and estimate
the importance of NLO corrections, we have also included
the LO results obtained beyond this approximation, i.e.,
leading twist and higher-twist results obtained in Ref. \cite{BraunLW06}.
We have considered here just the proton case.

For $F_1$
twist-4 LO contributions are dominant and positive,
and there are no NLO corrections in $M_N=0$
approximation.
When one considers the
size of various contributions to $F_2$, one realizes that 
twist-3 LO contributions are dominant and positive.
The twist-5 LO contributions are more pronounced than twist-4
LO contributions and for both the ratio to twist-3 LO contribution
is very sensitive to the shape of DAs.
The $x^2$-contributions are negative and 
the ratio of $x^2$-contributions
and LO twist-3 contributions does not change much for various DAs.
The twist-3 NLO corrections are positive and cca 60\%.
%for $\mu_R^2=\mu_F^2=\mu_{R,1}^2=Q^2$
%while in the range of cca 70-40\% 
%for $\mu_R^2=Q^2$ and $\mu_F^2=\mu_{R,1}^2=M_B^2$.
The NLO corrections to $F_2$ calculated at twist-3 taking $M_N=0$ are thus
large.
They vary for different DAs 
and depend on the choice of renormalization and factorization scales.
In contrast to the dependence of $F_2^{\rm NLO}/F_2^{\rm LO}$,
the dependence of the complete NLO prediction of $F_2$
on the choice of renormalization and factorization scales
is small.

The inclusion of NLO corrections calculated at twist-3 for
$M_N = 0$ introduces significant changes in
the LCSR predictions for
$G_M/(\mu_p G_D)$, $\mu_p G_E/G_M$ and $\sqrt{Q^2} F_2/(\kappa_p F_1)$.
It seems that NLO corrections, as well as the use of the corrected value
for $\lambda_1$
(\protect\ref{eq:lambda1corrA}), bring the predictions for
$\mu_p G_E/G_M$ and $\sqrt{Q^2} F_2/(\kappa_p F_1)$
in better agreement with the experimental data.
For these quantities, the results obtained using the asymptotic DAs
seem to describe well the experimental data obtained using
the Rosenbluth separation, while the NLO results obtained using the
BLW DAs seem to follow the slope of the preferred experimental data 
obtained via polarization transfer.

Further analysis and inclusion of NLO corrections at higher-twists
is needed to draw some more conclusive results.

%%%%%%%%%%%%%%%%%%%%%%%%%%%%%%%%%%%%%%%%%%%%%%%%%%%%%%%%%%%%%%%%%%%%%%%%%%%%%%%

\subsection*{Acknowledgements}

We are indebted to V. Braun for both proposing this subject
and for clarifying discussions.
K. P-K. would like to thank the theory group at the University 
of Regensburg for its warm hospitality
and 
would also like to thank 
G. Duplan\v{c}i\'{c},
K. Kumeri\v{c}ki,
A. Lenz, 
B. Meli\'{c}, 
D. M\"{u}ller,
and A. Peters, 
for taking the time to discuss some aspects of this work.
This project has been supported by 
the  German Research Foundation (DFG) 
under the contract no. 9209070, 
Croatian Ministry  of Science, Education and Sport 
under the contract no. 098-0982930-2864
and the Studienstiftung des deutschen Volkes.

%%%%%%%%%%%%%%%%%%%%%%%%%%%%%%%%%%%%%%%%%%%%%%%%%%%%%%%%%%%%%%%%%%%%%%%%%%%%%%%
%%%%%%%%%%%%%%%%%%%%%%%%%%%%%%%%%%%%%%%%%%%%%%%%%%%%%%%%%%%%%%%%%%%%%%%%%%%%%%%
\appendix

\renewcommand{\theequation}{\Alph{section}.\arabic{equation}}%

%%%%%%%%%%%%%%%%%%%%%%%%%%%%%%%%%%%%%%%%%%%%%%%%%%%%%%%%%%%%%%%%%%%%%%%%%%%%%%%
\section{Feynman rules}
\setcounter{equation}{0}
%%%%%%%%%%%%%%%%%%%%%%%%%%%%%%%%%%%%%%%%%%%%%%%%%%%%%%%%%%%%%%%%%%%%%%%%%%%%%%%
\label{sec:appFEYN}

In calculating 
$
(-i \, T_{\mu})
$
where $T_\mu$ is a correlator function \req{eq:correlator}
we use the standard Feynman rules for quark and gluon propagators
and vertices, as well as for the quark-photon vertex.
For loop integrals one has to introduce the usual integration over
loop momenta%
\footnote{We use dimensional regularization
in $D=4 - 2 \eps$ dimensions and for integral measure
we choose 
$\mu^{2 \eps} \int d^D l(2 \pi)^D$ -- see, for example, App. C in Ref. 
\cite{MelicNP01}
for some comments on this choice, i.e., on 
introduction of scale $\mu^2$ in Feynman integrals.}.

The vertex corresponding to the interpolating nucleon current
given by \req{eq:Ioffecurr} 
reads
\begin{equation}
\mbox{u-quark(} \alpha, a
\mbox{)--u-quark(} \beta, b
\mbox{)--d-quark(} \gamma, c
\mbox{)-- Ioffe current}: \; 
%\left|
\left( C \gamma_\lambda \right)_{\alpha \beta} \;
\left( \gamma_5 \gamma^\lambda \right)_{\delta \gamma}\;
\varepsilon^{abc}
%\right.
\, ,
\label{eq:IoffeFEYN}
\end{equation}
where all quark lines are going in the vertex
and the order of u,u, and d-quarks with corresponding
Lorentz ($\alpha, \beta, \gamma$)
and colour ($a,b,c$) indices is counterclockwise.
Furthermore, $(C \gamma_\lambda)^T=C \gamma_\lambda$, i.e.,
$\left( C \gamma_\lambda \right)_{\alpha \beta}=
\left( C \gamma_\lambda \right)_{\beta \alpha}$.

The (incoming) nucleon ``projector'' corresponding to \req{eq:gLdecomp}
and the first case of \req{eq:cases} is given by
\begin{eqnarray}
\lefteqn{\mbox{u-quark(} u_1 P, \alpha, a
\mbox{)--u-quark(} u_2 P, \beta, b
\mbox{)--d-quark(} u_3 P,\gamma, c\mbox{):}}
\nonumber \\[0.3cm] &{}\qquad & \hspace*{2cm} \qquad
\;
\frac{1}{4} \; \int {\cal D}u \,F^{(i)}(u_1,u_2,u_3) \;
\;
X^{(i)}_{\alpha \beta}
\;
Y^{(i)}_{\gamma}
\;
\frac{\varepsilon^{abc}}{6}
\, ,
\label{eq:case1FEYN}
\end{eqnarray}
where all quark lines are outgoing from the nucleon blob 
and the order is clockwise.
The trivial identity  
$X^{(i)}_{\alpha \beta}= \left(X^{(i)}\right)^T_{\beta \alpha}$
together with
\req{eq:XT} is useful in some cases.

The typical contribution obtained using the general Lorentz decomposition
\req{eq:gLdecomp} and Ioffe current \req{eq:Ioffecurr}, i.e., 
Feynman rules \req{eq:case1FEYN} and \req{eq:IoffeFEYN}, respectively, 
has two parts. 
For the $d-d$ quark line, 
by going, following the standard rule, in the opposite direction 
of the fermion line,
one obtains the product of $\gamma$ matrices with the nucleon spinor. 
The $u-u$ lines close the trace and obviously, in writing it
down, one goes opposite to the
direction of the one quark line and along the other one.
The latter case corresponds to
$\left(\gamma_{\mu 1} \gamma_{\mu 2} \ldots \gamma_{\mu_n}\right)^T$
(where 
$\gamma_{\mu 1}  \ldots \gamma_{\mu_n}$
is the order of $\gamma$ matrices opposite to the direction of the fermion line)
and one then makes use of
\begin{equation}
\left(
\gamma_{\mu 1} \gamma_{\mu 2} \ldots \gamma_{\mu_n}
\right)^T
=(-1)^n C
\gamma_{\mu n} \ldots \gamma_{\mu_1} C^{-1}
\, ,
\end{equation}
i.e. when going in the direction of fermion line
one puts the $\gamma$ matrices on the that line
between $C$ and $C^{-1}$.

Finally, let us mention that the usual relations for SU($N_C$)
algebra should be employed in calculating the colour factors.
Since we are dealing here with nucleon described by three quarks
we are actually already assuming $N_C=3$ and only this choice
leads to gauge invariant results.
Obviously,
\begin{equation}
\varepsilon^{abc} \; \varepsilon^{abc}=6
\, .
\end{equation}

%%%%%%%%%%%%%%%%%%%%%%%%%%%%%%%%%%%%%%%%%%%%%%%%%%%%%%%%%%%%%%%%%%%%%%%%%%%%%%%
\section{$\gamma_5$ ambiguity in dimensional regularization}
\setcounter{equation}{0}
%%%%%%%%%%%%%%%%%%%%%%%%%%%%%%%%%%%%%%%%%%%%%%%%%%%%%%%%%%%%%%%%%%%%%%%%%%%%%%%
\label{sec:gamma5-app}

When using dimensional regularization, one runs
into trouble with quantities that have the well-defined
properties only in $D=4$ space-time dimensions, that is, 
with the Levi-Civita tensor
$\eps_{\mu \nu \lambda \kappa}$, which is a genuine
4 dimensional object,
and consequently with the pseudoscalar $\gamma_5$ Dirac matrix.
Let us mention that the appearance and mixing with evanescent operators
\cite{evanesc},
as well as, the definition of Fierz transformation in $D$ dimensions 
are also connected to this problem.
We shall handle it similarly to Ref. \cite{MelicNP01} 
%(for details, see App. A in \cite{MelicNP01}) 
with some additional finesse concerning Chisholm identity 
(see Sec. \ref{sec:Dgamma5}).
Below we explain the general features of 
the ambiguities that we encounter in our calculation.
In order to resolve these one should generally use some
other input like the knowledge of the quantity that does
not suffer from ambiguities, condition of cancellation
of singularities, condition of preservation of gauge invariance, 
Ward identities etc.

\subsection{General remarks -- trace ambiguity}

The generalization of the $\gamma_5$ matrix
in $D$ dimensions represents a problem,
since it is not possible to simultaneously retain
its anticommuting and trace properties.
In practice, the ambiguity arises
when evaluating a trace
containing  a  $\gamma_5$
and pairs of
contracted $\gamma$ matrices
and/or pairs of Dirac slashed loop momenta
%\footnote{
%The presence of a pair of Dirac slashed loop momenta
%leads in fact  to the  appearance of
%a pair of contracted $\gamma$ matrices,
%since the loop integration
%\begin{displaymath}
%   \int \frac{d^D l}{(2 \pi)^D} \,
%     \frac{ l^{\kappa} l^{\tau} }{(Denominator)} =
%    g^{\kappa \tau} I_g + \cdots
%\end{displaymath}
%transforms
%$l^{\kappa} l^{\tau} \gamma_{\kappa} \cdots \gamma_{\tau}$ into
%$\gamma_{\kappa} \cdots \gamma^{\kappa} I_g + \cdots$.
%Apart from the contracting $\gamma$ matrices and pairs of
%Dirac slashed loop momenta, the rest of the trace elements
%could be treated as 4 dimensional, so their (anti)commutation
%with $\gamma_5$ does not make a difference.
%}%
.
To deal with a $\gamma_5$ matrix,
several possible schemes have been proposed
in the literature.

In the so-called naive-$\gamma_5$ scheme \cite{ChanowitzFH79},
the anticommutation
property of $\gamma_5$
%\begin{equation}
%   \left\{ \gamma_{\mu}, \gamma_5 \right\} = 0
%\label{eq:ng5sh}
%\end{equation}
is retained, while
the cyclicity of the trace
is abandoned.
%, so that, for example,
%\begin{subequations}
%\begin{eqnarray}
%  \text{Tr} \left[ \gamma_5
%          \not \! a  \gamma_{\mu}
%      \not \! b  \not \! c \not \! d \gamma^{\mu} \right] &=&
%        (D-6) \,
%  \text{Tr} \left[ \gamma_5
%       \not \! a  \not \! b  \not \! c  \not \! d   \right]
%           \, ,  \quad   \label{eq:Trexamp1} \\
%  \text{Tr} \left[ \gamma^{\mu} \gamma_5
%          \not \! a \gamma_{\mu}
%      \not \! b \not \! c \not \! d  \right] &=&
%        (2-D) \,
%  \text{Tr} \left[ \gamma_5
%        \not \! a  \not \! b  \not \! c  \not \! d  \right]
%           \, . \quad \label{eq:Trexamp2}
%\end{eqnarray}
The traces obtained by cyclic permutation of
the matrices
%$\text{$\gamma_5$,
%$ \not \! a$,
%$\gamma_{\mu}$, $\not \! b$,  $ \not \! c$, $\not \! d$,
%$ \gamma^{\mu}$}$
%can be  divided into two classes, depending on
%the location of $\gamma_5$ with respect to the contracted $\gamma$
%matrices:
%and, consequently, considering the results in
%naive-$\gamma_5$ scheme:
%those in which
%$\gamma_5$ is outside the contracted pair
%as in \req{eq:Trexamp1}, and those
%where $\gamma$s are contracted through $\gamma_5$ as
%in \req{eq:Trexamp2}.
%As is seen, the result \req{eq:Trexamp1}
%and the result \req{eq:Trexamp2},
%in which  the anticommuting property \req{eq:ng5sh}
%of $\gamma_5$ had to be used before the contraction
%of $\gamma$ matrices can be performed, 
differ by $D-4$.
Consequently, if the
trace is multiplied by a pole in $D-4$,
there appears a finite ambiguity in the result.
%\label{eq:Trexamp}
%\end{subequations}
An alternative scheme has been proposed
in the original paper on the dimensional regularization
by 't Hooft and Veltman \cite{tHooftV72}, and further
systematized by Breitenlohner and Maison \cite{BreitenlohnerM77}.
In this scheme, to which we refer as HV scheme,
the anticommutativity of
$\gamma_5$ is abandoned. 
%and replaced by
%\begin{eqnarray}
%   \left\{ \gamma_{\mu}, \gamma_5 \right\} &=& 0
%           \text{ for }  \mu=0, \cdots, 4
%       \nonumber \\
%   \left[ \gamma_{\mu}, \gamma_5 \right] &=& 0
%           \text{ for }  \mu=4, \cdots, D
%         \, .
%\label{eq:BMsh1}
%\end{eqnarray}
In contrast to naive-$\gamma_5$ scheme, this scheme is claimed to be
mathematically consistent but still not without drawbacks.
Namely, this prescription for $\gamma_5$ violates the Ward identities
and introduces ``spurious'' anomalies which violate chiral
symmetry.
% and hence gauge invariance
To restore the Ward identities,
finite counterterms should be added order by order
in perturbation theory \cite{Bonneau81}.
In this scheme,
the cyclicity of the trace is retained.
%and
%the traces given in \req{eq:Trexamp}
%become
%\begin{eqnarray}
%  \text{Tr} \left[ \gamma_5
%          \not \! a  \gamma_{\mu}
%      \not \! b  \not \! c \not \! d \gamma^{\mu} \right] &=&
%  \text{Tr} \left[ \gamma^{\mu} \gamma_5
%          \not \! a \gamma_{\mu}
%      \not \! b \not \! c \not \! d  \right]
%         \nonumber \\ &=&
%        (D-6) \,
%  \text{Tr} \left[ \gamma_5
%       \not \! a  \not \! b  \not \! c  \not \! d   \right]
%       \, . \quad
%\label{eq:TrexampBM}
%\end{eqnarray}
%As is seen,  the result \req{eq:TrexampBM}, obtained
%in the HV scheme, corresponds to the result \req{eq:Trexamp1},
%obtained using the naive-$\gamma_5$ scheme.

If a trace contains an even number of $\gamma_5$
matrices, then the property $\gamma_5^2=1$
can be used to eliminate $\gamma_5$'s from the trace,
and the Ward identities are preserved if the
naive-$\gamma_5$ scheme %\req{eq:ng5sh}
is used \cite{ChanowitzFH79}
(the cyclicity of the trace is restored and
the corresponding results are unambiguous).
On the other hand, in the HV scheme
the ``spurious'' anomalies can occur
owing to the
non-anticommuting property of $\gamma_5$.
As for the traces containing an odd number of
$\gamma_5$ matrices, we are left with the
above mentioned ambiguities in the results.

For details, we refer to App. A in Ref. \cite{MelicNP01}. 

\subsection{General remarks -- Chisholm identity}

Additionally, the Chisholm identity
that we need in our calculation
\begin{eqnarray}
\gamma^{\mu} \varepsilon_{\mu \alpha \beta \gamma}
=
i \left(
\gamma_{\alpha} \gamma_{\beta} \gamma_{\gamma}
- g_{\alpha \beta} \gamma_{\gamma}
+ g_{\gamma \alpha} \gamma_{\beta}
- g_{\beta \gamma} \gamma_{\alpha}
\right)
\label{eq:Chisholm}
\end{eqnarray}
is strictly speaking valid only in $D=4$ dimensions.
The modification for HV scheme can be found in the literature
(see, for example, {\it Tracer} \cite{JaminL93} manual) 
but we have not been able to find any recipe for the naive-$\gamma_5$ scheme.

Our analysis of the problem 
has shown that when applying the Chisholm identity on expressions
of the form
\begin{equation}
\gamma^\mu \gamma_\kappa \gamma^\nu \varepsilon_{\mu \nu \alpha \beta}
\end{equation}
different results appear in dependence of whether
one first contracts the Levi-Civita tensor with the $\gamma$
matrix on the left or the right side of the non-contracted matrix
($\gamma_\kappa$ in this case).
The difference is again, as expected, proportional to $D-4$.
For example,
\begin{subequations}
\begin{equation}
\gamma^\mu \gamma^{\alpha}
\gamma_\kappa \gamma^\nu \gamma^\beta \varepsilon_{\mu \nu \alpha \beta}
\end{equation}
has two sets of results
\begin{equation}
\left(\gamma^\mu \varepsilon_{\mu \nu \alpha \beta}\right) \gamma^{\alpha}
\gamma_\kappa \gamma^\nu \gamma^\beta =
\gamma^\mu \left(\gamma^{\alpha} \varepsilon_{\mu \nu \alpha \beta}\right) 
\gamma_\kappa \gamma^\nu \gamma^\beta = -i (D-4) (D-2) (D-1) \gamma_\kappa \gamma_5
\, ,
\end{equation}
while
\begin{equation}
\gamma^\mu \gamma^{\alpha}  \gamma_\kappa 
\left(\gamma^\nu \varepsilon_{\mu \nu \alpha \beta}\right) \gamma^\beta  =
\gamma^\mu \gamma^{\alpha}  \gamma_\kappa 
\gamma^\nu \left(\gamma^\beta \varepsilon_{\mu \nu \alpha \beta}\right)  =
i (D-4) (D-2) (D-1) \gamma_\kappa \gamma_5
\, .
\end{equation}
\end{subequations}
So, when using the Chisholm identity in its form as in $D=4$ dimensions 
(which should be in agreement with the ``philosophy`` of  
naive-$\gamma_5$ scheme) we again, as in the case of trace ambiguity, 
encounter the ambiguity proportional to $D-4$, 
which, when multiplied by pole in $D-4$, possibly leads to finite
ambiguity of the results. 

%%%%%%%%%%%%%%%%%%%%%%%%%%%%%%%%%%%%%%%%%%%%%%%%%%%%%%%%%%%%%%%%%%%%%%%%%%%%%%%
\section{NLO results for $M_N=0$ case}
\setcounter{equation}{0}
%%%%%%%%%%%%%%%%%%%%%%%%%%%%%%%%%%%%%%%%%%%%%%%%%%%%%%%%%%%%%%%%%%%%%%%%%%%%%%%
\label{sec:appBfunM0}

\subsection{Formalism and notation}

In $M_N=0$ case we cannot asses the ${\cal A}$ contribution
from Eq. \req{eq:projections},
while the corresponding ${\cal B}$ contribution 
we list in this section.

The ${\cal B}$ function can be written as a convolution
in terms of $V_1$ and $A_1$ nucleon DAs
\begin{eqnarray}
\lefteqn{{\mathcal B}_{M_N=0}(Q^2, P'^2)}
\nonumber \\ &=&
  T_{\mathcal B,V_1,M_N=0}(\{x_k\},Q^2, P'^2;\mu_F^2) \; 
\otimes \;  V_1(\{x_k\};\mu_F^2)
\nonumber \\ &&
+  T_{\mathcal B,A_1,M_N=0}(\{x_k\},Q^2, P'^2;\mu_F^2)
\otimes \;  A_1(\{x_k\},\mu_F^2)
\, ,
\label{eq:convolB-V1A1}
\end{eqnarray}
or in terms of twist-3 nucleon DA $\Phi_3=V_1-A_1$
\begin{eqnarray}
\lefteqn{{\mathcal B}_{M_N=0}(Q^2, P'^2)}
\nonumber \\ &=&
  T_{\mathcal B,\Phi,M_N=0}(\{x_k\},Q^2, P'^2;\mu_F^2) 
\; \otimes \;  \Phi_3(\{x_k\};\mu_F^2)
\, .
\label{eq:convolB-Phi}
\end{eqnarray}
Here $q=-Q^2$ is a photon virtuality and $P'=P-q$, while $P$ is incoming
nucleon momentum. The factorization scale is denoted by $\mu_F^2$,
and by $\{x_k\}$ the quark momentum fractions $x_1$, $x_2$ and $x_3$
are denoted. Note that $x_1$ and $x_2$ denote the momentum fractions
of $u$-quarks, while $x_3$ correspond to the momentum fraction 
of the $d$-quark. A stated in Eq. \req{eq:Fsym}, $V_1$ is symmetric and
$A_1$ antisymmetric under $x_1 \leftrightarrow x_2$ exchange. 

It is convenient to introduce a dimensionless quantity
\begin{equation}
W = \frac{P'^2 + Q^2}{Q^2}
\, ,
\label{eq:defW}
\end{equation}
and we will from now on express% 
\footnote{Although, the functional dependence on $W$ is different from the one
on $P'^2$ we retain the same nomenclature, i.e., from now on
we use ${\cal B'}(Q^2,W)\equiv{\cal B}(Q^2,W)$.}
the ${\cal B}$ function using $W$ instead of $P'^2$.

Generally, we write
\begin{eqnarray}
{\cal B}_{M_N=0}(Q^2,W)= 
\frac{1}{Q^2} \,
\sum_i \, T_{{\cal B},F^{(i)},M_N=0}(\{x_k\},W; \mu_F^2/Q^2) \otimes
                          F^{(i)}(\{x_k\};\mu_F^2)
\end{eqnarray}
with $F^{(i)}$ denoting nucleon DAs and $T_{{\cal B},F^{(i)},M_N=0}$
the corresponding ''hard-scattering'' part.
The nucleon DAs are intrinsically non-perturbative quantities but
their evolution to scale $\mu_F^2$ can be calculated perturbatively.
Nevertheless, we take into account only the LO evolution
or neglect the evolution completely.
The ''hard-scattering'' part is calculated perturbatively
and in this work the calculation to NLO is performed.
Hence, following  Sec.\ref{sec:firstcaseMN0} 
(see Eqs. \req{eq:finiteM-Phi} and \req{eq:FiniteM-NLO}, 
and Eqs. \req{eq:finiteM-V1A1} and \req{eq:FiniteM-V1A1-NLO} )
we can write the expansion of $T_{{\cal B},F^{(i)},M_N=0}$ as
\begin{eqnarray}
\lefteqn{T_{{\cal B},F^{(i)},M_N=0}(\{x_k\},W; \mu_F^2/Q^2)}
\nonumber \\ &=&
\left\{ 
T_{{\cal B},F^{(i)},M_N=0}^{\rm LO}(\{x_k\},W)
 + \frac{\alpha_s(\mu_R^2)}{4 \pi}  
T_{{\cal B},F^{(i)},M_N=0}^{\rm NLO}(\{x_k\},W; \mu_F^2/Q^2,\mu_{R,1}^2/Q^2)
    + \cdots \right\} 
\, ,
\nonumber \\
\label{eq:TBMN0}
\end{eqnarray}
where
\begin{eqnarray}
\lefteqn{T_{{\cal B},F^{(i)},M_N=0}^{\rm NLO}(\{x_k\},W; \mu_F^2/Q^2,\mu_{R,1}^2/Q^2)} 
\nonumber \\[0.3cm]
&=& 
T_{{\cal B},F^{(i)},M_N=0}^{\rm NLO,fin}(\{x_k\},W)
\nonumber \\[0.3cm] & &
+ 
T_{{\cal B},F^{(i)},M_N=0}^{\rm NLO,UV}(\{x_k\},W)
\; \ln(\mu_{R,1}^2/Q^2) 
\nonumber \\[0.3cm] & &
+  
T_{{\cal B},F^{(i)},M_N=0}^{\rm NLO,IR}(\{x_k\},W)
\; \ln(\mu_F^2/Q^2) 
\, ,
\label{eq:TBMN0nlo}
\end{eqnarray}
and $\mu_R^2$ and $\mu_{R,1}^2$ scales denote the
coupling constant and Ioffe current renormalization scales
(which are in practice often taken the same, and even same to 
the factorization scale $\mu_F^2$, but are in principle independent). 

Note that all order result for $T_{{\cal B},F^{(i)},M_N=0}$
would not depend on the choice of the renormalization scale,
but the truncation of the series to any finite order (in or case NLO)
introduces the residual dependence. This dependence would be stabilized by 
inclusion of higher-orders ($\alpha_s^n$, $n\ge 2$).
One is left also with the residual dependence of ${\cal B}$ on 
the factorization scale (see Ref. \cite{MelicNP01a} for details on that point).

In the following we summarize the LO and NLO results
for $T_{{\cal B},F^{(i)},M_N=0}$.
These are proportional either to $e_u$ or $e_d$ being $u$ and $d$-quark
charges (depending on where the photon coupled), respectively.
Remember that we are displaying here the proton case, while for the
neutron $e_u$ and $e_d$ have to be exchanged.
Hence, we burden our notation with one more index
\begin{equation}
T_{{\cal B},F^{(i)},M_N=0}^{\cdots}(\cdots)
=
T_{{\cal B},F^{(i)},M_N=0}^{\cdots,e_d}(\cdots)
+
T_{{\cal B},F^{(i)},M_N=0}^{\cdots,e_u}(\cdots)
\, .
\end{equation}

In order to simplify the expressions and write them in a form
most suitable for further calculation, we introduce the following functions
{\allowdisplaybreaks \begin{eqnarray}
g_{0}(x_i,W)&=& 
\frac{1}{(x_i W-1 +i \eta)}
\, ,
\nonumber \\[0.2cm]
g_{1}(x_i,W)&=& - 
\frac{\ln(1-x_i W -i \eta)}{(1-x_i W -i \eta)}
\, ,
\nonumber \\[0.2cm]
g_{2}(x_i,W)&=& -
\frac{\ln^2(1-x_i W -i \eta)}{(1-x_i W -i \eta)}
\, ,
\nonumber \\[0.2cm]
g_{3}(x_i,W)&=& 
\frac{\ln(1-x_i W -i \eta)}{(W +i \eta)}
\, ,
\nonumber \\[0.2cm]
g_{4}(x_i,W)&=& 
\frac{\ln(1-x_i W -i \eta)}{(W +i \eta)^2}
\, ,
\nonumber \\[0.2cm]
g_{5}(x_i,W)&=& 
\frac{\ln^2(1-x_i W -i \eta)}{(W +i \eta)}
\, ,
\nonumber \\[0.2cm]
g_{6}(x_i,W)&=& 
\frac{\ln^2(1-x_i W -i \eta)}{(W +i \eta)^2}
\, ,
\nonumber \\[0.2cm]
g_{7}(x_i,x_j,W)&=& - 
\frac{\ln(1-(x_i + x_j)W - i \eta)}{(1-x_i W -i \eta)}
\, ,
\nonumber \\[0.2cm]
g_{8}(x_i,x_j,W)&=& -
\frac{\ln^2(1-(x_i + x_j)W - i \eta)}{(1-x_i W -i \eta)}
\, ,
\nonumber \\[0.2cm]
g_{9}(x_i,x_j,W)&=& 
\frac{\ln(1-(x_i + x_j)W - i \eta)}{(W+i \eta)}
\, ,
\nonumber \\[0.2cm]
g_{10}(x_i,x_j,W)&=& 
\frac{\ln(1-(x_i + x_j)W - i \eta)}{(W+i \eta)^2}
\, ,
\nonumber \\[0.2cm]
g_{11}(x_i,x_j,W)&=& 
\frac{\ln^2(1-(x_i + x_j)W - i \eta)}{(W+i \eta)}
\, ,
\nonumber \\[0.2cm]
g_{12}(x_i,x_j,W)&=& 
\frac{\ln^2(1-(x_i + x_j)W - i \eta)}{(W+i \eta)^2}
\, .
\label{eq:g-functions}
\end{eqnarray}}
Note that we have kept $i \eta$ terms ($\eta > 0$ and $\eta \ll $)
coming from the Feynman diagram
calculation (quark and gluon propagators), which will enable the correct
determination of imaginary parts necessary for LCSR in Sec. \ref{sec:LCSR}.

\subsection{Complete list of results}

In previously introduced notation the LO contributions to
$T_{{\cal B},V_1,M_N=0}$ and $T_{{\cal B},A_1,M_N=0}$ read:
\begin{eqnarray}
T_{{\cal B},V_1,M_N=0}^{{\rm LO},e_d}(\{x_k\},W)
&=& -2 \, e_d \; g_{0}(x_3,W)
\, ,
\nonumber \\[0.3cm]
T_{{\cal B},A_1,M_N=0}^{{\rm LO},e_d}(\{x_k\},W)
&=& 0
\, ,
\label{eq:TBLOed}
\end{eqnarray}
and
\begin{eqnarray}
T_{{\cal B},V_1,M_N=0}^{{\rm LO},e_u}(\{x_k\},W)
&=& e_u \; \left[ g_{0}(x_1,W) + g_{0}(x_2,W) \right]
\, ,
\nonumber \\[0.3cm]
T_{{\cal B},A_1,M_N=0}^{{\rm LO},e_u}(\{x_k\},W)
&=& e_u \; \left[ - g_{0}(x_1,W) + g_{0}(x_2,W) \right]
\, .
\label{eq:TBLOeu}
\end{eqnarray}
Furthermore, the NLO contributions proportional to
$\ln(\mu_{R,1}^2/Q^2)$ take also the simple form% 
\footnote{They are as expected proportional to LO and Ioffe current 
renormalization factor $2$ -- as renormalization
of UV divergences demanded (see Sec.\ref{sec:firstcaseMN0}).}
\begin{eqnarray}
T_{{\cal B},V_1,M_N=0}^{{\rm NLO,UV},e_d}(\{x_k\},W)
&=& -4 \, e_d \; g_{0}(x_3,W)
\, ,
\nonumber \\[0.3cm]
T_{{\cal B},A_1,M_N=0}^{{\rm NLO,UV},e_d}(\{x_k\},W)
&=& 0
\, ,
\end{eqnarray}
and
\begin{eqnarray}
T_{{\cal B},V_1,M_N=0}^{{\rm NLO,UV},e_u}(\{x_k\},W)
&=& 2 e_u \; \left[ g_{0}(x_1,W) + g_{0}(x_2,W) \right]
\, ,
\nonumber \\[0.3cm]
T_{{\cal B},A_1,M_N=0}^{{\rm NLO,UV},e_u}(\{x_k\},W)
&=& 2 e_u \; \left[- g_{0}(x_1,W) + g_{0}(x_2,W) \right]
\, .
\end{eqnarray}
Next we turn to NLO contributions proportional to
$\ln(\mu_{F}^2/Q^2)$ which originate from the factorization
of collinear divergences% 
\footnote{For the details we again refer to
Sec.\ref{sec:firstcaseMN0}.}
and read
{\allowdisplaybreaks\begin{eqnarray}
T_{{\cal B},V_1,M_N=0}^{{\rm NLO,IR},e_d}(\{x_k\},W)
&=& \frac{4}{3} \, e_d \; 
\Big\{ 
6 \; g_{0}(x_3,W)
+8 \; g_{1}(x_3,W)
 \nonumber \\[0.2cm] & & \left.
- \frac{4 x_1 x_2 -x_1 x_3 -x_2 x_3}{x_1 x_2 x_3} \; g_{3}(x_3,W)
+ \frac{x_1 +x_2}{x_1 x_2 x_3} \; g_{4}(x_3,W)
\right. \nonumber \\[0.2cm] & & \left.
-2 \; g_{7}(x_3,x_1,W)
-2 \; g_{7}(x_3,x_2,W)
\right. \nonumber \\[0.2cm] & & \left.
-\frac{1}{x_1} \; g_{9}(x_3,x_1,W)
-\frac{1}{x_2} \; g_{9}(x_3,x_2,W)
\right. \nonumber \\[0.2cm] & & \left.
-\frac{1}{x_1 (x_1+x_3)} \; g_{10}(x_3,x_1,W)
-\frac{1}{x_2 (x_2+x_3)} \; g_{10}(x_3,x_2,W)
\right\} 
\, , \quad
\nonumber \\[0.5cm]
T_{{\cal B},A_1,M_N=0}^{{\rm NLO,IR},e_d}(\{x_k\},W)
&=& \frac{4}{3} \, e_d \; 
\Big\{ 
 \frac{x_1 -x_2}{x_1 x_2} \; g_{3}(x_3,W)
- \frac{x_1 -x_2}{x_1 x_2 x_3} \; g_{4}(x_3,W)
 \nonumber \\[0.2cm] & & \left.
+\frac{1}{x_1} \; g_{9}(x_3,x_1,W)
-\frac{1}{x_2} \; g_{9}(x_3,x_2,W)
\right. \nonumber \\[0.2cm] & & \left.
-\frac{1}{x_1 (x_1+x_3)} \; g_{10}(x_3,x_1,W)
+\frac{1}{x_2 (x_2+x_3)} \; g_{10}(x_3,x_2,W)
\right\} 
\, , \quad
\nonumber \\
\end{eqnarray}}
and
{\allowdisplaybreaks\begin{eqnarray}
T_{{\cal B},V_1,M_N=0}^{{\rm NLO,IR},e_u}(\{x_k\},W)
&=& - \frac{4}{3} \, e_u \; 
\Big\{ 
3 \; g_{0}(x_1,W)
+3 \; g_{0}(x_2,W)
\nonumber \\[0.2cm] & & 
+4 \; g_{1}(x_1,W)
+4 \; g_{1}(x_2,W)
 \nonumber \\[0.2cm] & & \left.
+ \frac{x_1 -2 x_3}{x_1 x_3} \; g_{3}(x_1,W)
+ \frac{x_2 -2 x_3}{x_2 x_3} \; g_{3}(x_2,W)
\right. \nonumber \\[0.2cm] & & \left.
+ \frac{1}{x_1 x_2} \; g_{4}(x_1,W)
+ \frac{1}{x_1 x_2} \; g_{4}(x_2,W)
\right. \nonumber \\[0.2cm] & & \left.
-\; g_{7}(x_1,x_2,W)
-\; g_{7}(x_2,x_1,W)
\right. \nonumber \\[0.2cm] & & \left.
-\; g_{7}(x_1,x_3,W)
-\; g_{7}(x_2,x_3,W)
\right. \nonumber \\[0.2cm] & & \left.
-\frac{1}{x_3} \; g_{9}(x_1,x_3,W)
-\frac{1}{x_3} \; g_{9}(x_2,x_3,W)
\right. \nonumber \\[0.2cm] & & \left.
-\frac{1}{x_1 x_2} \; g_{10}(x_1,x_2,W)
\right\} 
\, , \quad
\nonumber \\[0.5cm]
T_{{\cal B},A_1,M_N=0}^{{\rm NLO,IR},e_u}(\{x_k\},W)
&=& - \frac{4}{3} \, e_u \; 
\Big\{ 
-3 \; g_{0}(x_1,W)
+3 \; g_{0}(x_2,W)
\nonumber \\[0.2cm] & & 
-4 \; g_{1}(x_1,W)
+4 \; g_{1}(x_2,W)
 \nonumber \\[0.2cm] & & \left.
- \frac{x_1 -2 x_3}{x_1 x_3} \; g_{3}(x_1,W)
+ \frac{x_2 -2 x_3}{x_2 x_3} \; g_{3}(x_2,W)
\right. \nonumber \\[0.2cm] & & \left.
- \frac{1}{x_1 x_2} \; g_{4}(x_1,W)
+ \frac{1}{x_1 x_2} \; g_{4}(x_2,W)
\right. \nonumber \\[0.2cm] & & \left.
+\; g_{7}(x_1,x_2,W)
-\; g_{7}(x_2,x_1,W)
\right. \nonumber \\[0.2cm] & & \left.
+\; g_{7}(x_1,x_3,W)
-\; g_{7}(x_2,x_3,W)
\right. \nonumber \\[0.2cm] & & \left.
+\frac{1}{x_3} \; g_{9}(x_1,x_3,W)
-\frac{1}{x_3} \; g_{9}(x_2,x_3,W)
\right. \nonumber \\[0.2cm] & & \left.
+\frac{x_1-x_2}{x_1 x_2(x_1+x_2)} \; g_{10}(x_1,x_2,W)
\right\} 
\, . \quad
\end{eqnarray}}
Finally we give the lengthy expressions for ''finite''
NLO contributions
{\allowdisplaybreaks\begin{subequations}
\begin{eqnarray}
T_{{\cal B},V_1,M_N=0}^{{\rm NLO,fin},e_d}(\{x_k\},W)
&=& \frac{2}{3} e_d 
\Big\{ 
18 \; g_0(x_3,W)
+12 \; g_1(x_3,W)
-8 \; g_2(x_3,W)
\nonumber \\[0.2cm] &&
-\frac{12}{x_3} \; g_3(x_3,W)
+\frac{5(x_1+x_2)}{x_1 x_2 x_3}\; g_4(x_3,W)
\nonumber \\[0.2cm] &&
+\frac{4 x_1 x_2 - x_2 x_3 -x_1 x_3}{x_1 x_2 x_3} \; g_5(x_3, W) 
-\frac{x_1 + x_2}{x_1 x_2 x_3} \; g_6(x_3, W)
\nonumber \\[0.2cm] &&
-3 \; g_7(x_3,x_1,W) 
-3 \; g_7(x_3,x_2,W) 
\nonumber \\[0.2cm] &&
+ 2 \; g_8(x_3,x_1,W) 
+ 2 \; g_8(x_3,x_2,W) 
\nonumber \\[0.2cm] &&
- \frac{5}{x_1 (x_1 + x_3)} \; g_{10}(x_1,x_3,W)
- \frac{5}{x_2 (x_2 + x_3)} \; g_{10}(x_2,x_3,W)
\nonumber \\[0.2cm] &&
+ \frac{1}{x_1} \; g_{11}(x_1,x_3,W)
+ \frac{1}{x_2} \; g_{11}(x_2,x_3,W)
\nonumber \\[0.2cm] &&
\left.
+\frac{1}{x_1(x_1+x_3)} \; g_{12}(x_1,x_3,W)
+\frac{1}{x_2(x_2+x_3)} \; g_{12}(x_2,x_3,W)
\right\}
\, ,
\nonumber \\[0.5cm] & &
\\
T_{{\cal B},A_1,M_N=0}^{{\rm NLO,fin},e_d}(\{x_k\},W)
&=& \frac{2}{3} e_d 
\Big\{ 
-\frac{5(x_1-x_2)}{x_1 x_2 x_3}\; g_4(x_3,W)
\nonumber \\[0.2cm] &&
+\frac{x_2 - x_1}{x_1 x_2} \; g_5(x_3, W) 
+\frac{x_1 - x_2}{x_1 x_2 x_3} \; g_6(x_3, W)
\nonumber \\[0.2cm] &&
-3 \; g_7(x_3,x_1,W) 
+3 \; g_7(x_3,x_2,W) 
\nonumber \\[0.2cm] &&
- \frac{5}{x_1 (x_1 + x_3)} \; g_{10}(x_1,x_3,W)
+ \frac{5}{x_2 (x_2 + x_3)} \; g_{10}(x_2,x_3,W)
\nonumber \\[0.2cm] &&
- \frac{1}{x_1} \; g_{11}(x_1,x_3,W)
+ \frac{1}{x_2} \; g_{11}(x_2,x_3,W)
\nonumber \\[0.2cm] &&
\left.
+\frac{1}{x_1(x_1+x_3)} \; g_{12}(x_1,x_3,W)
-\frac{1}{x_2(x_2+x_3)} \; g_{12}(x_2,x_3,W)
\right\}
\, .
\nonumber \\
\end{eqnarray}
\end{subequations}}
and
{\allowdisplaybreaks\begin{subequations}
\begin{eqnarray}
\lefteqn{T_{{\cal B},V_1,M_N=0}^{{\rm NLO,fin},e_u}(\{x_k\},W)}
\nonumber \\[0.2cm]
&=& \frac{2}{3} e_u 
\left\{ 
-\frac{19}{2} \; g_0(x_1,W)
-\frac{19}{2} \; g_0(x_2,W)
\right. \nonumber \\[0.2cm] &&
-6 \; g_1(x_1,W)
-6 \; g_1(x_2,W)
+4 \; g_2(x_1,W)
+4 \; g_2(x_2,W)
\nonumber \\[0.2cm] &&
+\frac{2 x_1 x_2 + 6 x_2 x_3 -3 x_1 x_3}{x_1 x_2 x_3} \; g_3(x_1,W)
+\frac{2 x_1 x_2 + 6 x_1 x_3 -3 x_2 x_3}{x_1 x_2 x_3} \; g_3(x_2,W)
\nonumber \\[0.2cm] &&
-\frac{4}{x_1 x_2}\; g_4(x_1,W)
-\frac{4}{x_1 x_2}\; g_4(x_2,W)
\nonumber \\[0.2cm] &&
+\frac{x_1 - 2 x_3}{x_1 x_3} \; g_5(x_1, W) 
+\frac{x_2 - 2 x_3}{x_2 x_3} \; g_5(x_2, W) 
\nonumber \\[0.2cm] &&
+\frac{1}{x_1 x_2} \; g_6(x_1, W)
+\frac{1}{x_1 x_2} \; g_6(x_2, W)
\nonumber \\[0.2cm] &&
+3 \; g_7(x_1,x_2,W) 
+3 \; g_7(x_2,x_1,W) 
\nonumber \\[0.2cm] &&
- \; g_8(x_1,x_2,W) 
- \; g_8(x_2,x_1,W) 
- \; g_8(x_1,x_3,W) 
- \; g_8(x_2,x_3,W) 
\nonumber \\[0.2cm] &&
+ \frac{3(x_1+x_2)}{x_1 x_2} \; g_9(x_1,x_2,W) 
- \frac{2}{x_3} \; g_9(x_1,x_3,W) 
- \frac{2}{x_3} \; g_9(x_2,x_3,W) 
\nonumber \\[0.2cm] &&
+ \frac{4}{x_1 x_2} \; g_{10}(x_1,x_2,W)
\nonumber \\[0.2cm] &&
- \frac{1}{x_3} \; g_{11}(x_1,x_3,W)
- \frac{1}{x_3} \; g_{11}(x_2,x_3,W)
\nonumber \\[0.2cm] &&
\left.
-\frac{1}{x_1 x_2} \; g_{12}(x_1,x_2,W)
\right\}
\, ,
\end{eqnarray}
\begin{eqnarray}
\lefteqn{T_{{\cal B},A_1,M_N=0}^{{\rm NLO,fin},e_u}(\{x_k\},W)}
\nonumber \\[0.2cm]
&=& \frac{2}{3} e_u 
\left\{ 
\frac{23}{2} \; g_0(x_1,W)
-\frac{23}{2} \; g_0(x_2,W)
\right. \nonumber \\[0.2cm] &&
+6 \; g_1(x_1,W)
-6 \; g_1(x_2,W)
-4 \; g_2(x_1,W)
+4 \; g_2(x_2,W)
\nonumber \\[0.2cm] &&
+\frac{4 x_1 x_2 - 6 x_2 x_3 -7 x_1 x_3}{x_1 x_2 x_3} \; g_3(x_1,W)
+\frac{-4 x_1 x_2 + 6 x_1 x_3 +7 x_2 x_3}{x_1 x_2 x_3} \; g_3(x_2,W)
\nonumber \\[0.2cm] &&
-\frac{2 (x_2-5 x_3)}{x_1 x_2 x_3}\; g_4(x_1,W)
+\frac{2 (x_1-5 x_3)}{x_1 x_2 x_3}\; g_4(x_2,W)
\nonumber \\[0.2cm] &&
+\frac{2 x_3 - x_1}{x_1 x_3} \; g_5(x_1, W) 
+\frac{x_2-2 x_3}{x_2 x_3} \; g_5(x_2, W) 
\nonumber \\[0.2cm] &&
-\frac{1}{x_1 x_2} \; g_6(x_1, W)
+\frac{1}{x_1 x_2} \; g_6(x_2, W)
\nonumber \\[0.2cm] &&
-3 \; g_7(x_1,x_2,W) 
+3 \; g_7(x_2,x_1,W) 
\nonumber \\[0.2cm] &&
+ \; g_8(x_1,x_2,W) 
- \; g_8(x_2,x_1,W) 
+ \; g_8(x_1,x_3,W) 
- \; g_8(x_2,x_3,W) 
\nonumber \\[0.2cm] &&
+ \frac{7(x_1-x_2)}{x_1 x_2} \; g_9(x_1,x_2,W) 
- \frac{4}{x_3} \; g_9(x_1,x_3,W) 
+ \frac{4}{x_3} \; g_9(x_2,x_3,W) 
\nonumber \\[0.2cm] &&
- \frac{10(x_1-x_2)}{x_1 x_2(x_1+x_2)} \; g_{10}(x_1,x_2,W)
\nonumber \\[0.2cm] &&
+ \frac{2}{x_3(x_1+x_3)} \; g_{10}(x_1,x_3,W)
- \frac{2}{x_3(x_2+x_3)} \; g_{10}(x_2,x_3,W)
\nonumber \\[0.2cm] &&
+ \frac{1}{x_3} \; g_{11}(x_1,x_3,W)
- \frac{1}{x_3} \; g_{11}(x_2,x_3,W)
\nonumber \\[0.2cm] &&
\left.
+\frac{x_1-x_2}{x_1 x_2(x_1+x_2)} \; g_{12}(x_1,x_2,W)
\right\}
\, .
\end{eqnarray}
\end{subequations}}

%\newpage
%
The contributions to
$T_{{\cal B},\Phi,M_N=0}$
which, as can be seen, due to symmetry properties
of contributions 
$T_{{\cal B},V_1,M_N=0}$
and
$T_{{\cal B},A_1,M_N=0}$%
\footnote{The former is symmetric under $x_1 \leftrightarrow x_2$ exchange 
and the latter antisymmetric, analogous to symmetry properties of 
$V_1$ and $A_1$.},
correspond exactly to
\begin{equation}
T_{{\cal B},\Phi,M_N=0}
= T_{{\cal B},V_1,M_N=0} - T_{{\cal B},A_1,M_N=0}
\, .
\end{equation}

\subsection{Summary for proton case}

Finally in this section we list the expressions that we actually use in
our numerical calculations of the proton form factors
(see Eqs. (\ref{eq:convolB-Phi}-\ref{eq:g-functions}) 
for corresponding definitions).
We have already made use of $e_u=2/3$ and $e_d=-1/3$.

For $M_N=0$ twist-3 LO contribution reads:
\begin{eqnarray}
T_{{\cal B},\Phi,M_N=0}^{{\rm LO}}(\{x_k\},W) &=&
\frac{4}{3} \, g_0(x_1,W)+\frac{2}{3} \, g_0(x_3,W)
\, .
\label{eq:finresPhiLO}
\end{eqnarray}

NLO contribution proportional to $\ln(\mu_{R,1}^2/Q^2)$ is given by:
\begin{eqnarray}
T_{{\cal B},\Phi,M_N=0}^{{\rm NLO,UV}}(\{x_k\},W)
&=& 4 \, \left[ \frac{2}{3} \, g_0(x_1,W)+\frac{1}{3} \, g_0(x_3,W) \right]
\, .
\end{eqnarray}

NLO contribution proportional to $\ln(\mu_{F}^2/Q^2)$ reads:
{\allowdisplaybreaks\begin{eqnarray}
\lefteqn{T_{{\cal B},\Phi,M_N=0}^{{\rm NLO,IR}}(\{x_k\},W)} \nonumber\\
 &\qquad=&
4\Big[
-\frac{4}{3} \, g_0(x_1,W)-\frac{2}{3}\, g_0(x_3,W)
-\frac{16}{9} \, g_1(x_1,W)-\frac{8}{9}\, g_1(x_3,W)
\nonumber\\[0.2cm]
&&
+\frac{4(2 x_3-x_1)}{9 x_1 x_3} \, g_3(x_1,W)
+\frac{2(2 x_1-x_3)}{9 x_1 x_3} \, g_3(x_3,W)
\nonumber\\[0.2cm]
&&
-\frac{4}{9 x_1 x_2}\, g_4(x_1,W)
-\frac{2}{9 x_2 x_3}\, g_4(x_3,W)
\nonumber\\[0.2cm]
&&
+\frac{4}{9}\, g_7(x_1,x_2,W)
+\frac{4}{9}\,  g_7(x_1,x_3,W)
+\frac{2}{9}\, g_7(x_3,x_1,W)
+\frac{2}{9}\, g_7(x_3,x_2,W)
\nonumber\\[0.2cm]
&&
+\frac{2 (2 x_1+x_3)}{9 x_1 x_3}\, g_9(x_1,x_3,W)
\nonumber\\[0.2cm]
&&
+\frac{4}{9 x_2(x_1+x_2)}\, g_{10}(x_1,x_2,W)
+\frac{2}{9 x_2(x_2+x_3)}\, g_{10}(x_2,x_3,W)
\Big ]
\, . 
\end{eqnarray}}

''Finite'' NLO contribution reads:
{\allowdisplaybreaks\begin{eqnarray}
\lefteqn{T_{{\cal B},\Phi,M_N=0}^{{\rm NLO,fin}}(\{x_k\},W)}
\nonumber\\[0.2cm]
&\quad =& 
4 \Big[
-\frac{7}{3} \, g_{0}(x_1,W)
+\frac{2}{9} \, g_{0}(x_2,W)
-g_{0}(x_3,W)
\nonumber\\[0.2cm]
&& 
-\frac{4} {3} \, g_1(x_1,W)
-\frac{2}{3} \, g_1(x_3,W)
+\frac{8}{9} \, g_2(x_1,W)
+\frac{4}{9} \, g_2(x_3,W)
\nonumber\\[0.2cm]
&& 
+\frac{2}{9} \left( \frac{6} {x_1}+\frac{2}{x_2} -\frac{1}{x_3}\right) 
  g_3(x_1,W)
+\frac{2}{9} \left(\frac{3}{x_3}-\frac{5}{x_1}\right) g_3(x_2,W)
+\frac{2}{3 x_3} \, g_3(x_3,W)
\nonumber\\[0.2cm]
&& 
+\frac{2 (x_2-7 x_3)}{9 x_1 x_2 x_3} g_4(x_1,W)
-\frac{2 (x_1-3 x_3)}{9 x_1 x_2 x_3} g_4(x_2,W)
-\frac{5}{9 x_2 x_3} g_4(x_3,W)
\nonumber\\[0.2cm]
&& 
+\frac{2 (x_1-2 x_3)}{9 x_1 x_3} g_5(x_1,W)
+\frac{(x_3-2 x_1)}{9 x_1 x_3} g_5(x_3,W)
\nonumber\\[0.2cm]
&& 
+\frac{2}{9 x_1 x_2} g_6(x_1,W)
+\frac{1}{9 x_2 x_3} g_6(x_3,W)
+\frac{2}{3} \, g_7(x_1,x_2,W)
+\frac{1}{3} \, g_7(x_3,x_2,W)
\nonumber\\[0.2cm]
&& 
-\frac{2}{9} \, g_8(x_1,x_2,W)
-\frac{2}{9} \, g_8(x_1,x_3,W)
-\frac{1}{9} \, g_8(x_3,x_1,W)
-\frac{1}{9} \, g_8(x_3,x_2,W)
\nonumber\\[0.2cm]
&& 
+\frac{2}{9} \left(\frac{5}{x_1} -\frac{2}{x_2}\right) g_9(x_1,x_2,W)
+\frac{2}{9 x_3} g_9(x_1,x_3,W)
-\frac{2}{3 x_3} g_9(x_2,x_3,W)
\nonumber\\[0.2cm]
&& 
+\frac{(14 x_1-6 x_2)}{9 x_1 x_2 (x_1+x_2)} g_{10}(x_1,x_2,W)
-\frac{2}{9 x_3 (x_1+x_3)} g_{10}(x_1,x_3,W)
\nonumber\\[0.2cm]
&& 
+\frac{(2 x_2+5 x_3)}{9 x_2 x_3 (x_2+x_3)} g_{10}(x_2,x_3,W)
-\frac{(2 x_1+x_3)}{9 x_1 x_3} g_{11}(x_1,x_3,W)
\nonumber\\[0.2cm]
&& 
-\frac{2}{9 x_2 (x_1+x_2)} g_{12}(x_1,x_2,W)
-\frac{1}{9 x_2 (x_2+x_3)} g_{12}(x_2,x_3,W)
\Big ]
\, .
\label{eq:finresPhiNLO}
\end{eqnarray}}

%%%%%%%%%%%%%%%%%%%%%%%%%%%%%%%%%%%%%%%%%%%%%%%%%%%%%%%%%%%%%%%%%%%%%%%%%%%%%%%
\section{Imaginary parts of selected functions}
\setcounter{equation}{0}
%%%%%%%%%%%%%%%%%%%%%%%%%%%%%%%%%%%%%%%%%%%%%%%%%%%%%%%%%%%%%%%%%%%%%%%%%%%%%%%
\label{sec:appIm}

In this section we list the imaginary parts of selected functions
that appear in our calculation.

We start with the well-known result 
for the logarithmic function%
\begin{equation}
\ln(x-x_0 \pm i \eta)=\ln(|x-x_0|)\pm i \pi ~ \Theta(x_0-x)
\, .
\label{eq:ln}
\end{equation}
Here $x$, $x_0$, and $\eta$ are real, and $\eta >0, \eta \ll$. 
All other results that we list can be derived from \req{eq:ln}.

It follows trivially that
\begin{equation}
\ln^2(x-x_0\pm i\eta)=[\ln^2(|x-x_0|)- \pi^2 \Theta(x_0-x)] 
               \pm 2 i \pi \ln(|x-x_0|)\Theta(x_0-x)
\, ,
\label{eq:ln2}
\end{equation}
and higher exponents $\ln^n(x-x_0\pm i\eta)$, $n>2$ can be obtained similarly.

The other well-known result
\begin{eqnarray}
\frac{1}{x-x_0 \pm i \eta}&=&
   {\cal P} ~ \frac{1}{x-x_0} \mp i \pi \delta(x-x_0)
  \nonumber\\
 &=&\frac{1}{x-x_0}[\Theta(x-x_0)+\Theta(x_0-x)]\mp i \pi \delta(x-x_0)
\end{eqnarray}
can be obtained by taking a derivative of Eq. \req{eq:ln} with respect to $x$:
\begin{displaymath}
\frac{1}{x-x_0 \pm i \eta}= \frac{d}{d x}\ln(x-x_0 \pm i \eta)
\, .
\end{displaymath}
Furthermore, since
\begin{displaymath}
\frac{1}{(x-x_0 \pm i \eta)^n}= 
\frac{(-1)^{n-1}}{(n-1)!} \;  
\frac{d^{n-1}}{d x^{n-1}}\frac{1}{x-x_0 \pm i \eta}
\, ,
\end{displaymath}
it is easy to see that for $n \ge 2$
\begin{equation}
 \frac{1}{(x-x_0 \pm i \eta)^n}= 
\frac{1}{(x-x_0)^{n}}[\Theta(x-x_0)+\Theta(x_0-x)]
\mp i \pi
\frac{(-1)^{n-1}}{(n-1)!} 
\;  \delta^{(n-1)}(x-x_0)
\, .
\end{equation}

Finally, by making use of
\begin{displaymath}
\frac{\ln(X \pm i \eta)}{X \pm i \eta}= 
\frac{1}{2}
\frac{d}{d X}\ln^2(X \pm i \eta)
\qquad
\frac{\ln^2(X \pm i \eta)}{X \pm i \eta}= 
\frac{1}{3}
\frac{d}{d X}\ln^3(X \pm i \eta)
\, ,
\end{displaymath}
where $X=x-x_0$ or $X=x_0-x$,
we get
\begin{subequations}
\label{eq:ImlnD}
\begin{eqnarray}
 \frac{1}{\pi} ~ \mbox{Im}~\frac{\ln(x-x_0 \pm i \eta)}{x-x_0 \pm i \eta}
&=& 
\pm \left[ \frac{\Theta(x_0-x)}{x-x_0} 
- \delta(x-x_0) \ln(x_0-x)\right] 
\nonumber\\[0.2cm]
&=&
\pm \left[\left\lbrace \frac{\Theta(x_0-x)}{x-x_0}\right\rbrace _+ 
- \delta(x-x_0) \ln(|a-x_0|)\right] 
\, ,
\\[0.35cm]
 \frac{1}{\pi} ~ \mbox{Im}~\frac{\ln(x_0-x \pm i \eta)}{x_0-x \pm i \eta}
&=& 
\pm \left[ \frac{\Theta(x-x_0)}{x_0-x} 
- \delta(x-x_0) \ln(x-x_0)\right] 
\nonumber\\[0.2cm]
&=&
\pm \left[\left\lbrace \frac{\Theta(x-x_0)}{x_0-x}\right\rbrace _+ 
- \delta(x-x_0) \ln(|x_0-b|)\right] 
\, ,
\end{eqnarray}
\end{subequations}
and
\begin{subequations}
\label{eq:Imln2D}
\begin{eqnarray}
\lefteqn{
\frac{1}{\pi} ~ \mbox{Im}~ \frac{\ln^2(x-x_0 \pm i \eta)}{x-x_0 \pm i \eta}
} \nonumber \\
&=&\pm \left[ \Theta(x_0-x) \frac{2 \ln(x_0-x)}{x-x_0} 
+\delta(x-x_0)\left[ \frac{\pi^2}{3}-\ln^2(x_0-x)\right]\right] 
\nonumber\\
&=&\pm \left[\left\lbrace \Theta(x_0-x) \frac{2 \ln(x_0-x)}{x-x_0}
\right\rbrace _+ 
+\delta(x-x_0)\left[ \frac{\pi^2}{3}-\ln^2(|x_0-a|)\right]\right] 
\, ,
\end{eqnarray}
\begin{eqnarray}
\lefteqn{
\frac{1}{\pi} ~ \mbox{Im}~ \frac{\ln^2(x_0-x \pm i \eta)}{x_0-x \pm i \eta}
} \nonumber \\
&=&\pm \left[ \Theta(x-x_0) \frac{2 \ln(x-x_0)}{x_0-x} 
+\delta(x-x_0)\left[ \frac{\pi^2}{3}-\ln^2(x-x_0)\right]\right] 
\nonumber\\
&=&\pm \left[\left\lbrace \Theta(x-x_0) \frac{2 \ln(x-x_0)}{x_0-x}
\right\rbrace _+ 
+\delta(x-x_0)\left[ \frac{\pi^2}{3}-\ln^2(|x_0-b|)\right]\right] 
\, .
\end{eqnarray}
\end{subequations}
Note that right hand-side of the
first lines in \req{eq:ImlnD} and \req{eq:Imln2D}
consists of two terms which separately ''blow up'' for $x\to x_0$.
The sums are finite and in the following lines, 
using the usual $\{\}_+$ prescription
\begin{equation}
\left\lbrace F(x,x_0)\right\rbrace _+ = 
F(x,x_0) -\delta(x-x_0) \int_a^b dz F(z,x_0)
\, , 
\end{equation}
we express them as manifestly finite sum of two terms. 
It is easy to see that
\begin{eqnarray}
\left\lbrace \frac{\Theta(x_0-x)}{x-x_0}\right\rbrace_+ &=& 
\frac{\Theta(x_0-x)}{x-x_0}-\delta(x_0-x)\ln(x_0-x_0)
+\delta(x-x_0)\ln(|a-x_0|)
\, , \nonumber \\[0.2cm]
\left\lbrace \frac{\Theta(x-x_0)}{x_0-x}\right\rbrace_+ &=& 
\frac{\Theta(x-x_0)}{x_0-x}-\delta(x-x_0)\ln(x_0-x_0)
+\delta(x-x_0)\ln(|x_0-b|)
\, , \nonumber \\[0.2cm]
\end{eqnarray}
and similarly for
$\left\lbrace \Theta(x-x_0) \frac{2 \ln(x-x_0)}{x_0-x} \right\rbrace _+$ 
and 
$\left\lbrace \Theta(x_0-x) \frac{2 \ln(x_0-x)}{x-x_0} \right\rbrace _+$. 
Note that in this section we take $x$ as an integration variable of
the next step of the calculation and the variable with respect to
which the $\{\}_+$ prescription has been defined%
\footnote{Some similar results but not for general $\{\}_+$ prescription
($a=0$) can be found in App. A of Ref. \cite{SchmeddingY99}. 
Note the typo in Ref. \cite{SchmeddingY99}: $\Theta$-function is not written 
inside the $\{\}_+$-prescription.}.

%%%%%%%%%%%%%%%%%%%%%%%%%%%%%%%%%%%%%%%%%%%%%%%%%%%%%%%%%%%%%%%%%%%%%%%%%%%%%%%
\section{LO results and LCSRs for $M_N \ne 0$ case} 
\setcounter{equation}{0}
%%%%%%%%%%%%%%%%%%%%%%%%%%%%%%%%%%%%%%%%%%%%%%%%%%%%%%%%%%%%%%%%%%%%%%%%%%%%%%%
\label{sec:LOMN}

In this section we present 
LO twist-$3$ and twist-$4$ contributions  
to $\cal{A}$ and $\cal{B}$ functions
\req{eq:projections}
based on results listed in
Tables \ref{t:coeffLO1}, \ref{t:coeffLO3} and \ref{t:coeffLO2}
($C_{P_\mu M_N}^{{\rm LO},F^{(i)}}$ and 
$C_{P_\mu  \not q}^{{\rm LO},F^{(i)}}$
 coefficients, respectively)
which were calculated for $M_N\neq 0$.
Furthermore, using \req{eq:LCSRb} we determine the LCSR contributions
to form factors $F_1(Q^2)$ and $F_2(Q^2)$.

Analogously to Eq. \req{eq:g-functions}, we define
\begin{eqnarray}
g_0(x_i, W, M_N^2/Q^2)&=&
\frac{1}{(x_i W -1 -\frac{M_N^2}{Q^2} x_i (1-x_i) + i \eta) } 
\, .
\label{eq:g0gen}
\end{eqnarray}
Note that
$g_0(x_i, W)$ introduced in \req{eq:g-functions}
corresponds to $g_0(x_i, W, 0)$, i.e, in other words, in 
\req{eq:g0gen} we introduce the generalization to
$M_N^2 \ne 0$.
In this calculation we will also encounter
\begin{eqnarray}
g_0^2(x_i, W, M_N^2/Q^2)&=&
\frac{1}{(x_i W -1 -\frac{M_N^2}{Q^2} x_i (1-x_i) + i \eta)^2 } 
\, .
\end{eqnarray}

The twist-3 contributions 
and the twist-4 contributions 
corresponding to nucleon DAs $V_3$ and $A_3$
can be expressed in a form of a convolution:
\begin{eqnarray}
{\cal A}^{\rm LO}(Q^2,W, M_N^2; \mu_F^2)= 
\frac{1}{Q^2} \,
\sum_i \, T^{{\rm LO}}_{{\cal A},F^{(i)}}(\{x_k\},W, M_N^2/Q^2)\otimes
                          F^{(i)}(\{x_k\}; \mu_F^2)
\, ,
\nonumber\\
{\cal B}^{\rm LO}(Q^2,W, M_N^2; \mu_F^2)= 
\frac{1}{Q^2} \,
\sum_i \, T^{{\rm LO}}_{{\cal B},F^{(i)}}(\{x_k\},W, M_N^2/Q^2)\otimes
                          F^{(i)}(\{x_k\}; \mu_F^2)
\, ,
\label{eq:ABF1F3}
\end{eqnarray}
where $F^{(i)} \in \{ V_1, A_1, V_3, A_3 \}$.

The coefficients $C_{P_\mu M_N}^{{\rm LO},F^{(i)}}$
from Table \ref{t:coeffLO1}  
determine the ''hard-scattering'' 
twist-3 contributions to ${\cal A}$ function
\begin{eqnarray}
T_{{\cal A},V_1}^{{\rm LO}}(\{x_k\},W, M_N^2/Q^2)
&=& - 2 e_u \left[ x_1 \, g_0(x_1, W, M_N^2/Q^2) 
+ x_2  \, g_0(x_2, W, M_N^2/Q^2)\right]
\, ,
\nonumber\\[0.3cm]
T_{{\cal A},A_1}^{{\rm LO}}(\{x_k\},W, M_N^2/Q^2)&=&0
\, ,
\label{eq:calAV1A1}
\end{eqnarray}
while
the coefficients $C_{P_\mu  \not q}^{{\rm LO},F^{(i)}}$
determine the ''hard-scattering'' 
twist-3 contributions to ${\cal B}$ function:
\begin{eqnarray}
T_{{\cal B},V_1}^{{\rm LO}}(\{x_k\},W, M_N^2/Q^2)&=&
e_u \left[   g_0(x_1, W, M_N^2/Q^2)+  g_0(x_2, W, M_N^2/Q^2)\right]
\nonumber\\[0.2cm]
&&
-2 e_d \, g_0(x_3, M_N^2/Q^2, W) 
\, ,
\nonumber\\[0.3cm]
T_{{\cal B},A_1}^{{\rm LO}}(\{x_k\},W, M_N^2/Q^2)
&=&
e_u \left[ -g_0(x_1, W, M_N^2/Q^2)+  g_0(x_2, W, M_N^2/Q^2) \right]
\, .
\label{eq:calBV1A1}
 \end{eqnarray}
Similarly from Table \ref{t:coeffLO3},
the ''hard-scattering'' twist-4 contributions
to ${\cal A}$ function
are given by
\begin{eqnarray}
T_{{\cal A},V_3}^{{\rm LO}}(\{x_k\},W, M_N^2/Q^2)&=&
3 e_u \left[  x_1 \, g_0(x_1, W, M_N^2/Q^2)+x_2 \, g_0(x_2, W, M_N^2/Q^2)\right]
\nonumber\\[0.2cm]
&&
+2 e_d \, x_3 \, g_0(x_3, W, M_N^2/Q^2)
\, ,
\nonumber\\[0.3cm]
T_{{\cal A},A_3}^{{\rm LO}}(\{x_k\},W, M_N^2/Q^2)
&=&  e_u\left[-x_1 \, g_0(x_1, W, M_N^2/Q^2)+x_2 \, g_0(x_2, W, M_N^2/Q^2)\right]
\, , \quad
\label{eq:calAV3A3}
 \end{eqnarray}
while the contributions
to ${\cal B}$ function
read
\begin{eqnarray}
T_{{\cal B},V_3}^{{\rm LO}}(\{x_k\},W, M_N^2/Q^2)&=&0
\, ,
\nonumber\\[0.3cm]
T_{{\cal B},A_3}^{{\rm LO}}(\{x_k\},W, M_N^2/Q^2)&=&0
\,.
\label{eq:calBV3A3}
 \end{eqnarray}

The contributions listed in Table \ref{t:coeffLO2} 
contribute 
to functions ${\cal A}$ and ${\cal B}$ 
according to 
\begin{eqnarray}
{\cal A}^{\rm LO}(Q^2,W, M_N;\mu_F^2)= 
\frac{1}{Q^2} \,
\sum_i \sum_{k=1}^3
\int_{0}^{1} dx_k \, T^{{\rm LO}}_{{\cal A},F_{123}^{(i)}}(x_k,W,M_N^2/Q^2)
         \, \widetilde{F}_{123}^{(i)}(x_k)
\, ,
\nonumber\\
{\cal B}^{\rm LO}(Q^2,W, M_N;\mu_F^2)= 
\frac{1}{Q^2} \,
\sum_i \sum_{k=1}^3
\int_{0}^{1} dx_k \, T^{{\rm LO}}_{{\cal B},F_{123}^{(i)}}(x_k,W,M_N^2/Q^2)
         \, \widetilde{F}_{123}^{(i)}(x_k)
\, ,
\label{eq:ABF123}
\end{eqnarray}
where $\widetilde{F}_{123}^{(i)} \in \{ \widetilde{V}_{123}^{(i)},
\widetilde{A}_{123}^{(i)} \}$
-- see corresponding definitions (\ref{eq:wtilde}-\ref{tilde}).

The ''hard-scattering'' contributions to $\cal A$ then read
\begin{eqnarray}
T^{{\rm LO}}_{{\cal A},V_{123}}(x_1,W,M_N^2/Q^2)
&=&
e_u \left[ g_0^2(x_1,W,M_N^2/Q^2)-g_0(x_1,W,M_N^2/Q^2) \right]
\, ,
\nonumber \\[0.2cm]
T^{{\rm LO}}_{{\cal A},V_{123}}(x_2,W,M_N^2/Q^2)
&=&
e_u \left[ g_0^2(x_2,W,M_N^2/Q^2)-g_0(x_2,W,M_N^2/Q^2) \right]
\, ,
\nonumber \\[0.2cm]
T^{{\rm LO}}_{{\cal A},V_{123}}(x_3,W,M_N^2/Q^2)
&=&
2 e_d \left[ g_0^2(x_3,W,M_N^2/Q^2)+g_0(x_3,W,M_N^2/Q^2) \right]
\, , \quad
\label{eq:calAV123}
\end{eqnarray}
and
\begin{eqnarray}
T^{{\rm LO}}_{{\cal A},A_{123}}(x_1,W,M_N^2/Q^2)
&=&
 - e_u \left[ g_0^2(x_1,W,M_N^2/Q^2)+g_0(x_1,W,M_N^2/Q^2) \right]
\, , \quad
\nonumber \\[0.2cm]
T^{{\rm LO}}_{{\cal A},A_{123}}(x_2,W,M_N^2/Q^2)
&=&
  e_u \left[ g_0^2(x_2,W,M_N^2/Q^2)+g_0(x_2,W,M_N^2/Q^2) \right]
\, ,
\nonumber \\[0.2cm]
T^{{\rm LO}}_{{\cal A},A_{123}}(x_3,W,M_N^2/Q^2)
&=&
0
\, .
\label{eq:calAA123}
\end{eqnarray}
The ''hard-scattering'' contributions to $\cal B$ are given by
\begin{eqnarray}
T^{{\rm LO}}_{{\cal B},V_{123}}(x_1,W,M_N^2/Q^2)
&=&
e_u \frac{M_N^2}{Q^2} \, x_1 \, g_0^2(x_1,W,M_N^2/Q^2)
\, ,
\nonumber \\[0.2cm]
T^{{\rm LO}}_{{\cal B},V_{123}}(x_2,W,M_N^2/Q^2)
&=&
e_u \frac{M_N^2}{Q^2}\,  x_2 \, g_0^2(x_2,W,M_N^2/Q^2)
\, ,
\nonumber \\[0.2cm]
T^{{\rm LO}}_{{\cal B},V_{123}}(x_3,W,M_N^2/Q^2)
&=&
2 e_d \frac{M_N^2}{Q^2}\,  x_3 \, g_0^2(x_3,W,M_N^2/Q^2)
\, , \quad
\label{eq:calBV123}
\end{eqnarray}
and
\begin{eqnarray}
T^{{\rm LO}}_{{\cal B},A_{123}}(x_1,W,M_N^2/Q^2)
&=&
- e_u \frac{M_N^2}{Q^2}\,  x_1 \, g_0^2(x_1,W,M_N^2/Q^2)
\, , \quad
\nonumber \\[0.2cm]
T^{{\rm LO}}_{{\cal B},A_{123}}(x_2,W,M_N^2/Q^2)
&=&
 e_u \frac{M_N^2}{Q^2}\,  x_2 \, g_0^2(x_2,W,M_N^2/Q^2)
\, ,
\nonumber \\[0.2cm]
T^{{\rm LO}}_{{\cal B},A_{123}}(x_3,W,M_N^2/Q^2)
&=&
0
\, .
\label{eq:calBA123}
\end{eqnarray}

For $M_N = 0$ there are no contributions to ${\cal A}$ 
since in the decomposition of the correlation function  \req{eq:projections}
${\cal A}$ is multiplied by $M_N$.
Moreover, in the limit $M_N \to 0$ only the twist-3 contributions
to ${\cal B}$, given in \req{eq:calBV1A1}, ''survive'' 
and take the form (\ref{eq:TBLOed}-\ref{eq:TBLOeu}). 
By taking $M_N \ne 0$ but $M_N^2 = 0$
(corresponds to the first two terms in the expansion
\req{eq:Mexpansion}),
one is left with the same twist-3 contribution to
${\cal B}$ (\req{eq:calBV1A1} with $M_N^2=0$,
i.e., (\ref{eq:TBLOed}-\ref{eq:TBLOeu}))
and twist-3 and twist-4 contributions
to ${\cal A}$:
\req{eq:calAV1A1}, \req{eq:calAV3A3},
\req{eq:calAV123} and \req{eq:calAA123}
taken with $M_N^2=0$.
Note that in comparison to twist-3,
the twist-4 contribution to ${\cal B}$ are suppressed
by $M_N^2/Q^2$ factor 
(see \req{eq:calBV3A3}, \req{eq:calBV123}, \req{eq:calBA123}).
This is not the case for ${\cal A}$ 
(see \req{eq:calAV1A1}, \req{eq:calAV3A3}, \req{eq:calAV123}, 
\req{eq:calAA123}) where there is no such additional suppression
factor between twist-3 and twist-4 contributions.
For $M_N^2 \ne 0$ the higher twists also contribute
(twist-5, \ldots), which we do not consider here but just refer
to the results presented in, for example, Ref. \cite{BraunLW06}
(App. A).
From these results one can see that both twist-4 and twist-5
contributions to ${\cal B}$ are suppressed by $M_N^2/Q^2$
in comparison to twist-3.
For ${\cal A}$, twist-5 and twist-6 contributions
are suppressed by $M_N^2/Q^2$ in comparison
to twist-3 and twist-4 contributions (remember that there is an additional
factor $M_N$ infront of ${\cal A}$).

According to Eq. (\ref{eq:LCSRb}) and analogously to
(\ref{eq:substitution}), we can now formulate the 
''rules''
for separate terms contributing to
$T^{\rm LO}_{\cal A}$ and $T^{\rm LO}_{\cal B}$
leading to separate terms contributing to
$F_1$ and $F_2$, respectively.
The contributions (\ref{eq:calAV1A1} - \ref{eq:calBV3A3}) 
can be conveniently expressed as a sum of the terms of general form
\begin{equation}
g_0(\{x_k\},W,M_N^2/Q^2) \; \; f(\{x_k\})
\, ,
\end{equation}
and they then contribute as
\begin{eqnarray}
\left(\begin{array}{c}
 2 F_{1}^{{\rm LO}, \{V_1, A_1, V_3, A_3\}}(Q^2; \mu_F^2) \\[0.2cm] 
F_{2}^{{\rm LO},\{V_1, A_1, V_3, A_3\}}(Q^2; \mu_F^2) 
\end{array}
\right)
&:& 
\frac{1}{\lambda_1 \pi } \int {\cal D}x
\int_1^{(s_0+Q^2)/Q^2} dw \; e^{-(w-1) Q^2/M_B^2+M_N^2/M_B^2} \; 
\nonumber \\[0.2cm]&& \times \;
\; {\rm Im} \left[g_0(\{x_k\}, w, M_N^2/Q^2) \right] \, 
f(\{x_k\}) 
\nonumber \\[0.2cm]&& \times \;
\, F^{(i)}(\{x_k\}; \mu_F^2) 
\nonumber \\[0.5cm]&\to&
\frac{1}{\lambda_1} 
\int_{x_0}^{1}  dx_i \int_{0}^{1-x_i} dx_j \;  
e^{-(1-x_i)Q^2/(x_i M_B^2)+x_i M_N^2/M_B^2} \quad
\nonumber \\[0.2cm]&& \times \;
\left( - \frac{1}{x_i} \right) \, f(\{x_i,x_j,1-x_i-x_j\}) 
\nonumber \\[0.2cm]&& \times \;
\; \;   F^{(i)} (\{x_i,x_j,1-x_i-x_j\}; \mu_F^2)
\, .
\nonumber \\
\label{eq:subfortwist3withmasses}
\end{eqnarray}
Here $F^{(i)} \in \{ V_1, A_1, V_3, A_3 \}$ and 
\begin{eqnarray}
x_0=\frac{\sqrt{(Q^2+s_0 -M_N^2 )^2 +4 M_N^2 Q^2}-(Q^2+s_0-M_N^2) }{2 M_N^2} 
\,.
\end{eqnarray}
Note that $\lim_{M_N^2 \to 0} x_0= Q^2/(Q^2+s_0)$, i.e., 
one recovers the lower limit from Eq. \req{eq:substitution}.

The contributions (\ref{eq:calAV123}-\ref{eq:calBA123})
consist of the terms of the form
\begin{equation}
g_0(x_i,W,M_N^2/Q^2) \; \; f(x_i)
\qquad \mbox{and} \qquad
g_0^2(x_i,W,M_N^2/Q^2) \; \; f(x_i)
\, .
\end{equation}
Analogously to
\req{eq:subfortwist3withmasses}, the former 
contribute as
\begin{eqnarray}
\left(\begin{array}{c}
 2 F_{1}^{{\rm LO}, \{V_{123}, A_{123}\}}(Q^2; \mu_F^2) \\[0.2cm] 
F_{2}^{{\rm LO},\{V_{123}, A_{123}\}}(Q^2; \mu_F^2) 
\end{array}
\right)
&:& 
\frac{1}{\lambda_1 \pi } \int_0^1 dx_k
\int_1^{(s_0+Q^2)/Q^2} dw \; e^{-(w-1) Q^2/M_B^2+M_N^2/M_B^2} \; 
\nonumber \\[0.2cm]&& \times \;
\; {\rm Im} \left[g_0(x_k, w, M_N^2/Q^2) \right] \, 
f(x_k) 
\, \widetilde{F}^{(i)}_{123}(x_k; \mu_F^2) 
\nonumber \\[0.5cm]&\to&
\frac{1}{\lambda_1} 
\int_{x_0}^{1}  dx_k  \;  
e^{-(1-x_k)Q^2/(x_k M_B^2)+x_k M_N^2/M_B^2} \quad
\nonumber \\[0.2cm]&& \times \;
\left( - \frac{1}{x_k} \right) \, f(x_k) 
 \; \widetilde{F}^{(i)}_{123} (x_k; \mu_F^2)
\, .
\label{eq:subfortwist4123withmassesI}
\end{eqnarray}
The terms with $g_0^2$ take slightly more complicated form
\begin{eqnarray}
\left(\begin{array}{c}
 2 F_{1}^{{\rm LO}, \{V_{123}, A_{123}\}}(Q^2; \mu_F^2) \\[0.2cm] 
F_{2}^{{\rm LO},\{V_{123}, A_{123}\}}(Q^2; \mu_F^2) 
\end{array}
\right)
&:& 
\frac{1}{\lambda_1 \pi } \int_{0}^{1} dx_k
\int_1^{(s_0+Q^2)/Q^2} dw \; e^{-(w-1) Q^2/M_B^2+M_N^2/M_B^2} \; 
\nonumber \\[0.2cm]&& \times \;
\; {\rm Im} \left[ g_{0}^{2}(x_k, w, M_N^2/Q^2) \right] \, 
f(x_k) \; \widetilde{F}^{(i)}_{123}(x_k; \mu_F^2) 
\nonumber \\[0.5cm]&\to&
\frac{1}{\lambda_1} 
\Big [ 
\,e^{-(s_0-M_N^2)/M_{B}^2}
\, \frac{Q^2}{Q^2+x_{0}^2 M_N^2}
\, f({x_0})
\; \widetilde{F}_{123}^{(i)}(x_k=x_0;\mu_F^2)
\nonumber\\
&&
+\frac{Q^2}{M_{B}^2}\,\int_{x_0}^{1} dx_k
\; e^{-(1-x_k)Q^2/(x_k M_B^2)+x_k M_N^2/M_B^2} 
\nonumber \\[0.2cm]&& \times \;
\, \left( \frac{1}{x_{k}^2} \right)  
\, f(x_k)
\;\widetilde{F}_{123}^{(i)}(x_k; \mu_F^2)
\Big ]
\, ,
\label{eq:subfortwist4123withmassesII}
\end{eqnarray}
where $\widetilde{F}^{(i)}_{123} \in \{\widetilde{V}^{(i)}_{123},
\widetilde{A}^{(i)}_{123}\}$. 

We note that these results are in agreement with the somewhat 
differently derived expressions from Ref. \cite{BraunLW06},
Eqs. (A.15-A.18, \cite{BraunLW06}).
We refer to that paper for higher-twist contributions.

%%%%%%%%%%%%%%%%%%%%%%%%%%%%%%%%%%%%%%%%%%%%%%%%%%%%%%%%%%%%%%%%%%%%%%%%%%%%%%%

%%%%%%%%%%%%%%%%%%%%%%%%%%%%%%%%%%%%%%%%%%%%%%%%%%%%%%%%%%%%%%%%%%%%%%%%%%%%%%%%


\begin{thebibliography}{10}

\bibitem{ChernyakZ77}
  V.~L.~Chernyak and A.~R.~Zhitnitsky,
  %``Asymptotic Behavior Of Hadron Form-Factors In Quark Model. 
  % (In Russian),''
  JETP Lett.\  {\bf 25} (1977) 510
  [Pisma Zh.\ Eksp.\ Teor.\ Fiz.\  {\bf 25} (1977) 544].
  %%CITATION = ZFPRA,25,544;%%

\bibitem{ChernyakZ80}
  V.~L.~Chernyak and A.~R.~Zhitnitsky,
  %``Asymptotics Of Hadronic Form-Factors In The Quantum Chromodynamics. (In
  %Russian),''
  Sov.\ J.\ Nucl.\ Phys.\  {\bf 31} (1980) 544
  [Yad.\ Fiz.\  {\bf 31} (1980) 1053].
  %%CITATION = YAFIA,31,1053;%%

\bibitem{Radyushkin77}
  A.~V.~Radyushkin,
  %``Deep elastic processes of composite particles in field theory and
  %asymptotic freedom,''
  Eprint is English translation of 1977 Dubna preprint P2-10717,
  hep-ph/0410276.
  %%CITATION = HEP-PH/0410276;%%

\bibitem{EfremovR80}
  A.~V.~Efremov and A.~V.~Radyushkin,
  %``Asymptotical Behavior Of Pion Electromagnetic Form-Factor In QCD,''
  Theor.\ Math.\ Phys.\  {\bf 42} (1980) 97
  [Teor.\ Mat.\ Fiz.\  {\bf 42} (1980) 147].
  %%CITATION = TMFZA,42,147;%%

\bibitem{EfremovR80a}
  A.~V.~Efremov and A.~V.~Radyushkin,
  %``Factorization And Asymptotical Behavior Of Pion Form-Factor In QCD,''
  Phys.\ Lett.\  B {\bf 94} (1980) 245.
  %%CITATION = PHLTA,B94,245;%%

\bibitem{LepageB79}
  G.~P.~Lepage and S.~J.~Brodsky,
  %``Exclusive Processes In Quantum Chromodynamics: Evolution Equations For
  %Hadronic Wave Functions And The Form-Factors Of Mesons,''
  Phys.\ Lett.\  B {\bf 87} (1979) 359.
  %%CITATION = PHLTA,B87,359;%%

\bibitem{LepageB80}
  G.~P.~Lepage and S.~J.~Brodsky,
  %{\it Exclusive Processes In Perturbative Quantum Chromodynamics},
  Phys.\ Rev.\  D {\bf 22} (1980) 2157.
  %%CITATION = PHRVA,D22,2157;%%

\bibitem{DuncanM80}
  A.~Duncan and A.~H.~Mueller,
  %``Asymptotic Behavior Of Exclusive And Almost Exclusive Processes,''
  Phys.\ Lett.\  B {\bf 90} (1980) 159.
  %%CITATION = PHLTA,B90,159;%%

\bibitem{DuncanM80a}
  A.~Duncan and A.~H.~Mueller,
  %``Asymptotic Behavior Of Composite Particle Form-Factors And The
  %Renormalization Group,''
  Phys.\ Rev.\  D {\bf 21} (1980) 1636.
  %%CITATION = PHRVA,D21,1636;%%

\bibitem{MuellerRGDH94}
  D.~M\"{u}ller, D.~Robaschik, B.~Geyer, F.~M.~Dittes and J.~Ho\v{r}ej\v{s}i,
  %``Wave functions, evolution equations and evolution kernels from light-ray
  %operators of {QCD},''
  Fortsch.\ Phys.\  {\bf 42} (1994) 101
  [hep-ph/9812448].
  %%CITATION = FPYKA,42,101;%%

\bibitem{Radyushkin96}
  A.~V.~Radyushkin,
  %``Scaling Limit of Deeply Virtual Compton Scattering,''
  Phys.\ Lett.\  B {\bf 380} (1996) 417
  [hep-ph/9604317].
  %%CITATION = PHLTA,B380,417;%%

\bibitem{Ji96}
  X.~D.~Ji,
  %``Deeply-virtual Compton scattering,''
  Phys.\ Rev.\  D {\bf 55} (1997) 7114
  [hep-ph/9609381].
  %%CITATION = PHRVA,D55,7114;%%

\bibitem{Diehl03}
  M.~Diehl,
  %``Generalized parton distributions,''
  Phys.\ Rept.\  {\bf 388} (2003) 41
  [hep-ph/0307382].
  %%CITATION = PRPLC,388,41;%%

\bibitem{BelitskyR05}
  A.~V.~Belitsky and A.~V.~Radyushkin,
  %``Unraveling hadron structure with generalized parton distributions,''
  Phys.\ Rept.\  {\bf 418} (2005) 1
  [hep-ph/0504030].
  %%CITATION = PRPLC,418,1;%%

\bibitem{ShifmanVZ79}
  M.~A.~Shifman, A.~I.~Vainshtein and V.~I.~Zakharov,
  %``QCD And Resonance Physics. Sum Rules,''
  Nucl.\ Phys.\  B {\bf 147} (1979) 385.
  %%CITATION = NUPHA,B147,385;%%

\bibitem{ShifmanVZ79a}
  M.~A.~Shifman, A.~I.~Vainshtein and V.~I.~Zakharov,
  %``QCD And Resonance Physics: Applications,''
  Nucl.\ Phys.\  B {\bf 147} (1979) 448.
  %%CITATION = NUPHA,B147,448;%%

\bibitem{NesterenkoR82}
  V.~A.~Nesterenko and A.~V.~Radyushkin,
  %``Sum Rules And Pion Form-Factor In QCD,''
  Phys.\ Lett.\  B {\bf 115} (1982) 410.
  %%CITATION = PHLTA,B115,410;%%

\bibitem{IoffeS83}
  B.~L.~Ioffe and A.~V.~Smilga,
  %``Meson Widths And Form-Factors At Intermediate Momentum Transfer In
  %Nonperturbative QCD,''
  Nucl.\ Phys.\  B {\bf 216} (1983) 373.
  %%CITATION = NUPHA,B216,373;%%

\bibitem{BelyaevK93}
  V.~M.~Belyaev and I.~I.~Kogan,
  %``Hadron form-factors in QCD,''
  Int.\ J.\ Mod.\ Phys.\  A {\bf 8} (1993) 153.
  %%CITATION = IMPAE,A8,153;%%

\bibitem{CastilloDL03}
  H.~Castillo, C.~A.~Dominguez and M.~Loewe,
  %``Electromagnetic nucleon form factors from QCD sum rules,''
  JHEP {\bf 0503} (2005) 012
  [hep-ph/0302271].
  %%CITATION = JHEPA,0503,012;%%

\bibitem{BalitskyBK89}
  I.~I.~Balitsky, V.~M.~Braun and A.~V.~Kolesnichenko,
  %``Radiative Decay Sigma+ $\to$ p gamma in Quantum Chromodynamics,''
  Nucl.\ Phys.\  B {\bf 312} (1989) 509.
  %%CITATION = NUPHA,B312,509;%%

\bibitem{ChernyakZ90}
  V.~L.~Chernyak and I.~R.~Zhitnitsky,
  %``B Meson Exclusive Decays Into Baryons,''
  Nucl.\ Phys.\  B {\bf 345} (1990) 137.
  %%CITATION = NUPHA,B345,137;%%

\bibitem{Braun98}
  V.~M.~Braun,
  %``Light-cone sum rules,''
  hep-ph/9801222.
  %%CITATION = HEP-PH/9801222;%%

\bibitem{ColangeloKh00}
  P.~Colangelo and A.~Khodjamirian,
  %{\it QCD sum rules: A modern perspective},
  hep-ph/0010175.
  %%CITATION = HEP-PH/0010175;%%

\bibitem{BraunFMS00}
  V.~Braun, R.~J.~Fries, N.~Mahnke and E.~Stein,
  %{\it Higher twist distribution amplitudes of the nucleon in QCD},
  Nucl.\ Phys.\  B {\bf 589} (2000) 381
  [Erratum-ibid.\  B {\bf 607} (2001) 433]
  [hep-ph/0007279].
  %%CITATION = NUPHA,B589,381;%%

\bibitem{BraunLMS01}
  V.~M.~Braun, A.~Lenz, N.~Mahnke and E.~Stein,
  %{\it Light-cone sum rules for the nucleon form factors},
  Phys.\ Rev.\  D {\bf 65} (2002) 074011
  [hep-ph/0112085].
  %%CITATION = PHRVA,D65,074011;%%

\bibitem{LenzWS03}
  A.~Lenz, M.~Wittmann and E.~Stein,
  %``Improved light-cone sum rules for the electromagnetic form factors 
  % of  the nucleon,''
  Phys.\ Lett.\  B {\bf 581} (2004) 199
  [hep-ph/0311082].
  %%CITATION = PHLTA,B581,199;%%

\bibitem{BraunLW06}
  V.~M.~Braun, A.~Lenz and M.~Wittmann,
  %{\it Nucleon form factors in QCD},
  Phys.\ Rev.\  D {\bf 73} (2006) 094019
  [hep-ph/0604050].
  %%CITATION = PHRVA,D73,094019;%%

\bibitem{WangWY06}
  Z.~G.~Wang, S.~L.~Wan and W.~M.~Yang,
  %``Scalar form-factor of the proton with light-cone QCD sum rules,''
  Phys.\ Rev.\  D {\bf 73} (2006) 094011
  [hep-ph/0601025].
  %%CITATION = PHRVA,D73,094011;%%

\bibitem{WangWY06a}
  Z.~G.~Wang, S.~L.~Wan and W.~M.~Yang,
  %``Axial form-factor and induced pseudoscalar 
  % form-factor of the nucleons,''
  Eur.\ Phys.\ J.\  C {\bf 47} (2006) 375
  [hep-ph/0601060].
  %%CITATION = EPHJA,C47,375;%%

\bibitem{HuangW04}
  M.~q.~Huang and D.~W.~Wang,
  %``Light-cone QCD sum rules for the semileptonic 
  % decay Lambda/b --> p l %anti-nu,''
  Phys.\ Rev.\  D {\bf 69} (2004) 094003
  [hep-ph/0401094].
  %%CITATION = PHRVA,D69,094003;%%

\bibitem{BraunLPR05}
  V.~M.~Braun, A.~Lenz, G.~Peters and A.~V.~Radyushkin,
  %``Light cone sum rules for gamma* N --> Delta transition form factors,''
  Phys.\ Rev.\  D {\bf 73} (2006) 034020
  [hep-ph/0510237].
  %%CITATION = PHRVA,D73,034020;%%


\bibitem{BraunKM99}
  V.~M.~Braun, A.~Khodjamirian and M.~Maul,
  %``Pion form factor in QCD at intermediate momentum transfers,''
  Phys.\ Rev.\  D {\bf 61} (2000) 073004
  [hep-ph/9907495].
  %%CITATION = PHRVA,D61,073004;%%

\bibitem{BijnensK02}
  J.~Bijnens and A.~Khodjamirian,
  %``Exploring light-cone sum rules for pion and kaon form factors,''
  Eur.\ Phys.\ J.\  C {\bf 26} (2002) 67
  [hep-ph/0206252].
  %%CITATION = EPHJA,C26,67;%%


\bibitem{Passek04}
  K.~Passek-Kumeri\v{c}ki,
  %``Hard exclusive processes and higher-order QCD corrections,''
  Springer Proc.\ Phys.\  {\bf 98} (2005) 399
  [hep-ph/0407122].
  %%CITATION = SPPPE,98,399;%%

\bibitem{IvanovSK04}
  D.~Y.~Ivanov, L.~Szymanowski and G.~Krasnikov,
  %``Vector meson electroproduction at next-to-leading order,''
  JETP Lett.\  {\bf 80} (2004) 226
  [Pisma Zh.\ Eksp.\ Teor.\ Fiz.\  {\bf 80} (2004) 255]
  [hep-ph/0407207].
  %%CITATION = ZFPRA,80,255;%%

\bibitem{Mueller05}
  D.~M\"{u}ller,
  %``Next-to-next-to leading order corrections to deeply virtual Compton
  %scattering: The non-singlet case,''
  Phys.\ Lett.\  B {\bf 634} (2006) 227
  [hep-ph/0510109].
  %%CITATION = PHLTA,B634,227;%%

\bibitem{KumerickiMPS06}
  K.~Kumeri\v{c}ki, D.~M\"{u}ller, K.~Passek-Kumeri\v{c}ki and A.~Sch\"{a}fer,
  %``Deeply virtual Compton scattering beyond next-to-leading order: 
  % The flavor singlet case,''
  Phys.\ Lett.\  B {\bf 648} (2007) 186
  [hep-ph/0605237].
  %%CITATION = PHLTA,B648,186;%%

\bibitem{SchmeddingY99}
  A.~Schmedding and O.~I.~Yakovlev,
%  {\it Perturbative effects in the form factor gamma gamma* $\rightarrow$ $\pi^0$ 
   % and extraction of the pion wave function from CLEO data},
  Phys.\ Rev.\  D {\bf 62} (2000) 116002
  [hep-ph/9905392].
  %%CITATION = PHRVA,D62,116002;%%

\bibitem{BaganBB98}
  E.~Bagan, P.~Ball and V.~M.~Braun,
  %``Radiative corrections to the decay B --> pi e nu and the heavy quark
  %limit,''
  Phys.\ Lett.\  B {\bf 417} (1998) 154
  [hep-ph/9709243].
  %%CITATION = PHLTA,B417,154;%%

\bibitem{BallZ01}
  P.~Ball and R.~Zwicky,
  %``Improved analysis of B --> pi e nu from QCD sum rules on the  
  % light-cone,''
  JHEP {\bf 0110} (2001) 019
  [hep-ph/0110115].
  %%CITATION = JHEPA,0110,019;%%

\bibitem{BelyaevKR93}
  V.~M.~Belyaev, A.~Khodjamirian and R.~R\"{u}ckl,
  %``QCD calculation of the B $\to$ pi, K form-factors,''
  Z.\ Phys.\  C {\bf 60} (1993) 349
  [hep-ph/9305348].
  %%CITATION = ZEPYA,C60,349;%%

\bibitem{KhodjamirianRWY97}
  A.~Khodjamirian, R.~Ruckl, S.~Weinzierl and O.~I.~Yakovlev,
  %``Perturbative QCD correction to the B --> pi transition form factor,''
  Phys.\ Lett.\  B {\bf 410} (1997) 275
  [hep-ph/9706303].
  %%CITATION = PHLTA,B410,275;%%

\bibitem{BallZ04}
  P.~Ball and R.~Zwicky,
  %``New results on B --> pi, K, eta decay formfactors from light-cone sum
  %rules,''
  Phys.\ Rev.\  D {\bf 71} (2005) 014015
  [hep-ph/0406232].
  %%CITATION = PHRVA,D71,014015;%%

\bibitem{DuplancicKMMO08}
  G.~Duplan\v{c}i\'{c}, A.~Khodjamirian, T.~Mannel, B.~Meli\'{c} and N.~Offen,
  %``Light-cone sum rules for $B \to \pi$ form factors revisited,''
  JHEP {\bf 0804} (2008) 014
  [arXiv:0801.1796 [hep-ph]].
  %%CITATION = JHEPA,0804,014;%%

\bibitem{BallB98}
  P.~Ball and V.~M.~Braun,
  %``Exclusive semileptonic and rare B meson decays in {QCD},''
  Phys.\ Rev.\  D {\bf 58} (1998) 094016
  [hep-ph/9805422].
  %%CITATION = PHRVA,D58,094016;%%

\bibitem{BallZ04a}
  P.~Ball and R.~Zwicky,
  %``B/(d,s) --> rho, omega, K*, Phi decay form factors 
  % from light-cone sum %rules revisited,''
  Phys.\ Rev.\  D {\bf 71} (2005) 014029
  [hep-ph/0412079].
  %%CITATION = PHRVA,D71,014029;%%


\bibitem{Ioffe81}
  B.~L.~Ioffe,
  %``Calculation Of Baryon Masses In Quantum Chromodynamics,''
  Nucl.\ Phys.\  B {\bf 188} (1981) 317
  [Erratum-ibid.\  B {\bf 191} (1981) 591].
  %%CITATION = NUPHA,B188,317;%%

\bibitem{Wittmann}
  M.~Wittmann,
  {\it Light-Cone Sum Rules for Nucleon Form Factors},
  Diplomarbeit,  Universit\"{a}t Regensburg, 2005.

\bibitem{Melic02}
  B.~Meli\'{c},
  %{\it Exclusive nonleptonic B decays from QCD light-cone sum ruleis},
  Lect.\ Notes Phys.\  {\bf 647} (2004) 287
  [hep-ph/0209265].
  %%CITATION = LNPHA,647,287;%%

\bibitem{Bakulev06}
  A.~P.~Bakulev,
  %{\it QCD sum rules: From quantum-mechanical oscillator to pion structure in QCD},
  Acta Phys.\ Polon.\  B {\bf 37} (2006) 3603
  [hep-ph/0610266].
  %%CITATION = APPOA,B37,3603;%%

\bibitem{MelicNP01}
  B.~Meli\'{c}, B.~Ni\v{z}i\'{c} and K.~Passek,
  %{\it BLM scale setting for the pion transition form factor},
  Phys.\ Rev.\ D {\bf 65} (2002) 053020 
  [hep-ph/0107295].
  %%CITATION = HEP-PH 0107295;%%

\bibitem{Smirnov04}
  V.~A.~Smirnov,
  %{\it Evaluating Feynman integrals},
  Springer Tracts Mod.\ Phys.\  {\bf 211} (2004) 1.
  %%CITATION = STPHB,211,1;%%

\bibitem{ChanowitzFH79}
M.~Chanowitz, M.~Furman and I.~Hinchliffe,
%{\it The Axial Current In Dimensional Regularization},
Nucl.\ Phys.\  {\bf B159}, 225 (1979).
%%CITATION = NUPHA,B159,225;%%

\bibitem{tHooftV72}
G.~'t Hooft and M.~Veltman,
%{\it Regularization And Renormalization Of Gauge Fields},
Nucl.\ Phys.\  {\bf B44}, 189 (1972).
%%CITATION = NUPHA,B44,189;%%

\bibitem{BreitenlohnerM77}
P.~Breitenlohner and D.~Maison,
%{\it Dimensionally Renormalized Green's Functions For Theories
%With Massless Particles. 1},
Commun.\ Math.\ Phys.\  {\bf 52}, 39 (1977).
%%CITATION = CMPHA,52,39;%%

\bibitem{Nyeo92}
  S.~L.~Nyeo,
  %{\it Anomalous dimensions of nonlocal baryon operators},
  Z.\ Phys.\  C {\bf 54} (1992) 615.
  %%CITATION = ZEPYA,C54,615;%%

\bibitem{KrollPK02}
  P.~Kroll and K.~Passek-Kumeri\v{c}ki,
  %{\it The two-gluon components of the eta and eta' mesons to leading-twist
%  accuracy},
  Phys.\ Rev.\ D {\bf 67} (2003) 054017 
  [hep-ph/0210045].
  %%CITATION = HEP-PH 0210045;%%

\bibitem{ChernyakOZ87}
  V.~L.~Chernyak, A.~A.~Ogloblin and I.~R.~Zhitnitsky,
  %``ON THE NUCLEON WAVE FUNCTION,''
  Sov.\ J.\ Nucl.\ Phys.\  {\bf 48} (1988) 536
  [Yad.\ Fiz.\  {\bf 48} (1988) 841].
  %%CITATION = YAFIA,48,841;%%

\bibitem{SadovnikovaDR05}
  V.~A.~Sadovnikova, E.~G.~Drukarev and M.~G.~Ryskin,
  %``Nucleon QCD sum rules with the radiative corrections,''
  Phys.\ Rev.\  D {\bf 72} (2005) 114015
  [hep-ph/0508240].
  %%CITATION = PHRVA,D72,114015;%%

\bibitem{Walkeretal94}
  R.~C.~Walker {\it et al.},
  %``Measurements of the proton elastic form-factors for 1-GeV/c**2 <= Q**2 <=
  %3-GeV/C**2 at SLAC,''
  Phys.\ Rev.\  D {\bf 49} (1994) 5671.
  %%CITATION = PHRVA,D49,5671;%%

\bibitem{Andivahisetal94}
  L.~Andivahis {\it et al.},
  %``Measurements of the electric and magnetic form-factors of the proton from
  %Q**2 = 1.75-GeV/c**2 to 8.83-GeV/c**2,''
  Phys.\ Rev.\  D {\bf 50} (1994) 5491.
  %%CITATION = PHRVA,D50,5491;%%

\bibitem{Christyetal04}
  M.~E.~Christy {\it et al.}  [E94110 Collaboration],
  %``Measurements of electron proton elastic cross sections for  0.4-(GeV/c)**2
  %< Q**2 < 5.5-(GeV/c)**2,''
  Phys.\ Rev.\  C {\bf 70} (2004) 015206
  [nucl-ex/0401030].
  %%CITATION = PHRVA,C70,015206;%%

\bibitem{Qattanetal04}
  I.~A.~Qattan {\it et al.},
  %``Precision Rosenbluth measurement of the proton elastic form factors,''
  Phys.\ Rev.\ Lett.\  {\bf 94} (2005) 142301
  [nucl-ex/0410010].
  %%CITATION = PRLTA,94,142301;%%

\bibitem{Littetal70}
  J.~Litt {\it et al.},
  %``Measurement Of The Ratio Of The Proton Form-Factors, G(E) / G(M), At High
  %Momentum Transfers And The Question Of Scaling,''
  Phys.\ Lett.\  B {\bf 31} (1970) 40.
  %%CITATION = PHLTA,B31,40;%%

\bibitem{Bergeretal71}
  C.~Berger, V.~Burkert, G.~Knop, B.~Langenbeck and K.~Rith,
  %``Electromagnetic form-factors of the proton at squared four momentum
  %transfers between 10-fm**-2 and 50 fm**-2,''
  Phys.\ Lett.\  B {\bf 35} (1971) 87.
  %%CITATION = PHLTA,B35,87;%%

\bibitem{Arrington03}
  J.~Arrington,
  %``How well do we know the electromagnetic form factors of the proton?,''
  Phys.\ Rev.\  C {\bf 68} (2003) 034325
  [nucl-ex/0305009].
  %%CITATION = PHRVA,C68,034325;%%

\bibitem{Gayouetal01}
  O.~Gayou {\it et al.}  [Jefferson Lab Hall A Collaboration],
  %``Measurement of G(E(p))/G(M(p)) in e(pol.) p --> e p(pol.) to Q**2 =
  %5.6-GeV**2,''
  Phys.\ Rev.\ Lett.\  {\bf 88} (2002) 092301
  [nucl-ex/0111010].
  %%CITATION = PRLTA,88,092301;%

\bibitem{Jonesetal99}
  M.~K.~Jones {\it et al.}  [Jefferson Lab Hall A Collaboration],
  %``G(E(p))/G(M(p)) ratio by polarization transfer in  e(pol.) p --> e
  %p(pol.),''
  Phys.\ Rev.\ Lett.\  {\bf 84} (2000) 1398
  [nucl-ex/9910005].
  %%CITATION = PRLTA,84,1398;%%

\bibitem{evanesc}
  M.~J.~Dugan and B.~Grinstein,
  %{\it On the vanishing of evanescent operators},
  Phys.\ Lett.\  B {\bf 256} (1991) 239;
  %%CITATION = PHLTA,B256,239;%%
  S.~Herrlich and U.~Nierste,
  %{\it Evanescent operators, scheme dependences and double insertions},
  Nucl.\ Phys.\  B {\bf 455} (1995) 39
  [hep-ph/9412375].
  %%CITATION = NUPHA,B455,39;%%

\bibitem{Bonneau81}
  G.~Bonneau,
  %{\it Preserving Canonical Ward Identities In Dimensional Regularization With A
%  Nonanticommuting Gamma(5)},
  Nucl.\ Phys.\  B {\bf 177} (1981) 523.
  %%CITATION = NUPHA,B177,523;%%

\bibitem{JaminL93}
  M.~Jamin and M.~E.~Lautenbacher,
  %``Tracer: Version 1.1: A Mathematica Package For Gamma Algebra In Arbitrary
  %Dimensions,''
  Comput.\ Phys.\ Commun.\  {\bf 74} (1993) 265.
  %%CITATION = CPHCB,74,265;%%

\bibitem{MelicNP01a}
  B.~Meli\'{c}, B.~Ni\v{z}i\'{c} and K.~Passek,
  %{\it A note on the factorization scale independence of the PQCD predictions  for exclusive processes},
  Eur.\ Phys.\ J.\  C {\bf 36} (2004) 453
  [hep-ph/0107311].
  %%CITATION = EPHJA,C36,453;%%

\end{thebibliography}
\end{document}